\documentclass[pra,aps,amssymb,twocolumn,superscriptaddress,noshowpacs]{revtex4-1}
\usepackage{color}
\usepackage{graphicx}
\usepackage{hyperref}
\usepackage{amsmath}
\usepackage{latexsym}
\usepackage{amssymb}
\usepackage{mathrsfs}
\usepackage{mathtools}
\usepackage{layout}
\usepackage{verbatim}
\usepackage{multirow}
\usepackage{bm}
\usepackage{amsfonts,epsfig}
\usepackage{multirow}
\usepackage{rotating}
\usepackage{textcomp}

\usepackage{tikz}

\begin{document}

\title{Coherence enhancement of Rydberg polaritons}
\date{\today}
\author{Xiao-Feng Shi}
\email[Contact author: ]{xshi@hainanu.edu.cn}
\affiliation{Center for Theoretical Physics and School of Physics and Optoelectronic Engineering, Hainan University, Haikou 570228, China}
\affiliation{School of Physics, Xidian University, Xi’an 710071, China}
\author{Yan Lu}
\affiliation{Center for Theoretical Physics and School of Physics and Optoelectronic Engineering, Hainan University, Haikou 570228, China}
\affiliation{School of Physics, Xidian University, Xi’an 710071, China}

\author{Yuechun Jiao}
\email[Contact author: ]{ycjiao@sxu.edu.cn}
\affiliation{State Key Laboratory of Quantum Optics Technologies and Devices, Institute of Laser Spectroscopy, Shanxi University, Taiyuan 030006, China}
\affiliation{Collaborative Innovation Center of Extreme Optics, Shanxi University, Taiyuan 030006, China}
\author{Jianming Zhao}
\affiliation{State Key Laboratory of Quantum Optics Technologies and Devices, Institute of Laser Spectroscopy, Shanxi University, Taiyuan 030006, China}
\affiliation{Collaborative Innovation Center of Extreme Optics, Shanxi University, Taiyuan 030006, China}

\begin{abstract}
  Quantum nonlinear optics by Rydberg polaritons can enable single-photon transistor and switch, single-photon source, and deterministic quantum information processing. A major hindrance in this study is the fast motional decoherence. Here, we devise a scheme to significantly enhance the coherence of Rydberg polariton by letting the atoms {\it remember} their velocities, or, alternatively, by {\it changing} the phase of Rydberg polariton according to its storage time. After the Rydberg polariton is prepared with a Rydberg state $|r_1\rangle$, i.e., during the storage time, two laser fields induce a transition between $|r_1\rangle$ and a nearby Rydberg state $|r_2\rangle$ via a low-lying intermediate state $\lvert f\rangle$ which is largely detuned. In particular, we find that either a $2\pi\mathbb{N}$ protocol, a $\pi$-wait-$\pi$ protocol, or a wait-$\pi$ protocol, along with an appropriate choice of $\lvert f\rangle$ can lead to a phase-coherent Rydberg polariton upon its retrieval. Importantly, the coherent transition between $|r_1\rangle$ and $|r_2\rangle$ ensures that the Rydberg polariton can block the Rydberg excitation of nearby atoms as in usual applications of Rydberg polaritons. Numerics show that the theory can nearly completely eliminate the motional dephasing, leaving Rydberg-state decay as the only fundamental channel of decoherence. This sheds light on a broad application of Rydberg-mediated quantum nonlinear optics.

\end{abstract}
\maketitle

\newpage
\section{introduction}\label{sec01}
Rydberg polariton, a superposition of $N$-atom states
\begin{eqnarray}
\frac{1}{\sqrt{N}} \sum_{j=1}^{N}|gg\cdots r_1^{(j)}\cdots ggg\rangle,\label{ss1}\nonumber
\end{eqnarray} 
with $\lvert g\rangle$ ($\lvert r_1\rangle$) a ground (Rydberg) state,
can induce quantum nonlinear optical effect on the single-photon level~\cite{Firstenberg2016,Adams2020} due to the strong blockade interaction~\cite{Saffman2010,Shi2021qst} between Rydberg atoms~\cite{Gallagh2005}, enabling optical switch and transistor~\cite{Tiarks2014,Baur2014,Gorniaczyk2014,Gorniaczyk2016}, photonic quantum gate~\cite{Tiarks2019} and entanglement~\cite{Ye2023} on the single photon level. Because of the possibility to coherently convert Rydberg polaritons to photons, single photons can be generated and manipulated on demand~\cite{Dudin2012,Peyronel2012,Firstenberg2013,Tresp2016,Distante2016,Distante2017,Busche2017,Paris-Mandoki2017,Murray2017,Ripka446,Schmidt-Eberle2020,Cantu2020,Jiao2020,Chen2021,Padron-Brito2021-2,Spong2021,Padron-Brito2021,Xu2021,PhysRevLett.128.123601,Fan2023,Fan2023oe2}. However, atomic motion generates a rapid dephasing and wipes out the quantum nature of Rydberg polariton on a microsecond timescale~\cite{Dudin2012,Baur2014,Schmidt-Eberle2020,Padron-Brito2021}.

The puzzle of the fast dephasing of Rydberg polariton originates from the motion of atoms with unpredictable velocities in real space. In an atomic gas with $N$ atoms, the creation operator of the prepared polariton is $\hat{S}^\dag = \frac{1}{\sqrt{N}}\sum_{j=1}^{N}e^{ikz_j} (|r_1\rangle\langle g|)_j,$ where $k$ is the wavevector of the polariton~\cite{Jenkins2012}, $z_j$ is the initial location of the $j$th atom, and $|r_1\rangle$ and $|g\rangle$ are the Rydberg and ground states, respectively. By using $v_j$ to label the velocity of the atom along $\mathbf{z}$ for the $j$th atom, the creation operator of  the ``phase-coherent'' Rydberg polariton is
$\hat{\mathscr{S}}^\dag =\frac{1}{\sqrt{N}} \sum_{j=1}^{N}e^{ik(z_j+v_jt)} (|r_1\rangle\langle g|)_j$. $v_j$ is random, and, it is difficult to know $v_j$ {\it a priori} for each atom $j$ in the ensemble, so inhomogeneous broadening occurs, leading to dephasing. Such dephasing has been well studied in literature. For a signal field whose transverse amplitude is a Gaussian with mode diameter $2\Sigma$, the efficiency to convert the stored Rydberg polariton to a single photon, $\eta=|\langle \hat{S}^\dag(z,t)  \hat{\mathscr{S}}(z,t) \rangle|^2$, decays in a timescale characterized by two parameters~\cite{Jenkins2012},
\begin{eqnarray}
T_2^\ast = 1/(k v), ~\zeta = 2\Sigma/v, \label{twopar}
\end{eqnarray}
where $v=\sqrt{k_\text{B}T/m}$ with $k_\text{B}$ the Boltzmann constant, $T$ is the motional temperature of the atom, and $m$ the atomic mass. For typical experimental setups with Rydberg polariton, we have $T_2^\ast\ll\zeta$, so that the retrieval efficiency decays according to~\cite{Jenkins2012}
\begin{eqnarray}
\eta &=& e^{- (t/T_2^\ast)^2}\label{eta01}
\end{eqnarray}
which has been experimentally verified in Ref.~\cite{Schmidt-Eberle2020}. Because the value of $T_2^\ast$ is usually orders of magnitude smaller than the lifetime of Rydberg states, the motional effect inducing Eq.~(\ref{eta01}) becomes the major factor determining the dephasing of Rydberg polariton though other residual issues do exist~\cite{Schmidt-Eberle2020}. This rapid dephasing has motivated alternative methods of trapping both ground and Rydberg atoms in a magic lattice~\cite{Li2013,Lampen2018,PhysRevLett.128.123601}, mapping the Rydberg state to another ground state during the storage time~\cite{Li2016}, or using BEC to create stationary polaritons~\cite{PhysRevLett.131.133001}. Recently, an interesting method by ac Stark lattice modulation was used to suppress this dephasing with an upper bound $0.06$ in its retrieval efficiency~\cite{kurzyna2024}.

The random nature of the thermal motion of atoms means that if there were similar dynamical decoupling protocols as spin-echo~\cite{Hahn1950}, quantum bang-bang~\cite{Viola1998}, or other dynamical decoupling methods~(e.g., the one in Ref.~\cite{Du2009}), i.e., if one were able to devise a similar control sequence to let each of the atoms reverse its specific velocity, then it would be possible to have the atoms return to their original position at a certain time, resulting in a coherence revival so as to quench the dephasing of Rydberg polariton. Unfortunately, it seems a forbidden task to use a global control pulse to reverse the real-space thermal motion of each atom in the ensemble.

\begin{figure}
\includegraphics[width=3.2in]
{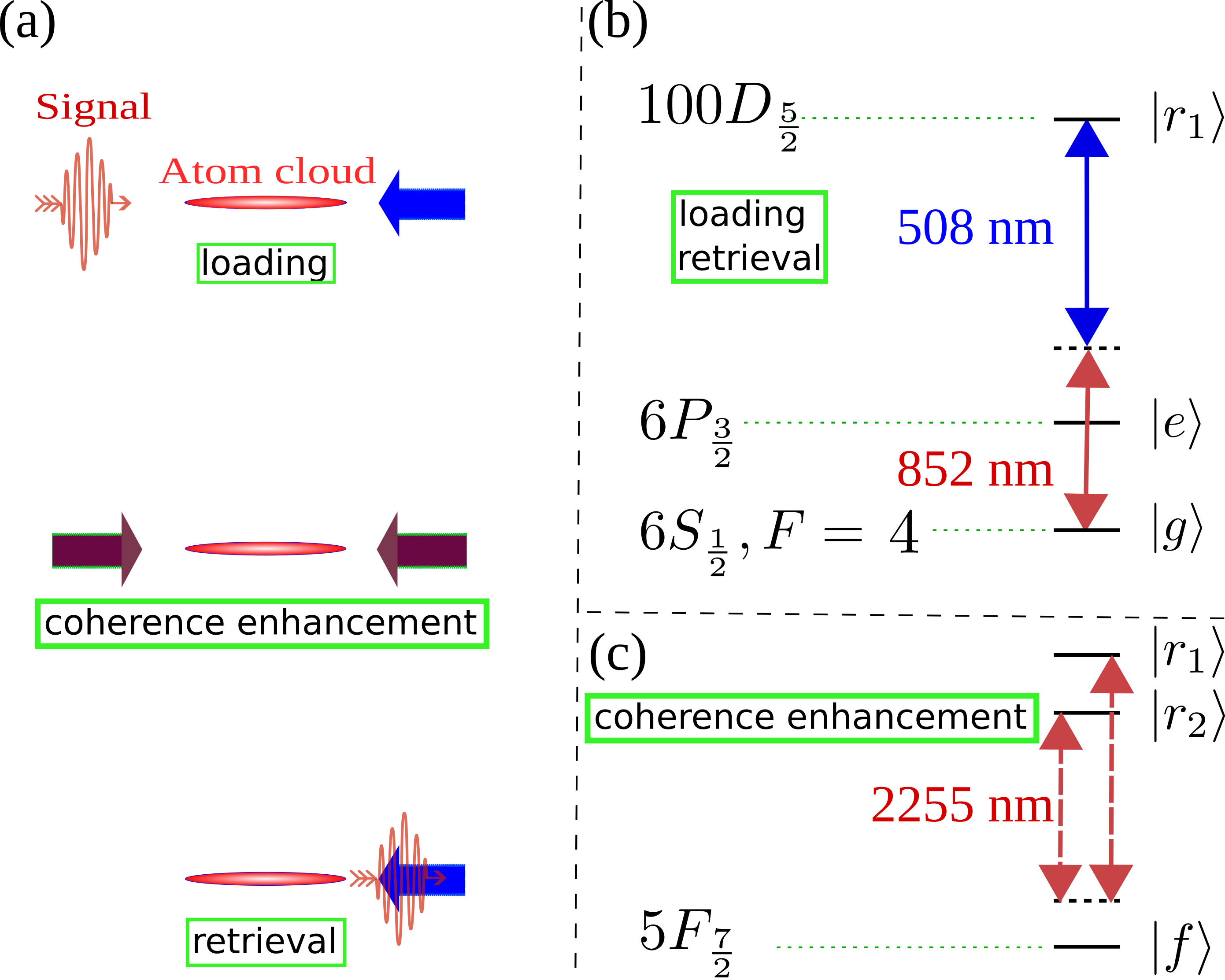}
\caption{ (a) Loading, storage, and retrieval of a single optical signal photon by using Rydberg EIT. All the fields are circularly polarized; the quantization axis is along the longitudinal axis of the atomic medium. (b) An atomic level diagram with cesium-133~(see case vii of Table~\ref{table2}). The signal field is near resonant with the transition between $|6S_{1/2},F=4,m_F=4\rangle$ and $|e\rangle\equiv|6P_{3/2},F=5,m_F=5\rangle$, and the control field is near resonant with the transition between $|r_1\rangle=|100D_{5/2},m_J=5/2,m_I=7/2\rangle$ and $|e\rangle$. (c) During the storage of the Rydberg polariton, CE laser fields are used to induce a transition between $|r_1\rangle$ and a nearby Rydberg state $|r_2\rangle$ via $5F_{7/2}$ by a $\pi$-wait-$\pi$ protocol as detailed in Sec.~\ref{sec03B}. The dephasing is suppressed when the transition between $|r_1\rangle$ and $|r_2\rangle$ is fast compared to the motion of the atoms. \label{figure01} }
\end{figure}

Here, instead of looking for a dynamical decoupling scheme as spin-echo~\cite{Hahn1950} or quantum bang-bang~\cite{Viola1998}, i.e., instead of trying to change the external state, namely, the real-space position of the Rydberg atom, we attack the dephasing by letting each Rydberg atom adjust their internal state according to their external state. This strategy can nearly completely remove the motion-induced dephasing of Rydberg polariton. The dephasing of Rydberg polariton arises because the phase of the Rydberg state does not change according to the change of the real-space position of the Rydberg atom. To let the Rydberg atom adjust its internal state according to its external state, we employ a two-photon coherence enhancement~(CE) Raman transition of wavevector $k_{\text{\tiny{CE}}}$ and Rabi frequency $\Omega_{\text{\tiny{CE}}}$ between two nearby Rydberg states via a low-lying, largely detuned intermediate state $\lvert f\rangle$, shown in Fig.~\ref{figure01}(c). By creating a velocity-dependent phase to the Rydberg atom in free flight, we devise three protocols, a $2\mathbb{N}\pi$ protocol, a $\pi$-wait-$\pi$ protocol, and a wait-$\pi$ protocol, where the latter two protocols were recently verified experimentally~\cite{Jiao2024,Li2025}. These protocols are conditioned on $\Omega_{\text{\tiny{CE}}}\gg k_{\text{\tiny{CE}}}v$, i.e., with such a condition the motional dephasing will be completely removed, leaving the radiation decay of Rydberg atoms as the only fundamental channel of dephasing. However, numerical simulation shows that with a moderate, easily realizable $\Omega_{\text{\tiny{CE}}}/2\pi=2$~MHz, the coherence time can still be significantly elongated, even to hundreds of microseconds. This enables Rydberg polariton as a practical route toward single photon quantum nonlinear optics, all-optical quantum information, and quantum networking~\cite{Chang2014,Firstenberg2016,Adams2020,shao2024}.

The theoretical protocols have two features, (i) the Rydberg component, though transitions back and forth between two different Rydberg states, preserves the Rydberg character which is a key factor for Rydberg-mediated quantum nonlinear optics, (ii) our  protocols
are compatible with various setups of neutral-atom Rydberg media. For example, in alkaline-earth-atom based Rydberg atom quantum technology~\cite{Shi2021,Shi2021pra,Chen2022,Shi2024,PhysRevA.110.012610,Omanakuttan2024}, the theory in this work can be useful by appropriately chosen intermediate state for the CE Raman transitions.

The remainder of this paper is organized as follows. In Sec.~\ref{sec02}, we review the dephasing of Rydberg polariton. In Sec.~\ref{sec03}, we detail the theory of suppressing the motional dephasing of Rydberg polariton in terms of the $2\pi\mathbb{N}$ protocol, the $\pi$-wait-$\pi$ protocol, and the wait-$\pi$ protocol. In Sec.~\ref{sec04}, we show atomic configurations suitable for our theory with $^{133}$Cs and $^{87}$Rb. In Sec.~\ref{sec4-5}, we study the feasibility of the protocols in experiments. In Sec.~\ref{sec05}, we use numerical simulation to validate the theory. In Sec.~\ref{sec06}, we give a simple photonic gate protocol based on the CE theory. Section~\ref{sec07} gives discussions and compares the theory with previous works. Section~\ref{sec08} gives a brief summary.


\section{The dephasing of Rydberg polariton and its storage}\label{sec02}
To clarify why the atomic motion leads to the dephasing of the Rydberg polariton, we will briefly review electromagnetically induced transparency (EIT)~\cite{Fleischhauer2002} in a cold atomic gas and Rydberg polariton~\cite{Gorshkov2011,Gorshkov2013,ShiJPB2016} in Sec.~\ref{sec02A}. In Sec.~\ref{sec02B}, we explain why the storage of a Rydberg polariton is necessary.

\subsection{The dephasing}\label{sec02A}
We focus on the rapid motional dephasing of a single Rydberg polariton. For the study of their dephasing from the interaction between multiple Rydberg polaritons, see, e.g.,  Refs.~\cite{Dudin2012,Bariani2012,Bariani2012pra,Gorshkov2013,Tresp2015,Li2015,Murray2016,Tian2018,Yang2019}. There has also been experimental study on the dissipative dynamics when the pulse contains a large number of photons~\cite{Bienias2020}.

We consider an adiabatic loading~\cite{Fleischhauer2002} of the signal photon with the configuration of Fig.~\ref{figure01}(b) with $^{133}$Cs as an example. For brevity, we consider the expansion of the atomic gas to be spatially uniform, assume the beam waist infinite, and ignore collisions between atoms. The longitudinal axis $\mathbf{z}$ of the atomic gas is along the quantization axis. The $^{133}$Cs atoms are prepared in the ground state $|g\rangle\equiv|6S_{1/2}, F = 4, m_F = 4 \rangle$, and coupled to a low-lying state $\lvert e\rangle$ by the signal field sent along $\mathbf{z}$, while the control field travels along $-\mathbf{z}$, as shown in the upper panel of Fig.~\ref{figure01}(a); in practice, however, the laser beams are not necessarily collinear, but can tilt away a little bit~\cite{Zhao2009}.
We assume that the atomic gas is in the interval $[0,~L]$ with a total number of atoms $N=\int_0^L n_{\text{d}}(z)dz$. The light propagation in the atomic gas can be described by a set of Heisenberg equation,
\begin{eqnarray}
\partial_t \hat{\mathcal{E}} (z,t)&=& -c \partial_z \hat{\mathcal{E}}  (z,t)+i\kappa^\ast\sqrt{n_{\text{d}}}\hat{p} (z,t) ,\nonumber\\
\partial_t \hat {p}  (z,t) &=& -\Gamma \hat {p}  (z) +i\Omega  \hat  {s}   (z,t) +i\kappa \sqrt{n_{\text{d}}}\hat {\mathcal{E}}  (z,t) + \sqrt{2\Gamma}\hat{F} ,\nonumber\\
\partial_t \hat {s}   (z,t) &=& i\Omega ^\ast \hat {p}   (z,t) , \label{Heisenberg01}
\end{eqnarray}
where $\hat{(\cdotp)}(z)$ is a coarse-grained field at $z$ ~\cite{Gorshkov2007} which is defined in Appendix~\ref{appendixA0}. Here, $\hat{\mathcal{E}}(z)$ is the annihilation operator of the light field, $\hat{p}(z)$ is the annihilation operator of the collective coherence between the ground state and $|e\rangle\equiv |6P_{1/2}, F = 4, m_F = 4\rangle$, and $\hat{s}(z)$ is the annihilation operator for the collective coherence between the ground state and the Rydberg state $|r_1\rangle$. In Eq.~(\ref{Heisenberg01}), $2\kappa \sqrt{n_{\text{d}}}$ and $2\Omega$ are the Rabi frequency for the transition between the coarse-grained states generated by $\hat{\mathcal{E}}(z)$,~$\hat{p}(z)$, and $\hat{s}(z)$, $2\Gamma$ is the linewidth of $|e\rangle$, and $\hat{F}$ is a Langevin noise operator defined in Appendix~\ref{appendixA0}. The equation of motion in Eq.~(\ref{Heisenberg01}) leads to the propagation of the signal photon with a reduced group velocity $c|\Omega|^2/\sqrt{|\Omega|^2+|\kappa|^2n_{\text{d}}}$ as detailed in Ref.~\cite{Fleischhauer2002}.

In EIT~\cite{Fleischhauer2002}, the propagation of the light field in Eq.~(\ref{Heisenberg01}) reduces to the propagation of the Rydberg dark-state polariton defined by
\begin{eqnarray}
\hat{\Psi} &=& \frac{\Omega\hat{\mathcal{E}}- \kappa \sqrt{n_{\text{d}}}\hat {s} }{\sqrt{ |\Omega |^2+|\kappa \sqrt{n_{\text{d}}}|^2 }}.
\end{eqnarray}
Thus, by using EIT, the signal photon converts to the Rydberg dark-state polariton, which has both matter part and photon part. The control field can be adiabatically switched off so that the polariton is stored as a Rydberg spin wave, or Rydberg polariton. This polariton, a Rydberg excitation distributed in the whole atomic gas, can block the transmission of another Rydberg polariton so that optical switch and transistor can be realized on the single photon level~\cite{Tiarks2014,Baur2014,Gorniaczyk2014}. It is also possible to use such a polariton to realize a phase change to a single photon~\cite{Tiarks2016,Thompson2017} and realize photon-photon entanglement~\cite{Ye2023}.

The atomic motion induces a fast dephasing of the Rydberg polariton. This is because although the prepared Rydberg polariton can be described by the creation operator
\begin{eqnarray}
\hat{S}^\dag&=& \frac{1}{\sqrt{N}}\sum_{j=1}^{N}e^{ikz_j} (|r_1\rangle\langle g|)_j ,\label{State-S1}
\end{eqnarray}
 no atom in the atomic medium is frozen in real space. In the correct creation operator of the Rydberg polariton at the moment $t>0$,
\begin{eqnarray}
\hat{\mathscr{S}}^\dag(t) &=&\frac{1}{\sqrt{N}} \sum_{j=1}^{N}e^{ik(z_j+v_jt)} (|r_1\rangle\langle g|)_j,
\end{eqnarray}
the atomic velocity $v_j$ is random, i.e., different atoms have different $v_j$, so the Rydberg polariton is no longer phase coherent, namely, $\langle\hat{S} (t) \hat{\mathscr{S}}^\dag(t) \rangle\neq1$. Such a dephasing has been analytically investigated in Ref.~\cite{Jenkins2012}, which shows that for a signal field whose transverse amplitude is a Gaussian with mode diameter $2\Sigma$, the efficiency to read out
the Rydberg polariton, $\eta=|\langle \hat{{S}} (t) \hat{\mathscr{S}}^\dag(t) \rangle|^2$, is determined by the two parameters in Eq.~(\ref{twopar}). If $T_2^\ast\ll\zeta$, the decay is with Eq.~(\ref{eta01}), but if $T_2^\ast\gg\zeta$, the Rydberg polariton decays as
\begin{eqnarray}
\eta &=& [1+ (t/\zeta)^2]^{-2}. 
\end{eqnarray}
For typical experimental setups, one has $T_2^\ast\ll\zeta$ and thus the coherence decays as in Eq.~(\ref{eta01}). In quantum nonlinear optics based on Rydberg atoms as an example, the coherence time was around $2.5~\mu$s~\cite{Dudin2012}, $0.32~\mu$s~\cite{Peyronel2012}, and $2~\mu$s~\cite{Maxwell2013} measured in early experiments, or $1~\mu$s~\cite{Tiarks2018}, $1.3~\mu$s~\cite{PhysRevLett.131.133001} in later measurements. Of course, some factors sensitive to particular experimental setups also cause extra dephasing. For example, the observed dephasing time was $2.5~\mu$s in Ref.~\cite{Dudin2012}, which was about half of the value predicted by Eq.~(\ref{eta01}); this mismatch between the experiment and the estimate in Eq.~(\ref{eta01}) possibly comes from the loss of atoms during experiments~\cite{Jiao2024}, stray electric fields~\cite{Booth2017,SaffmanElectri2023}, or Rydberg interactions between multiple-photon Rydberg polaritons~\cite{Dudin2012,Ye2023} because there could be more than one photon in the signal pulse of~\cite{Dudin2012}. But with cold enough media and when the signal photon pulse has one photon on average, Ref.~\cite{Schmidt-Eberle2020} showed that Eq.~(\ref{eta01}) can describe the dephasing with a high accuracy; see, e.g., Fig.~3 of Ref.~\cite{Schmidt-Eberle2020}.

\subsection{Storage of Rydberg polariton}\label{sec02B}
Here, we clarify the role played by the storage time. In the single photon switch studied in Ref.~\cite{Baur2014}, the signal photon is stored as a Rydberg polariton in the atomic medium, then a target signal is sent through the medium where Rydberg blockade occurs, and finally the signal photon is retrieved. This type of sequence is relevant to all-optical quantum information processing as well, where the single photon should be sent into an atomic medium and stored as a Rydberg polariton~\cite{Paredes-Barato2014,Busche2017,Ye2023}. The reason for the storage to be necessary is that it needs time for the blockade to put the interaction-induced information into the quantum system. Another reason to separate the loading stage and the interaction stage, i.e., storage stage, is that during loading of the photon as a Rydberg polariton, Rydberg interaction should be avoided so that high-fidelity loading can be achieved. If there is Rydberg interaction during the transform from light wave to Rydberg collective excitation, dissipation dominates~\cite{Gorshkov2013}.

The theoretical protocol outlined in Fig.~\ref{figure01} involves two Rydberg states $|r_1\rangle$ and $|r_2\rangle$ during the storage, but it doesn't alter the application of Rydberg polariton. Whether the Rydberg state is $|r_1\rangle$ or $|r_2\rangle$ in the Rydberg polariton, it can block the transmission of another Rydberg polariton. Thus, we can let the Rydberg state of the Rydberg polariton oscillate between two nearby Rydberg states $|r_1\rangle$ and $|r_2\rangle$ so that the Rydberg blockade between the stored Rydberg polariton and another signal photon remains. The extra thing we should bear in mind is that (i) at the end of the storage of the signal in the $2\pi\mathbb{N}$ protocol or the $\pi$-wait-$\pi$ protocol, the Rydberg state of the Rydberg polariton should be $|r_1\rangle$ so that a phase-matched retrieval of the signal photon can proceed as usual, (ii) for certain quantum information processing task, the interaction between a stored $|r_1\rangle$ or $|r_2\rangle$ Rydberg polariton and another flying photon should be of similar value when the flying photon is excited to a third Rydberg state $|R\rangle$; in principle, this can be easily satisfied by choosing Rydberg states of appropriate quantum numbers for $|r_1\rangle$ and $|r_2\rangle$ like $|r_1\rangle$ and $|r_2\rangle$ having the same principal and orbital angular momentum quantum numbers but different magnetic quantum numbers.

\section{Suppressing the dephasing}\label{sec03}
The suppression of motional dephasing of Rydberg polariton depends on choosing an appropriate low-lying intermediate state so that another Rydberg state $|r_2\rangle$ can serve as a special shelving state, and hence a rapid excitation between $|r_1\rangle$ and $|r_2\rangle$ imprints a phase change of the atomic state that is near to the correct value. As a result, the phase of each Rydberg atom in the Rydberg polariton follows after the desired value in a way as if each atom knows its velocity. The actual implementation of the theory can be with three protocols, a $2\pi\mathbb{N}$ protocol, a $\pi$-wait-$\pi$ protocol, and a wait-$\pi$ protocol. The first protocol is relatively simple in that no switch of the coherence-enhancement~(CE) laser fields is needed. The second protocol has an extra switch on and off for the CE fields, but incurs less noise of laser fields compared to the $2\pi\mathbb{N}$ protocol. The third one, namely, the wait-$\pi$ protocol, differs from the first two in that it requires retrieving the signal photon by coupling Rydberg state $|r_2\rangle$ instead of the original Rydberg state $|r_1\rangle$. The suppression of the motional dephasing depends on a condition that the phase of the Rydberg polariton upon retrieval has a value as if it is just loaded there.

\subsection{A $2\pi\mathbb{N}$ protocol}\label{sec03A}
The protocol in 
Fig.~\ref{figure01} is understood as follows. In an atomic medium, a Rydberg polariton is prepared in an s- or d-orbital Rydberg state $|r_1\rangle$ and has a wavevector of $k=2\pi/\Lambda$, where $k=2\pi|1/\lambda_2- 1/\lambda_1|$, and $\lambda_{1(2)}$ is the wavelength of the signal (control) field as in Fig.~\ref{figure01}. During the storage, CE fields are used for a Raman transition between $|r_1\rangle$ and another even-parity Rydberg state $|r_2\rangle$ that is near $|r_1\rangle$, shown in Fig.~\ref{figure01}(c). The two Raman laser fields
counter-propagate with each other. The Raman transition $|r_1\rangle \leftrightarrow |r_2\rangle$ has a wavevector $k_{\text{\tiny{CE}}}=4\pi/\lambda_3$ via a low-lying, largely detuned intermediate state $|f\rangle$, where $\lambda_3$ is the wavelength of the transition $|f\rangle\leftrightarrow|r_{1(2)}\rangle$.
Because we assume that $|r_1\rangle$ and $|r_2\rangle$ are nearby, the wavelengths for the two Raman lasers are of similar value. After performing dipole approximation and rotating wave approximation in the rotation frame, the Hamiltonian of the Raman fields is
\begin{eqnarray}
\hat{H}_{\text{\tiny{CE}}} &=& \Omega_{\text{\tiny{CE}}}e^{ik_{\text{\tiny{CE}}}( z_j+v_jt)}|r_1\rangle\langle r_2|/2+\text{H.c.},\label{infraH}
\end{eqnarray}
where we set $t=0$ at the beginning of the storage stage, and H.c. denotes Hermitian conjugate. Here the effective Rabi frequency $\Omega_{\text{\tiny{CE}}}$ is assumed to be real for brevity.

As analytically derived in Appendix A of Ref.~\cite{Shi2020}, if $|\Omega_{\text{\tiny{CE}}}|\gg k_{\text{\tiny{CE}}}v$, the Rydberg state of a certain atom $j$ in the atomic medium oscillates rapidly between $|r_1\rangle$ and $|r_2\rangle$, where the quantum oscillation follows such a pattern: when the transition $|r_1\rangle\rightarrow|r_2\rangle$ occurs from $t=0$ to $t=\pi/\Omega_{\text{\tiny{CE}}}$, the change of the phase of the Rydberg state is about $-\pi/2-k_{\text{\tiny{CE}}}[z_j+v_j\pi/(2\Omega_{\text{\tiny{CE}}})]$; during the subsequent transition $|r_2\rangle\rightarrow|r_1\rangle$ from $t=\pi/\Omega_{\text{\tiny{CE}}}$ to $t=2\pi/\Omega_{\text{\tiny{CE}}} $, the change of the phase of the Rydberg state is $-\pi/2+k_{\text{\tiny{CE}}}v_j\pi/(2\Omega_{\text{\tiny{CE}}} )+k_{\text{\tiny{CE}}}[z_j+v_j\pi/\Omega_{\text{\tiny{CE}}}]$, where the last term is due to that the phase for the initial Rabi frequency has a position-dependent term at the beginning of second half of the Rabi cycle. So, at the end of the Rabi cycle at $t=t_{\text{\tiny{CE}}}\equiv 2\pi/\Omega_{\text{\tiny{CE}}}$, the phase change of atom $j$ is $k_{\text{\tiny{CE}}}v_jt_{\text{\tiny{CE}}}/2-\pi$. If we set the following condition between the time for the CE laser excitation and the time of the Rydberg polariton storage,
\begin{eqnarray}
t_{\text{s}}/t_{\text{\tiny{CE}}} &=& k_{\text{\tiny{CE}}}/(2k),\label{condition01}
\end{eqnarray}
 then $k_{\text{\tiny{CE}}}v_jt_{\text{\tiny{CE}}}/2-\pi$ can be written as $kv_jt_{\text{s}}-\pi $. Note that the phase $-\pi$ is present for any $j$, thus is trivial, and the above result is valid for any $j$. In other words, the CE fields induce a uniform, $j$-dependent phase to each atom $j$ in a way that the atoms know their specific velocities and accurately adjust the phases of their wavefunctions accordingly. Since 
the CE laser fields should 
 be switched off before the retrieval of the Rydberg polariton, it is necessary to impose  $t_{\text{s}}/t_{\text{\tiny{CE}}}\geq1$, so that
\begin{eqnarray}
 k_{\text{\tiny{CE}}}\geq 2k,
\end{eqnarray}
or, in other words, we must make sure
\begin{eqnarray}
1/\lambda_3\geq|1/\lambda_1- 1/\lambda_2|,
\end{eqnarray}
which is a condition that can be easily satisfied as shown later.

The above analyses are based on one Rabi cycle. If $t_{\text{\tiny{CE}}} =2\pi\mathbb{N}/\Omega_{\text{\tiny{CE}}} $, where $\mathbb{N}$ is an integer, then the phase of the Rydberg-atom wavefunction in the Rydberg polariton at the end of the excitation of the CE fields, or at the end of the storage of Rydberg polariton, becomes $kz_j+k_{\text{\tiny{CE}}}v_jt_{\text{\tiny{CE}}}/2-\mathbb{N}\pi$, which can be written as $kz_j+ kv_jt_{\text{s}}$ plus a trivial phase in the condition of Eq.~(\ref{condition01}). In conclusion, the $2\pi\mathbb{N}$ protocol is implemented with the following steps.\newline\newline
(1) Load the Rydberg polariton. At the end of this step, we suppose $t=0$ and we have
\begin{eqnarray}
\lvert s_1\rangle &=&\frac{1}{\sqrt{N}} \sum_{j=1}^{N}e^{ik z_j } |gg\cdots r_1^{(j)}\cdots ggg\rangle,\label{protocol1-s1}
\end{eqnarray}
where $j$ labels the atoms, with $j=1,2,3,\cdots, N-2,N-1,N$, and $N$ is the total number of atoms in the ensemble.\newline
(2) Leave a gap time from $t=0$ to $t=t_1$ so as to switch the laser fields. This step essentially does nothing but may exist in experiments since the switch of laser fields needs time. \newline
(3) Switch on $\Omega{\text{\tiny{CE}}}$ for a $2\mathbb{N}\pi$ pulse to excite the Rydberg state between $\lvert r_1\rangle$ and $\lvert r_2\rangle$ for $\mathbb{N}$ cycles, i.e., with a pulse duration of $t_{\text{\tiny{CE}}} = \frac{2\mathbb{N}\pi}{\Omega_{\text{\tiny{CE}}}}$. This step spans the duration $[t_1,~t_1+t_{\text{\tiny{CE}}}]$. At the end of this step, the Rydberg component is still at $|r_1\rangle$ and the state is
\begin{eqnarray}
\lvert s_2\rangle &=&\frac{1}{\sqrt{N}} \sum_{j=1}^{N}e^{ik z_j+ik_{\text{\tiny{CE}}}v_jt_{\text{\tiny{CE}}}/2-i\mathbb{N}\pi } |gg\cdots r_1^{(j)}\cdots ggg\rangle,\nonumber\\\label{protocol1-s2}
\end{eqnarray}
where the detailed information about the phase change in each of the $\mathbb{N}$ Rabi cycle is given around Eq.~(\ref{condition01}). \newline
(4) Leave a gap time of
\begin{eqnarray}
&&t_{\text{s}} - \left(t_1+\frac{2\mathbb{N}\pi}{\Omega_{\text{\tiny{CE}}}}\right)\nonumber\\
&=&  \left( 1- \frac{2k} {k_{\text{\tiny{CE}}}}  \right) t_{\text{s}} -t_1,
\end{eqnarray}
where the first and second lines above are related to each other via
\begin{eqnarray}
&&kt_{\text{s}} =  k_{\text{\tiny{CE}}}\frac{\mathbb{N}\pi}{\Omega_{\text{\tiny{CE}}}}.\label{p1-condition}
\end{eqnarray}
This step spans the interval $[t_1+\frac{2\mathbb{N}\pi}{\Omega_{\text{\tiny{CE}}}},~t_{\text{s}}]$, and the internal state of the atomic state is still Eq.~(\ref{protocol1-s2}), which can also be written as
\begin{eqnarray}
\lvert s_2\rangle &=&\frac{1}{\sqrt{N}} \sum_{j=1}^{N}e^{ik z_j+ikv_jt_{\text{s}}-i\mathbb{N}\pi } |gg\cdots r_1^{(j)}\cdots ggg\rangle,\nonumber\label{protocol1-s3}
\end{eqnarray}
which shows that the Rydberg polariton won't be phase coherent until the moment $t=t_{\text{s}}$, which is why we shall have this gap time. \newline
(5) Retrieve the Rydberg polariton at $t=t_{\text{s}}$.

Notice that the Rabi frequency $\Omega_{\text{\tiny{CE}}}$ is built from a largely detuned state, thus can not be very large. However, our theory depends on the condition $|\Omega_{\text{\tiny{CE}}}|\gg |k_{\text{\tiny{CE}}}v|$, thus we prefer a large Rabi frequency. This is why we prefer to use higher $|f\rangle$ states since the dipole matrix element between $|r_{1(2)}\rangle$ and $|f\rangle$ decays rapidly along with the energy separation between them. To show that the theory is applicable even if $|\Omega_{\text{\tiny{CE}}}|\gg |k_{\text{\tiny{CE}}}v|$ is not strictly satisfied, however, we suppose $\Omega_{\text{\tiny{CE}}}/2\pi \sim 2$~MHz for later assessment with typical atomic levels.

As studied later, numerical simulation shows that for $\Omega_{\text{\tiny{CE}}}/2\pi \sim 2$~MHz, application of our protocols can elongate the $1/e$ coherence time to several times $100~\mu$s in typical setups, which is 
comparable to the Rydberg-state decay time. So, the complete coherence time should also include the Rydberg-state decay. It is noted that both $|r_1\rangle$ and $|r_2\rangle$ are Rydberg states, so the Rydberg polariton retains the character of a collective Rydberg excitation, and thus our scheme is compatible with most applications of Rydberg polariton in single-photon quantum optics~\cite{Tiarks2014,Baur2014,Gorniaczyk2014,Schmidt-Eberle2020,Spong2021,Padron-Brito2021,Xu2021,Busche2017,Tiarks2019,Jiao2020,PhysRevLett.128.123601,Ye2023,Srakaew2023}.

Note that Eq.~(\ref{condition01}) is a necessary condition for the $2\pi\mathbb{N}$ protocol, yet the protocol also needs $|\Omega_{\text{\tiny{CE}}}|/|k_{\text{\tiny{CE}}}v|\rightarrow\infty$, which is impossible. In practice, the population oscillation between $|r_1\rangle$ and $|r_2\rangle$ in each Rabi cycle is not complete, and, more importantly, more Rabi cycles incur heavier laser noise, so that a larger $\mathbb{N}$ is accompanied with a larger error. Motivated by this, below, we study protocols in Sec.~\ref{sec03B} to boost the coherence of Rydberg polariton in which effectively only one Rabi cycle between $|r_1\rangle$ and $|r_2\rangle$ is needed in the compromise that a gap time between the first half Rabi cycle and the second is used. Further, in Sec.~\ref{sec03C} we will show that only half a Rabi cycle is necessary if two sets of coupling lasers are available, one for loading the signal photon and another for retrieving the signal photon.

\subsection{A $\pi$-wait-$\pi$ protocol}\label{sec03B}
Note that our protocol can be of various forms, depending on specific experimental circumstances. For example, when laser noise or electric field noise come in, more Rabi cycles between $\lvert r_1\rangle$ and $\lvert r_2\rangle$ can bring unnecessary noise which hampers the applicability of the theory~\cite{Jiao2024}. Therefore, an alternative means is to use effectively only one Rabi cycle between $\lvert r_1\rangle$ and $\lvert r_2\rangle$ in the following way.\newline\newline
(1) Load the Rydberg polariton. At the end of this step, we suppose $t=0$ and the state is the same to Eq.~(\ref{protocol1-s1}).\newline
(2) There may be a gap time from $t=0$ to $t=t_1$ so as to switch the laser fields. This step essentially does nothing and in principle can be as short as possible.\newline
(3) Switch on $\Omega{\text{\tiny{CE}}}$ for a $\pi$ pulse to excite the Rydberg state from $\lvert r_1\rangle$ to $\lvert r_2\rangle$ during the time $[t_1,~t_1+\frac{\pi}{\Omega_{\text{\tiny{CE}}}}]$. The state becomes,
\begin{eqnarray}
-i\frac{1}{\sqrt{N}} \sum_{j=1}^{N}e^{ik z_j-ik_{\text{\tiny{CE}}}(z_j+v_jt_1+v_j\frac{\pi}{2\Omega_{\text{\tiny{CE}}}} )   } |gg\cdots r_2^{(j)}\cdots ggg\rangle,\nonumber\label{protocol2-s3}
\end{eqnarray}
where the factor $-i$ is due to the phase change $-\pi/2$ in a $\pi$ pulse.\newline
(4) Wait for the following time,
\begin{eqnarray}
t_{\text{\tiny{wait}}}=\frac{t_{\text{s}} k}{k_{\text{\tiny{CE}}}} - \frac{\pi}{\Omega_{\text{\tiny{CE}}}},\label{protocol2}
\end{eqnarray}
i.e., wait from $t=t_1+\frac{\pi}{\Omega_{\text{\tiny{CE}}}} $ to $t=t_1+\frac{\pi}{\Omega_{\text{\tiny{CE}}}}+t_{\text{\tiny{wait}}}$.\newline
(5) Use the reverse of step (3), namely, switch on $\Omega{\text{\tiny{CE}}}$ for a $\pi$ pulse to excite the Rydberg state from $\lvert r_2\rangle$ to $\lvert r_1\rangle$. This step spans the duration $[t_1+\frac{\pi}{\Omega_{\text{\tiny{CE}}}}+t_{\text{\tiny{wait}}},~ t_1+\frac{2\pi}{\Omega_{\text{\tiny{CE}}}}+t_{\text{\tiny{wait}}} ]$, and the wavefunction becomes,
\begin{eqnarray}
-\frac{1}{\sqrt{N}} \sum_{j=1}^{N}e^{ik z_j+ik_{\text{\tiny{CE}}} v_j\left(t_{\text{\tiny{wait}}}+\frac{\pi}{\Omega_{\text{\tiny{CE}}}} \right) } |gg\cdots r_1^{(j)}\cdots ggg\rangle,\nonumber\label{protocol2-s3}
\end{eqnarray}
where the phase term can also be written as $e^{ik z_j+ik_{\text{\tiny{CE}}} v_jt_{\text{s}}   }$ according to Eq.~(\ref{protocol2})
\newline
(6) Leave a gap time
\begin{eqnarray}
t_{\text{\tiny{gap}}}&=&t_{\text{s}} - \left( t_1+\frac{2\pi}{\Omega_{\text{\tiny{CE}}}}+t_{\text{\tiny{wait}}}\right)\nonumber\\
&=& \frac{ k_{\text{\tiny{CE}}}-k}{k_{\text{\tiny{CE}}}}t_{\text{s}}- \frac{\pi}{\Omega_{\text{\tiny{CE}}}} -t_1.\label{prot2gap}
\end{eqnarray}
(7) Retrieve the Rydberg polariton at $t=t_{\text{s}}$.
\newline\newline
The above seven-step $\pi$-wait-$\pi$ protocol works in a similar way as in the $2\pi\mathbb{N}$ protocol of Sec.~\ref{sec03A}. In brief, in the case of $|\Omega_{\text{\tiny{CE}}}|\gg k_{\text{\tiny{CE}}}v$, during the transition $|r_1\rangle\rightarrow|r_2\rangle$ in step (3), the change of the phase of the Rydberg state is about $-\pi/2-k_{\text{\tiny{CE}}}[z_j+v_j\pi/(2\Omega_{\text{\tiny{CE}}})]$; during step (4), the phase of the atomic state does not change, but the atom flies for a distance $v_jt_{\text{\tiny{wait}}}$ along the propagation direction of the CE fields; during the subsequent transition $|r_2\rangle\rightarrow|r_1\rangle$ in step (5), the change of the phase of the Rydberg state is $-\pi/2+k_{\text{\tiny{CE}}}v_j\pi/(2\Omega_{\text{\tiny{CE}}} )+k_{\text{\tiny{CE}}}[z_j+v_j(t_{\text{\tiny{wait}}}+\pi/\Omega_{\text{\tiny{CE}}})]$, where the last term arises from the fact that the phase of the Rabi frequency at the beginning of step (5) has a different term $k_{\text{\tiny{CE}}}v_j(t_{\text{\tiny{wait}}}+ \pi/\Omega_{\text{\tiny{CE}}})$ compared to that of step (3). At the end of step (5), the state restores to $\lvert r_1\rangle$ with a net phase change
\begin{eqnarray}
&&k_{\text{\tiny{CE}}}v_j\left(t_{\text{\tiny{wait}}}+\frac{\pi}{\Omega_{\text{\tiny{CE}}}}\right)-\pi,
\end{eqnarray}
which can be written as $kv_jt_{\text{s}}-\pi $ iff $kt_{\text{s}} = k_{\text{\tiny{CE}}}\left(t_{\text{\tiny{wait}}}+\frac{\pi}{\Omega_{\text{\tiny{CE}}}}\right) $, i.e., iff Eq.~(\ref{protocol2}). Because the storage time should be larger than the sum of the durations of steps (2,3,4,5,6), it is necessary to have the following condition
\begin{eqnarray}
t_{\text{s}}&\geq& \frac{\pi}{\Omega_{\text{\tiny{CE}}}} \frac{ k_{\text{\tiny{CE}}} }{k},\nonumber\\
t_{\text{s}}&\geq& (t_1+\frac{\pi}{\Omega_{\text{\tiny{CE}}}} )\frac{k_{\text{\tiny{CE}}} }{k_{\text{\tiny{CE}}}-k},
\end{eqnarray}
which, because $t_1$ can in principle approach zero, can alternatively be written as,
\begin{eqnarray}
1-\frac{1}{t_{\text{s}}} \frac{\pi}{\Omega_{\text{\tiny{CE}}}} &\geq& \frac{k}{ k_{\text{\tiny{CE}}} }\geq \frac{1}{t_{\text{s}}} \frac{\pi}{\Omega_{\text{\tiny{CE}}}}.
\end{eqnarray}
Because $t_{\text{s}}$ can be as small as $\frac{2\pi}{\Omega_{\text{\tiny{CE}}}}$, the above condition means that a necessary condition for the $\pi$-wait-$\pi$ protocol to work is to have
\begin{eqnarray}
 k_{\text{\tiny{CE}}} \geq k .\label{protocol2condition}
\end{eqnarray}

Compared to the $2\pi\mathbb{N}$ protocol, (i) the $\pi$-wait-$\pi$ protocol can have a large $t_{\text{s}}$ with only two $\pi$ pulses, or, effectively one Rabi cycle between $\lvert r_1\rangle$ and $\lvert r_2\rangle$, which can potentially suppress noise from lasers and noise fields such as unknown magnetic or electric fields; (ii) Eq.~(\ref{protocol2condition}) means that it is possible to use configurations with smaller $k_{\text{\tiny{CE}}}$ compared to the $2\pi\mathbb{N}$ protocol, so that a smaller $|\Omega_{\text{\tiny{CE}}}|$ is enough to satisfy $|\Omega_{\text{\tiny{CE}}}|\gg k_{\text{\tiny{CE}}}v$. However, the atomic medium always has atoms with large velocities, so that if a level configuration is available to have a smaller $k_{\text{\tiny{CE}}}$, it can exhibit stronger coherence enhancement effect.

\begin{figure*}
\includegraphics[width=6.5in]
{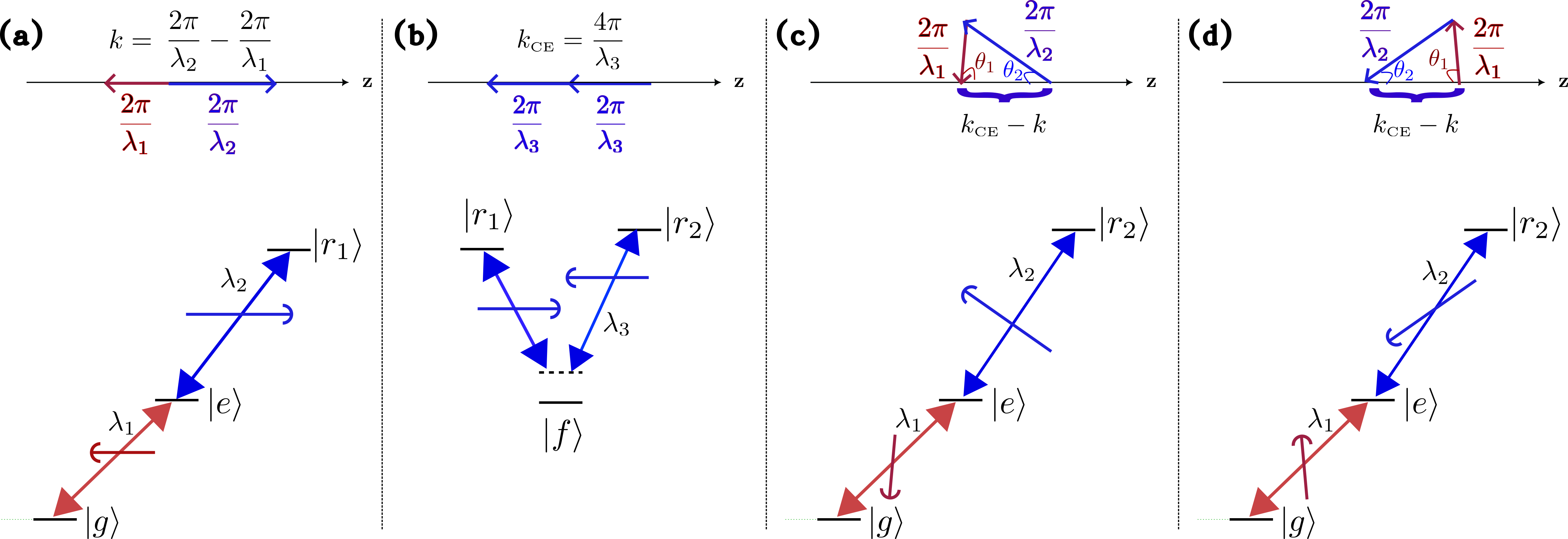}
\caption{ Illustration of the direction of light propagation and wavevectors for the wait-$\pi$ protocol if $\lambda_2<\lambda_1$. (a) During the loading of the signal photon of wavelength $\lambda_1$, the coupling laser of wavelength $\lambda_2$ counterpropagates with respect to the signal. The wave vector of the Rydberg polariton is $2\pi(1/\lambda_2-1/\lambda_1)\mathbf{z}$ when $\lambda_2<\lambda_1$. Arrows with round~(sharp) heads denote directions of the laser fields~(wavevectors). (b) After the wait time $t_{\text{\tiny{wait1}}}$, a $\pi$ pulse for the transition from $|r_1\rangle$ to $|r_2\rangle$ is realized by two laser fields, one addressing $|r_1\rangle\rightarrow|f\rangle$ with laser propagating along $\mathbf{z}$, the other addressing $|f\rangle\rightarrow|r_2\rangle$ with laser propagating along $-\mathbf{z}$. When $\Omega_{\text{\tiny CE}}\gg k_{\text{\tiny CE}}v$~\cite{Shi2020}, the Rydberg state $e^{ik z_j } | r_1^{(j)}\rangle$ of the $j$-th atom in the ensemble becomes $-i e^{i(k-k_{\text{\tiny{CE}}})z_j-ik_{\text{\tiny{CE}}} v_j(\frac{\pi}{×2 \Omega_{\text{\tiny{CE}}}} +t_{\text{\tiny{wait1}}})}| r_2^{(j)}\rangle$. (c,d) After another wait time denoted in Eq.~(\ref{protocol3-2}), a control laser for addressing $|e\rangle\leftrightarrow|r_2\rangle$ along a direction denoted in (c), i.e., tilted from the $-\mathbf{z}$ by $\theta_2$, retrieves the signal field which will propagate along a direction denoted by the red arrow in (c), or alternatively (d). The angles $\theta_1$ and $\theta_2$ are determined by the three vectors of lengths $2\pi/\lambda_1, 2\pi/\lambda_2, 2\pi(2/\lambda_3-1/\lambda_2+1/\lambda_1)$. A case that can be easily understood is when $\lvert e\rangle=\lvert f\rangle$, then $\lambda_2=\lambda_3$, leading to $\theta_1=\theta_2=0$, which means that the coupling laser during the retrieval shall be opposite to that during the loading, while the signal field comes out in the same direction as the incoming signal. When $\theta_1$ is near $\pi$, the signal photon will come out in a direction nearly opposite to that of the loaded signal photon.      \label{figure-wait-pi-direction} }
\end{figure*}

\subsection{A wait-$\pi $ protocol}\label{sec03C}
A third protocol for CE is to use only one $\pi$ pulse to excite $\lvert r_1\rangle$ to $\lvert r_2\rangle$, where $\lvert r_2\rangle$ differs from $\lvert r_1\rangle$ only by their principal quantum numbers, and then retrieve the Rydberg polariton at an appropriate time. The wait-$\pi$ protocol differs from those in Secs.~\ref{sec03A} and~\ref{sec03B} in (i) the coupling laser field at the retrieval shall couple
the intermediate state with $\lvert r_2\rangle$ instead of with the $\lvert r_1\rangle$ state; (ii) the direction of the retrieved signal in general will not be along the direction of the incoming signal, and, as shown below, the retrieved signal can even come out in a direction nearly opposite to that of the incoming signal, which means that the wait-$\pi$ protocol can potentially enable single-photon routers. For brevity, however, in Fig.~\ref{figure-wait-pi-direction} we will use a relatively exaggerated example to denote this where we have defined angles $\theta_{1(2)}$ to highlight the propagation directions for the fields. Usually, however, a case easy to handle in experiment is with $(\theta_1,\theta_2)\approx (0, 0)$, which is easily satisfied as shown later in Table~\ref{table3}.

\begin{figure*}
\includegraphics[width=6.5in]
{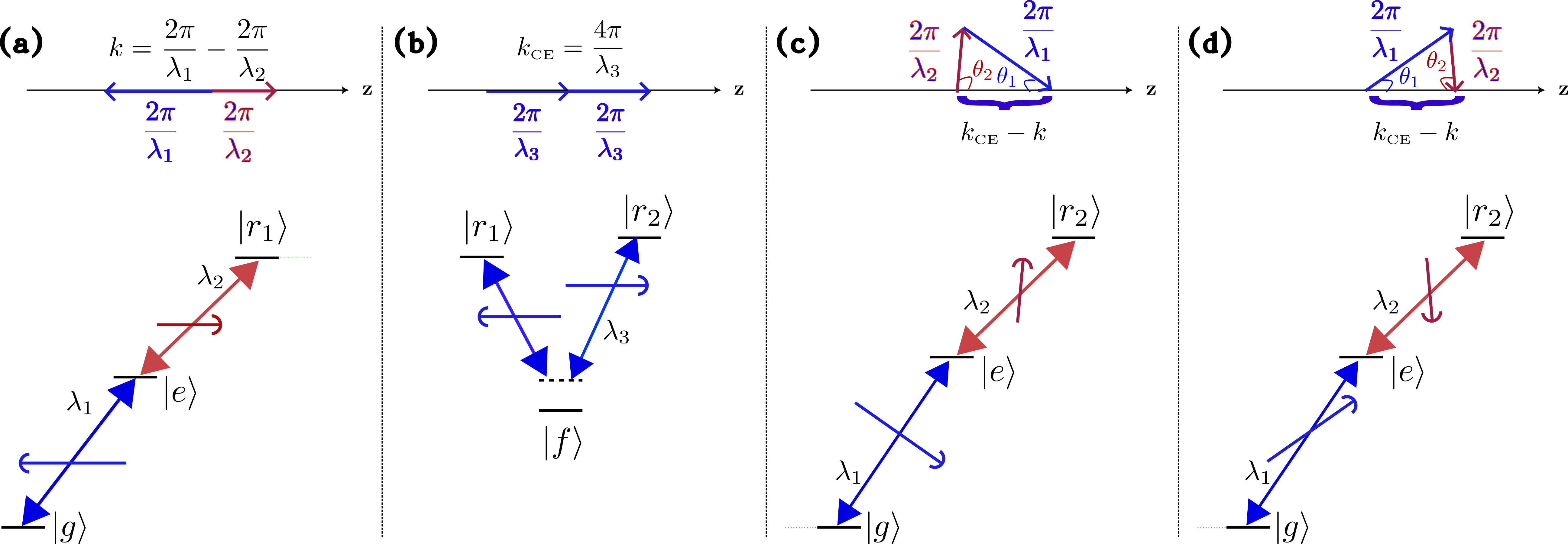}
\caption{ Illustration of the direction of light propagation and wavevectors for the wait-$\pi$ protocol if $\lambda_2>\lambda_1$. The meanings of the symbols here are identical to the corresponding ones in Fig.~\ref{figure-wait-pi-direction}. When $\theta_1$ is near $0$, the signal photon will come out in a direction nearly opposite to that of the loaded signal photon.   \label{figure-wait-pi-direction-2} }
\end{figure*}

In particular, the $\pi$ protocol proceeds in the following way.\newline
(1) Load the Rydberg polariton. At the end of this step, we suppose $t=0$.\newline
(2) Wait for the following time,
\begin{eqnarray}
t_{\text{\tiny{wait1}}}=t_{\text{s}}\left(1- \frac{  k}{k_{\text{\tiny{CE}}}}\right) - \frac{\pi}{2\Omega_{\text{\tiny{CE}}}},\label{protocol3-1}
\end{eqnarray}
This step essentially does nothing though it is a necessary step; in other words, we must ensure such an idle time.\newline
(3) Switch on $\Omega{\text{\tiny{CE}}}$ for a $\pi$ pulse to excite the Rydberg state from $\lvert r_1\rangle$ to $\lvert r_2\rangle$ during the time $[t_{\text{\tiny{wait1}}},~t_{\text{\tiny{wait1}}}+\frac{\pi}{\Omega_{\text{\tiny{CE}}}}]$.\newline
(4) Wait for the following time,
\begin{eqnarray}
t_{\text{\tiny{wait2}}}=\frac{t_{\text{s}} k}{k_{\text{\tiny{CE}}}}  - \frac{\pi}{2\Omega_{\text{\tiny{CE}}}},\label{protocol3-2}
\end{eqnarray}
i.e., wait from $t=t_{\text{s}} - t_{\text{\tiny{wait2}}} $ to $t= t_{\text{\tiny{s}}}$. Note that this second wait time is used to ensure that we do the retrieval at the predetermined moment. \newline
(5) Retrieve the Rydberg polariton at $t=t_{\text{s}}$.
\newline\newline
The wait-$\pi$ protocol is understood as follows. At the end of step (1), the wavefunction of the prepared Rydberg polariton in an $N$-atom ensemble is
\begin{eqnarray}
\lvert s_1\rangle &=&\frac{1}{\sqrt{N}} \sum_{j=1}^{N}e^{ik z_j } |gg\cdots r_1^{(j)}\cdots ggg\rangle.\label{protocol3-3}
\end{eqnarray}
At the end of step (3), the wavefunction of the Rydberg polariton becomes,
\begin{eqnarray}
\lvert s_2\rangle &=&\frac{-i}{\sqrt{N}} \sum_{j=1}^{N}e^{i(k-k_{\text{\tiny{CE}}})z_j-ik_{\text{\tiny{CE}}} v_j(\frac{\pi}{×2\Omega_{\text{\tiny{CE}}}} +t_{\text{\tiny{wait1}}})}\nonumber\\
&&\times|gg\cdots r_2^{(j)}\cdots ggg\rangle. \label{protocol3-4}
\end{eqnarray}
The state in Eq.~(\ref{protocol3-4}) does not change even to the onset of step (5); here we only write down the internal state of the atom while the position of the atom in real space is not specified in the wavefunction. A coherent retrieval of the signal photon can occur when
\begin{eqnarray}
 k-k_{\text{\tiny{CE}}}&=&-k_{\text{\tiny{CE}}} \frac{ \frac{\pi}{2\Omega_{\text{\tiny{CE}}}} +t_{\text{\tiny{wait1}}} }{×t_{\text{\tiny{s}}} },\nonumber\\
\frac{2\pi}{×\lambda_1}+ \frac{2\pi}{×\lambda_2}  &\geq&  |k-k_{\text{\tiny{CE}}}|\geq k,
 \label{protocol3-5}
\end{eqnarray}
where the second condition above is due to that we would use a retrieval coupling laser whose frequency is very near to that used during the loading stage, so when we tilt the coupling laser a little bit, the effective wavevector of the Rydberg polariton during the retrieval is in the interval $[k,~\frac{2\pi}{×\lambda_1}+ \frac{2\pi}{×\lambda_2} ]$. When the conditions in Eq.~(\ref{protocol3-5}) are met, Eq.~(\ref{protocol3-4}) can be written as
\begin{eqnarray}
\lvert s_2\rangle &=&\frac{-i}{\sqrt{N}} \sum_{j=1}^{N}e^{i(k-k_{\text{\tiny{CE}}}) (z_j + v_j t_{\text{\tiny{s}}})} |gg\cdots r_2^{(j)}\cdots ggg\rangle. \nonumber\\
\label{protocol3-6}
\end{eqnarray}
Thus, at the retrieval, the control laser field shall be sent along a certain direction specified by the three sides of the triangle in Fig.~\ref{figure-wait-pi-direction}(c). In this case, due to the collective radiation, the retrieved signal will come out from a direction determined by the three vectors in Fig.~\ref{figure-wait-pi-direction}(c). Alternatively, the configuration in Fig.~\ref{figure-wait-pi-direction}(d) can also be used. However, the illustration in Fig.~\ref{figure-wait-pi-direction} is based on Fig.~\ref{figure01} where $\lambda_2<\lambda_1$. If $\lambda_2>\lambda_1$, then the situation changes to Fig.~\ref{figure-wait-pi-direction-2} for the wait-$\pi$ protocol.

The first condition in Eq.~(\ref{protocol3-5}) means that $k_{\text{\tiny{CE}}}>k$, and the second one requires
\begin{eqnarray}
k+\frac{2\pi}{×\lambda_1}+ \frac{2\pi}{×\lambda_2} \geq k_{\text{\tiny{CE}}}&\geq&2k ,
\end{eqnarray}
which can be expanded as the following two conditions
\begin{eqnarray}
k_{\text{\tiny{CE}}}&\geq&2k ,
\end{eqnarray}
and
\begin{eqnarray}
\frac{1}{×\lambda_1}+ \frac{1}{×\lambda_2} +\left|  \frac{1}{×\lambda_1}- \frac{1}{×\lambda_2}\right|  &\geq&\frac{2}{×\lambda_3} .\label{waitPiCondition}
\end{eqnarray}
The above equations mean that the general condition for the wait-$\pi$ protocol to be realizable is the same as the $2\mathbb{N}\pi$ protocol if the condition in Eq.~(\ref{waitPiCondition}) is met.

\subsection{Fine-tuning of times for a high-efficiency readout}\label{sec03D}
The protocols shown in Secs.~\ref{sec03A},~\ref{sec03B}, and~\ref{sec03C} depend on fine tuning of the pulse duration and wait duration. In practice, however, laser pulses have about 10-nm
rise and fall edges,
which effectively makes the picture of the constant Rabi frequencies assumed in the analyses inaccurate.

The most important timing in the protocols is about the population transfer between $|r_1\rangle$ and $|r_2\rangle$ because a correct $\pi$ pulse causing full population transfer can result in the desired phase accumulation. The duration of laser pulses shall be adjusted to the correct value for a complete population transfer, which can be examined and adjusted by observing whether there is an undesired signal readout from the Rydberg state which shouldn't be populated in the ideal case after a laser pulse. This means that the values $\pi/\Omega_{\text{\tiny{CE}}}$ in Eq.~(\ref{p1-condition}) of
Sec.~\ref{sec03A}, Eqs.~(\ref{protocol2}) and (\ref{prot2gap}) of Sec.~\ref{sec03B}, and Eqs.~(\ref{protocol3-1}) and~(\ref{protocol3-2}) of
Sec.~\ref{sec03C} should be updated by an appropriate value derived from experimental measurement where time accuracy of 10~ns is achievable.

We note that the $\pi$ pulses assumed in the analyses for the two protocols in Secs.~\ref{sec03B} and Sec.~\ref{sec03C} can also be replaced by adiabatic pulses. Then, the value $\pi/\Omega_{\text{\tiny{CE}}}$ in Eqs.~(\ref{protocol2}) and (\ref{prot2gap}) of Sec.~\ref{sec03B}, and Eqs.~(\ref{protocol3-1}) and~(\ref{protocol3-2}) of
Sec.~\ref{sec03C} shall be updated by the duration of the adiabatic pulse.

The wait durations in the protocols, i.e., the time when no CE Raman laser fields are sent to the atomic gas, shall be adjusted according to the update of pulse durations via experiment. Finally, after all these have been fixed in the experimental setup, the moment for switching on the readout laser field shall be fine tuned. In practice, we collect the retrieved photon in a finite time window of order 100~ns~\cite{Jiao2024,Li2025}, which means that the center of the time window shall be carefully chosen so as to collect the most of the photons.

\section{Level schemes for rubidium and cesium  }\label{sec04}
Equations~(\ref{condition01}) and~(\ref{waitPiCondition}) indicate that
$k_{\text{\tiny{CE}}}\geq2k$ is required for the $2\mathbb{N}\pi$ and the wait-$\pi$ protocols, and Eq.~(\ref{protocol2condition}) means that $k_{\text{\tiny{CE}}}\geq k$ should be satisfied in the $\pi$-wait-$\pi$ protocol, therefore each atomic species has its own level scheme to preserve the coherence. Since rubidium-87 and cesium-133 are two frequently used elements for Rydberg polariton~\cite{Dudin2012,Peyronel2012,Firstenberg2013,Tresp2016,Distante2016,Distante2017,Busche2017,Paris-Mandoki2017,Murray2017,Ripka446,Tiarks2019,Adams2020,Schmidt-Eberle2020,Spong2021,Padron-Brito2021,Xu2021,Jiao2020,PhysRevLett.128.123601,Ye2023}, we use the data from Refs.~\cite{Sansonetti2006,Sansonetti2009} to discuss the cases for $^{87}$Rb and $^{133}$Cs successively below.

Because the dipole matrix element between $\lvert g\rangle$ and $\lvert e\rangle$ decreases quickly when $\lvert e\rangle$ goes high, we restrict the principal quantum level of $\lvert e\rangle$ to be, e.g., $5, 6, 7$ for rubidium, and $6, 7, 8$ for cesium.

\subsection{Cases for $^{87}$Rb}\label{sec04A}
For rubidium-87, we have shown 28 cases for $k_{\text{\tiny{CE}}}/(2k)>0.5$ in Table~\ref{table1}. Cases (1-12) are useful for the $2\mathbb{N}\pi$ protocol of Sec.~\ref{sec03A}, the $\pi$-wait-$\pi$ protocol in Sec.~\ref{sec03B}, and the wait-$\pi $ protocol in Sec.~\ref{sec03C}, while cases (i-xvi) can be used only for the $\pi$-wait-$\pi$ protocol in Sec.~\ref{sec03B}.

There are 12 cases for $k_{\text{\tiny{CE}}}/(2k)>1$, shown in Table~\ref{table1}, where the smallest $k_{\text{\tiny{CE}}}/(2k)$ is about $1.16$ with case 6, and thus it is in principle the best case for meeting the condition $|\Omega_{\text{\tiny{CE}}}|\gg k_{\text{\tiny{CE}}}v$. Both the $5p$ and $6p$ states can be used as $|f\rangle$ when the 5$p$ state is used for the $|e\rangle$ state. Also, if $6p$ state is used for the preparation of the Rydberg polariton, then only the 5$p$ state can be used as the $|f\rangle$ state. Because the dipole matrix element between a high-lying Rydberg state and $|f\rangle$ is smaller if we choose $5p$~(instead of $6p$) for it, the CE effect will be stronger if we choose cases 5-8 for the suppression of motional dephasing.

Table~\ref{table1} also indicates that there are 16 cases with $0.5<k_{\text{\tiny{CE}}}/(2k)<1$, which are useful only for the implementation of the $\pi$-wait-$\pi$ protocol in Sec.~\ref{sec03B}. Importantly, cases (vii) and (viii) use the $5f$ orbital state as the intermediate state so that the CE laser of wavelength about $2272$~nm can be used for the CE fields. The relatively high level of the $5f$ state means that the coupling between $\lvert r_{1,2}\rangle$ and the $5f$ state can be strong, so that one can achieve a larger $\Omega_{\text{\tiny{CE}}}$ so as to fulfill the condition $|\Omega_{\text{\tiny{CE}}}|\gg k_{\text{\tiny{CE}}}v$. Moreover, cases (vii) and (viii) have a small $k$, which means that the dephasing of the Rydberg polariton during its loading and retrieval~(which is not suppressed by our protocols) will be relatively weak.

Note that we have restricted the principal quantum level of $\lvert e\rangle$ to be $5, 6$, and $7$ because the dipole matrix element between $\lvert g\rangle$ and $\lvert e\rangle$ decreases quickly when the energy level $\lvert e\rangle$ shoots up. If we don't have this constraint, and since ultraviolet lasers are available for Rydberg atom quantum technology~\cite{Hines2023}, there are actually more cases with $\lambda_1\leq 335$~nm that are useful for realizing the $\pi$-wait-$\pi$ protocol.

\begin{table}
  \centering
  \begin{tabular}{c|c|c|ccccc}
       \hline
    \hline  & Case &       $\frac{k_{\text{\tiny{CE}}}}{2k}$ &  $\begin{array}{c} k\\(/\mu \text{m})\end{array}$    &  $\begin{array}{c} \lambda_1\\(\text{nm})\end{array}$  & $\begin{array}{c} \lambda_3\\(\text{nm})\end{array}$  & $|e\rangle$  & $|f \rangle $  \\ \hline
 \multirow{12}{*} {\begin{turn}{90}{Cases applicable to all the three protocols}\end{turn}} &1&	    2.48&  5.35 & 795.0 &   473.9  &$5P_{\frac{1}{2}}$  &   $5P_{\frac{1}{2}}$ 	 \\
&2&	    2.45&  5.35 & 795.0 &  479.3   &$5P_{\frac{1}{2}}$  &   $5P_{\frac{3}{2}}$ 	 \\
&3&	    2.62&  5.06 & 780.2 &  473.9   &$5P_{\frac{3}{2}}$ &   $5P_{\frac{1}{2}}$  \\
&4&	    2.59&  5.06 & 780.2 &   479.3   &$5P_{\frac{3}{2}}$  &   $5P_{\frac{3}{2}}$ \\ 
\cline{2-8}
&5&	    1.17& 5.35 & 795.0  &  1003.6   &$5P_{\frac{1}{2}}$  &   $6P_{\frac{1}{2}}$ 	 \\
&6&	    1.16&  5.35 &  795.0  & 1011.4   &$5P_{\frac{1}{2}}$  &   $6P_{\frac{3}{2}}$ 	 \\
&7&	    1.24&  5.06 & 780.2 &  1003.6  &$5P_{\frac{3}{2}}$ &   $6P_{\frac{1}{2}}$  \\
&8&	    1.23&  5.06 &  780.2 &  1011.4   &$5P_{\frac{3}{2}}$  &   $6P_{\frac{3}{2}}$ \\  \cline{2-8}
&9&	    1.53&  8.64 & 421.7   & 473.9  &$6P_{\frac{1}{2}}$  &   $5P_{\frac{1}{2}}$ 	 \\
&10&	    1.52&  8.64 &  421.7  & 479.3   &$6P_{\frac{1}{2}}$  &   $5P_{\frac{3}{2}}$ 	 \\
&11&	    1.52&  8.74 & 420.3   & 473.9   &$6P_{\frac{3}{2}}$ &   $5P_{\frac{1}{2}}$  \\
&12&	    1.50&  8.74 &  420.3  &  479.3   &$6P_{\frac{3}{2}}$  &   $5P_{\frac{3}{2}}$	\\ \hline
\multirow{12}{*} {\begin{turn}{90}{These cases are applicable only to the $\pi$-wait-$\pi$ protocol}\end{turn}}& i&	    0.686&  5.35 & 795.0   & 1711.0  &$5P_{\frac{1}{2}}$  &   $7P_{\frac{1}{2}}$ 	 \\
&ii&	    0.682&  5.35 & 795.0   & 1721.3   &$5P_{\frac{1}{2}}$  &   $7P_{\frac{3}{2}}$ 	 \\
&iii&	    0.726&  5.06 &  780.2  & 1711.0   &$5P_{\frac{3}{2}}$ &   $7P_{\frac{1}{2}}$  \\
&iv&	    0.722&  5.06 &  780.2  &  1721.3    &$5P_{\frac{3}{2}}$  &   $7P_{\frac{3}{2}}$
\\\cline{2-8}
&v&	    0.808&  5.35 &  795.0  & 1451.9  &$5P_{\frac{1}{2}}$  &   $4F_{\frac{5}{2}(\frac{7}{2})}$ 	\\
&vi&	    0.856&  5.06 &  780.2  & 1451.9  &$5P_{\frac{3}{2}}$  &   $4F_{\frac{5}{2}(\frac{7}{2})}$ \\\cline{2-8}
&vii&	    0.517&  5.35 &  795.0  & 2271.8  &$5P_{\frac{1}{2}}$  &   $5F_{\frac{5}{2}(\frac{7}{2})}$ 	\\
&viii&	    0.547&  5.06 & 780.2  & 2271.8  &$5P_{\frac{3}{2}}$  &   $5F_{\frac{5}{2}(\frac{7}{2})}$\\ \cline{2-8}
&ix&	    0.725&  8.64 & 421.7  & 1003.6 &$6P_{\frac{1}{2}}$  &   $6P_{\frac{1}{2}}$ 	 \\
&x&	    0.719 &  8.64 &  421.7    & 1011.4   &$6P_{\frac{1}{2}}$  &   $6P_{\frac{3}{2}}$ 	 \\
&xi&	    0.717 &  8.74 & 420.3  & 1003.6   &$6P_{\frac{3}{2}}$ &   $6P_{\frac{1}{2}}$  \\
&xii&	   0.711&  8.74 &  420.3  &   1011.4    &$6P_{\frac{3}{2}}$  &   $6P_{\frac{3}{2}}$\\ \cline{2-8}
&xiii&	    0.960&  13.8 & 359.3  & 473.9 &$7P_{\frac{1}{2}}$  &   $5P_{\frac{1}{2}}$ 	 \\
&xiv&	    0.949&  13.8 & 359.3  & 479.3 &$7P_{\frac{1}{2}}$  &   $5P_{\frac{3}{2}}$ 	 \\
&xv&	    0.956&  13.9 & 358.8  & 473.9 &$7P_{\frac{3}{2}}$  &   $5P_{\frac{1}{2}}$ 	 \\
&xvi&	    0.946&  13.9 & 358.8  & 479.3 &$7P_{\frac{3}{2}}$  &   $5P_{\frac{3}{2}}$
\\\hline \hline
  \end{tabular}
  \caption{  \label{table1} Level configurations for achieving $k_{\text{\tiny{CE}}}/(2k)\geq0.5$ with rubidium-87. Cases (1-12) are useful both for the $2\mathbb{N}\pi$ protocol of Sec.~\ref{sec03A}, the $\pi$-wait-$\pi$ protocol in Sec.~\ref{sec03B}, and the wait-$\pi $ protocol in Sec.~\ref{sec03C}, while the cases (i-xvi) can be used only for the $\pi$-wait-$\pi$ protocol in Sec.~\ref{sec03B}. The symbols $|f\rangle$ is the intermediate state of the CE transition. $k$ is the wavevector of the Rydberg polariton, $\lambda_3$ is the wavelength of the CE field for the transition $|r_{1(2)}\rangle\leftrightarrow|f\rangle$, where the wavevector of the transition of the CE laser field, $k_{\text{\tiny{CE}}}$, is given by $4\pi/\lambda_3$. $|r_{1}\rangle$ and $|r_{2}\rangle$ are different Rydberg states~(for example, $D_{\frac{3(5)}{2}}$-Rydberg state), and the principal quantum number $n$ for $|r_{1(2)}\rangle$ is around $100$ for the calculation here. All the cases here are with $|g\rangle$ from, e.g., the $5S_{\frac{1}{2}} $ ground state. The evaluation here assumes a Rydberg constant $109736.605$cm$^{-1}$~\cite{Li2013}, ionization energy $33690.81$cm$^{-1}$ for the $^{87}$Rb I series~\cite{Sansonetti2006}, and quantum defects from Ref.~\cite{Li2013}. For cases v, vi, the two fine states $4F_{\frac{5}{2}(\frac{7}{2})}$ are nearly degenerate for they are separated by only $0.026$cm$^{-1}$~\cite{Sansonetti2006}, so we list them together; similar for cases with $5F_{\frac{5}{2}(\frac{7}{2})}$. Cases with higher $\lvert e\rangle$ do exist but aren't shown here for the dipole matrix element between $\lvert g\rangle$ and $\lvert e\rangle$ is too small for them.  }
  \end{table}

\begin{table}
  \centering
  \begin{tabular}{c|c|c|ccccc}
    \hline
    \hline  & Case &       $\frac{k_{\text{\tiny{CE}}}}{2k}$ &  $\begin{array}{c} k\\(/\mu \text{m})\end{array}$    &  $\begin{array}{c} \lambda_1\\(\text{nm})\end{array}$  & $\begin{array}{c} \lambda_3\\(\text{nm})\end{array}$  & $|e\rangle$  & $|f \rangle $  \\ \hline
\multirow{14}{*} {\begin{turn}{90}{These are applicable to all the three protocols}\end{turn}}&1&	    2.24&  5.69 & 894.6 & 494.4   &$6P_{\frac{1}{2}}$  &   $6P_{\frac{1}{2}}$ 	 \\
&	2&	    2.17&  5.69 & 894.6 & 508.3   &$6P_{\frac{1}{2}}$  &   $6P_{\frac{3}{2}}$ 	 \\
&	3&	    2.55&  4.99 & 852.3  & 494.4   &$6P_{\frac{3}{2}}$ &   $6P_{\frac{1}{2}}$  \\
&4&	    2.48&  4.99 & 852.3  & 508.3   &$6P_{\frac{3}{2}}$  &   $6P_{\frac{3}{2}}$ \\  \cline{2-8}
&5&	    1.07&  5.69 & 894.6 & 1037.2   &$6P_{\frac{1}{2}}$  &   $7P_{\frac{1}{2}}$ 	 \\
&6&	    1.05&  5.69 & 894.6 & 1057.1   &$6P_{\frac{1}{2}}$  &   $7P_{\frac{3}{2}}$ 	 \\
&7&	    1.21&  4.99 & 852.3 & 1037.2   &$6P_{\frac{3}{2}}$ &   $7P_{\frac{1}{2}}$  \\
&8&	    1.19&  4.99 &  852.3 & 1057.1   &$6P_{\frac{3}{2}}$  &   $7P_{\frac{3}{2}}$  \\ \cline{2-8}
&9&	    1.67&  7.62 & 459.4 & 494.4   &$7P_{\frac{1}{2}}$  &   $6P_{\frac{1}{2}}$ 	 \\
&10&	    1.62&  7.62 & 459.4 & 508.3   &$7P_{\frac{1}{2}}$  &   $6P_{\frac{3}{2}}$ 	 \\
&11&	    1.62& 7.85  & 455.7 & 494.4   &$7P_{\frac{3}{2}}$ &   $6P_{\frac{1}{2}}$  \\
&12&	    1.58&  7.85 & 455.7  & 508.3  &$7P_{\frac{3}{2}}$  &   $6P_{\frac{3}{2}}$\\ \cline{2-8}
&13&	    1.01&  12.6 & 389.0 & 494.4   &$8P_{\frac{1}{2}}$  &   $6P_{\frac{1}{2}}$ 	 \\
&14&	    1.00& 12.7  & 387.7 & 494.4   &$8P_{\frac{3}{2}}$ &   $6P_{\frac{1}{2}}$\\ \hline
\multirow{15}{*} {\begin{turn}{90}{These are applicable only to the $\pi$-wait-$\pi$ protocol}\end{turn}}&i&	    0.766&  5.69 & 894.6 & 1442.1   &$6P_{\frac{1}{2}}$  &   $4F_{\frac{5}{2}(\frac{7}{2})}$ 	 \\
&ii&	    0.873&  4.99 & 852.3 & 1442.1   &$6P_{\frac{3}{2}}$  &   $4F_{\frac{5}{2}(\frac{7}{2})}$ 	 \\
&iii&	    0.630&  5.69 & 894.6 & 1755.1   &$6P_{\frac{1}{2}}$  &   $8P_{\frac{1}{2}}$ 	 \\
&iv&	    0.620&  5.69 & 894.6 & 1781.0   &$6P_{\frac{1}{2}}$  &   $8P_{\frac{3}{2}}$ 	 \\
&v&	    0.717& 4.99 & 852.3 & 1755.1   &$6P_{\frac{3}{2}}$  &   $8P_{\frac{1}{2}}$ 	 \\
&vi&	    0.707& 4.99 & 852.3  & 1781.0   &$6P_{\frac{3}{2}}$  &   $8P_{\frac{3}{2}}$ 	 \\
&vii&	  0.558&  4.99 & 852.3  & 2254.7(6)  &$6P_{\frac{3}{2}}$ &   $5F_{\frac{5}{2}(\frac{7}{2})}$ \\ \cline{2-8}
&viii&	   0.795&  7.62 & 459.4 & 1037.2   &$7P_{\frac{1}{2}}$  &   $7P_{\frac{1}{2}}$ 	 \\
&ix&	    0.780 &  7.62 & 459.4 & 1057.1  &$7P_{\frac{1}{2}}$  &   $7P_{\frac{3}{2}}$ 	 \\
&x&	    0.772 & 7.85  & 455.7 & 1037.2   &$7P_{\frac{3}{2}}$ &   $7P_{\frac{1}{2}}$  \\
&xi&	    0.758&  7.85 & 455.7  & 1057.1  &$7P_{\frac{3}{2}}$  &   $7P_{\frac{3}{2}}$\\
&xii & 0.572&  7.62 & 459.4 & 1442.1   &$7P_{\frac{1}{2}}$  &   $4F_{\frac{5}{2}(\frac{7}{2})}$ 	 \\
&xiii&	    0.555&  7.85 & 455.7 & 1442.1   &$7P_{\frac{3}{2}}$  &   $4F_{\frac{5}{2}(\frac{7}{2})}$
 \\  \cline{2-8}
&xiv&  0.983 &  12.6 & 389.0 & 508.3  &$8P_{\frac{1}{2}}$  &   $6P_{\frac{3}{2}}$ 	 \\
&xv&	  0.975 & 12.7  & 387.7 & 508.3   &$8P_{\frac{3}{2}}$ &   $6P_{\frac{3}{2}}$
\\ \hline \hline
  \end{tabular}
  \caption{  \label{table2} Level configurations for achieving $k_{\text{\tiny{CE}}}/(2k)\geq0.5$ with cesium-133. Cases (1-14) are useful both for the $2\mathbb{N}\pi$ protocol of Sec.~\ref{sec03A}, the $\pi$-wait-$\pi$ protocol in Sec.~\ref{sec03B}, and the wait-$\pi $ protocol in Sec.~\ref{sec03C}, while the cases (i-xv) can be used only for the $\pi$-wait-$\pi$ protocol in Sec.~\ref{sec03B}. Similar symbols as in Table~\ref{table1} are used here. All the cases here are with $|g\rangle$ from, e.g., the $6S_{\frac{1}{2}} $ ground state, and Rydberg states of principal quantum number around $100$ are used. The evaluation here assumes a Rydberg constant $  109737.31568525$cm$^{-1}$, ionization energy $  31406.46766$cm$^{-1}$ for the $^{133}$Cs I series~\cite{Sansonetti2009}, and quantum defects from Refs.~\cite{Lorenz1984,PhysRevA.35.4650}. For cases i, ii, xii, and xiii, two fine states $4F_{\frac{5}{2}(\frac{7}{2})}$ are nearly degenerate for they are separated by only $0.1814$cm$^{-1}$~\cite{Sansonetti2009}. The states $5F_{\frac{5}{2}(\frac{7}{2})}$ at (26971.3030) 26971.1535 are near to each other, so we list them as one case, case vii, with their $\lambda_3$ a little different, about 2254.7 and 2254.6, respectively. Cases with $\lvert e\rangle$ being $9P_{\frac{1(3)}{2}}$ or even higher states do exist but aren't shown here for the dipole matrix element between $\lvert g\rangle$ and $\lvert e\rangle$ is too small for them. }
  \end{table}


\subsection{Cases for $^{133}$Cs}\label{sec04B}
Table~\ref{table2} shows 29 possible configurations for $^{133}$Cs. Cases (1-14) are useful for the $2\mathbb{N}\pi$ protocol of Sec.~\ref{sec03A}, the $\pi$-wait-$\pi$ protocol in Sec.~\ref{sec03B}, and the wait-$\pi $ protocol in Sec.~\ref{sec03C}, while cases (i-xv) can be used only for the $\pi$-wait-$\pi$ protocol in Sec.~\ref{sec03B}.

As in the study for rubidium, here we have restricted the principal quantum number of $\lvert e\rangle$ to $ 6, 7, 8$ because the dipole matrix element between $\lvert g\rangle$ and $\lvert e\rangle$ decreases quickly when $\lvert e\rangle$ goes higher. So we don't consider states over $9P_{\frac{1(3)}{2}}$. For the 14 cases that are useful for all the three protocols, there are in general four classes of intermediate states $\{|e\rangle,|f\rangle\}$ in Table~\ref{table2}, namely, $\{6p,6p\}$,~$\{6p,7p\}$,~$\{7p,6p\}$, and $\{8p,6p\}$ states.

In Tables~\ref{table1} and~\ref{table2}, $k$ and $k_{\text{\tiny{CE}}}$ are calculated by using Rydberg states of principal quantum around 100. But if we use lower Rydberg states then the value of $k_{\text{\tiny{CE}}}/(2k)$ can change a little bit. When external fields like electric field or magnetic fields are used, the values of $k_{\text{\tiny{CE}}}/(2k)$ can change, too. Nevertheless, the change of $k_{\text{\tiny{CE}}}/(2k)$ is marginal in these circumstances, and, hence, not important unless extreme precision is needed. We have also assumed that the detunings for the intermediate states $\{|e\rangle,|f\rangle\}$ are much smaller than the energy spacing between the involved atomic states. Thus the predicted values in Tables~\ref{table1} and~\ref{table2} can be slightly different compared to those in real experiments. However, the actual value of Rabi frequency $\Omega_{\text{\tiny{CE}}}$ used in Eq.~\ref{infraH} can be adjusted according to Eqs.~(\ref{condition01}), (\ref{protocol2}), and~(\ref{protocol3-1}) for the $2\mathbb{N}\pi$ protocol of Sec.~\ref{sec03A}, the $\pi$-wait-$\pi$ protocol in Sec.~\ref{sec03B}, and the wait-$\pi $ protocol in Sec.~\ref{sec03C}, respectively.

\subsection{Comparison between $^{87}$Rb and $^{133}$Cs}\label{sec04C}
It will be interesting to compare the different data between Table~\ref{table1} and Table~\ref{table2}. One can notice that the cases 1-12 in Table~\ref{table1} are very similar to the cases 1-12 in Table~\ref{table2}. This is because the outermost electron in $^{133}$Cs or $^{87}$Rb is like that in a hydrogen atom, so that their behaviors are similar. But because they have different ion cores with different numbers of inner electrons, there is some minor difference between them. For example, the difference between the $k_{\text{\tiny{CE}}}/(2k)$'s in the $i$th case in Table~\ref{table1} and that in Table~\ref{table2} is about $0.1$ for $i=5-12$. The last two cases in Table~\ref{table2} have a $k_{\text{\tiny{CE}}}/(2k)$ that is barely over $1$, and thus their counterparts in Table~\ref{table1} have a $k_{\text{\tiny{CE}}}/(2k)$ that is smaller than 1. This is why there are only 12 cases in Table~\ref{table1} with $k_{\text{\tiny{CE}}}/(2k)>1$.

For our theory to work well, it is necessary to have the condition $\Omega_{\text{\tiny{CE}}}\gg k_{\text{\tiny{CE}}} v$~\cite{Shi2020}, which means that a smaller $k_{\text{\tiny{CE}}}$ is better. Further, smaller $k_{\text{\tiny{CE}}}$ indicates that the state $|f\rangle$ is higher, and $|f\rangle$ being higher will lead to larger dipole matrix elements between $|f\rangle$ and the Rydberg states, resulting in larger $\Omega_{\text{\tiny{CE}}}$. Also, higher $|f\rangle$ usually has a longer lifetime, thus the detuning at it can be not as high as in a lower $|f\rangle$, and, hence, larger $\Omega_{\text{\tiny{CE}}}$ can occur with higher $|f\rangle$. So, choosing cases with smaller $k_{\text{\tiny{CE}}}$ results in larger $\Omega_{\text{\tiny{CE}}}/k_{\text{\tiny{CE}}}v$. Thus, one can see that to employ the $2\mathbb{N}\pi$ protocol of Sec.~\ref{sec03A} or the wait-$\pi$ protocol of Sec.~\ref{sec03C} which require $k_{\text{\tiny{CE}}}/(2k)>1$, the best case for rubidium is case 8 in Table~\ref{table1} for it has the smallest $k_{\text{\tiny{CE}}}$ and the smallest $k$, where $k$ being small means that the dephasing of the Rydberg polariton is slow compared to other cases. Similarly, if we consider the $\pi$-wait-$\pi$ protocol in Sec.~\ref{sec03B}, the best case for rubidium is case (viii) in Table~\ref{table1} for it has the smallest $k_{\text{\tiny{CE}}}$ and the smallest $k$ when meanwhile $k_{\text{\tiny{CE}}}/(2k)>0.5$. In a similar way, one can find that in Table~\ref{table2}, the best case is case 8 for the $2\mathbb{N}\pi$ protocol of Sec.~\ref{sec03A} or the wait-$\pi$ protocol of Sec.~\ref{sec03C}, and case (vii) for the $\pi$-wait-$\pi$ protocol in Sec.~\ref{sec03B}.

When we try to consider the best case for preserving the coherence of Rydberg polariton, it appears that among all the cases in Tables~\ref{table1} and~\ref{table2}, the best choice is case (viii) of Table~\ref{table1} for rubidium, or case (vii) of Table ~\ref{table2} for cesium because they have the smallest $k$ as well as the smallest $k_{\text{\tiny{CE}}}$. In other words, it appears that using the $\pi$-wait-$\pi$ protocol may enhance the coherence better. Finally, the mass of cesium is larger than that of rubidium, so there is an intrinsic advantage in cesium. So case (vii) of Table ~\ref{table2} may lead to better performance.

\section{Experimental prospect}\label{sec4-5}
\subsection{Large enough $\Omega_{\text{\tiny{CE}}}$}
The protocols hinge on the condition $\Omega_{\text{\tiny{CE}}}\gg k_{\text{\tiny{CE}}} v$, where $v=\sqrt{k_{\text{B}}T/m}$~\cite{Jenkins2012}. So the experimental feasibility relies on whether such condition can be fulfilled. So, large $\Omega_{\text{\tiny{CE}}}$ is needed. The dephasing is suppressed by using the CE lasers which result in Eq.~(\ref{infraH}), which is derived from a two-photon transition via a low-lying intermediate state that is largely detuned by $\Delta$. In other words, Eq.~(\ref{infraH}) is derived via adiabatic approximation where one example was given in Ref.~\cite{Shi2014}, during which ac Stark shift appears. However, the ac Stark shift gives rise to the same phase to each state component in the wavefunction of the Rydberg polariton, therefore is trivial and not included in Eq.~(\ref{infraH}). Large $\Omega_{\text{\tiny{CE}}}$ means that $\Delta$ can't be too large, but small $\Delta$ will increase the scattering at the intermediate state $\lvert f\rangle$.

The scattering at $\lvert f\rangle$ can be be made negligible when we want large enough $\Omega_{\text{\tiny{CE}}}$. For brevity, we assume that the two one-photon Rabi frequencies for $\lvert r_1\rangle\leftrightarrow\lvert f\rangle\leftrightarrow\lvert r_2\rangle$ are equal to $\Omega$. Then the probability of spontaneous emission at $\lvert f\rangle$ in a $\pi$ pulse for the transition $\lvert r_1\rangle\leftrightarrow \lvert r_2\rangle$ is $\frac{\pi}{2}\frac{1}{\tau_{f}\Delta }$~\cite{PhysRevLett.123.230501}, where $\tau_{f}$ is the lifetime of $\lvert f\rangle$. If we do estimation based on state-of-the-art lasers of available wavelengths reported in experiments on Rydberg excitations, we can take, e.g., Ref.~\cite{Evered2023}, where $\Delta/2\pi=7.8$~GHz was used, leading to a scattering $0.029\%$ during a $\pi$ pulse with the $6P_{3/2}$ state as $\lvert f\rangle$ if the laser of Ref.~\cite{Evered2023} is used. Even if we decrease $\Delta$ to $2\pi\times1$~GHz, the scattering probability is about 0.0023. Compared to the Rydberg-state decay which will be shown later in Sec.~\ref{sec05} for storage times on the order of $10~\mu$s, this error is negligible.

If a laboratory doesn't have high-power lasers, it is still useful to use the theory in this work for long-time storage of Rydberg polariton. For example, Ref.~\cite{Li2025} used $\Delta/2\pi=335$~MHz with case 4 of Table~\ref{table2}, which is a hard case to suppress the scattering at $\lvert f\rangle$ whose lifetime is short. This leads to a scattering $\frac{\pi}{2}\frac{1}{\tau_{f}\Delta }\approx2.4\%$ for the $\pi$ pulse used in Ref.~\cite{Li2025}. In comparison, the Rydberg-state decay $e^{-t_{\text{s}}/\tau_{\text{r}}}$ for $t_{\text{s}}=10~\mu$s is about $8\%$, which is more than three times the error from the scattering at $\lvert f\rangle$. But without using the theory in this work, the motional error is larger than 99\% with the atomic medium of Ref.~\cite{Li2025}. This means that even with weak lasers so that smaller $\Delta$ shall be used, the theory can be useful.

\subsection{Correct $k_{\text{\tiny{CE}}}$}
Special attention shall be paid to the phase term in Eq.~(\ref{infraH}). Because the transition between the two Rydberg states is a `V' type transition, the two laser fields shall counter-propagate in an appropriate configuration as specified in Figs.~\ref{figure-wait-pi-direction} and~\ref{figure-wait-pi-direction-2}. If the two laser fields in the Raman transition are parallel to each other, then no phase term will appear in Eq.~(\ref{infraH}) because $k_{\text{\tiny{CE}}}$ will become zero. On the other hand, if the two laser fields are opposite to each other but in the wrong configuration, then the phase in Eq.~(\ref{infraH}) will reverse its sign, namely, $k_{\text{\tiny{CE}}}$ will become $-k_{\text{\tiny{CE}}}$, which, however, still can't be useful for our purpose as experimentally verified in Refs.~\cite{Jiao2024,Li2025}

For the theory to work, the directions of the two fields in the CE transition should obey the following rule. To be concrete, we take the configuration of Fig.~\ref{figure01} as an example, where the wavelength of the control field is smaller than that of the signal field during the loading stage; then, the direction of laser 
for addressing the transition $|r_1\rangle \leftrightarrow |f\rangle$ should be parallel to the control field. But if the wavelength for the signal field is shorter, then the field for addressing the transition $|r_1\rangle \leftrightarrow |f\rangle$ should be aligned parallel with the signal field. In all cases of the three different protocols explored in this work, the directions of the laser fields addressing the transition $|r_1\rangle \leftrightarrow |f\rangle$ and the transition $|r_2\rangle \leftrightarrow |f\rangle$ should be opposite to each other.

\subsection{Feasibility to have $\Omega_{\text{\tiny{CE}}}/k_{\text{\tiny{CE}}}v\gg1$}
Larger $\Omega_{\text{\tiny{CE}}}$ will have more chance to fulfill the condition $\Omega_{\text{\tiny{CE}}}\gg k_{\text{\tiny{CE}}} v$. So, strong lasers are necessary. One can estimate $\Omega_{\text{\tiny{CE}}}$ based on laser powers of available wavelengths reported in recent experiments. In Ref.~\cite{Evered2023}, a single-photon Rabi frequency $\Omega/2\pi=303~$MHz was realized for the 1013~nm transition $6P_{3/2}\leftrightarrow 53S_{1/2}$ of $^{87}$Rb for realizing quantum gates between individual atoms. In
Table~\ref{table1}, there are altogether four cases using the $6P_{3/2}$ state as $|f\rangle$, namely, cases 6, 8, x, and xii of Table~\ref{table1}. For these cases, the two-photon Rabi frequency $\Omega_{\text{\tiny{CE}}}$ is adiabatically derived from a V-type three-level transition
\begin{eqnarray}
\lvert r_1\rangle \xleftrightarrow[\Omega_{r_1f}]{1011.4~\text{nm}} \lvert f\rangle \xleftrightarrow[\Omega_{r_2f}]{1011.4~\text{nm}}\lvert r_2\rangle,\label{1011transition}
\end{eqnarray}
where the wavelength 1011.4~nm in cases 6, 8, x, and xii of Table~\ref{table1} is shown because Rydberg states with $n~\sim100$ is assumed. Here $\lvert r_1\rangle$ and $\lvert r_2\rangle$ can be two Rydberg states with principal quantum numbers differing by one or a little more~\cite{Jiao2024,Li2025}. If we choose principal quantum numbers of $\lvert r_1\rangle$ and $\lvert r_2\rangle$ to be around $n=53$ as used for the ground-Rydberg transition of Ref.~\cite{Evered2023}, then the wavelengths for the two transitions of  Eq.~(\ref{1011transition}) will both be around 1013~nm, a little longer than the 1011.4~nm shown on the arrows of Eq.~(\ref{1011transition}).

To realize Eq.~(\ref{1011transition}), in principle one laser source tuned to one of the two transitions can be used. Because the energy difference between two nearby Rydberg states is small, one laser beam can be split into two, where one of the two resulting beams can be frequency-modulated by an acousto-optic modulator to match the other of the two transitions of Eq.~(\ref{1011transition}). If a 1013~nm laser of high power as Ref.~\cite{Evered2023} is available, then one can consider $n\sim53$. The frequency difference of two s-orbital rubidium Rydberg states with two principal quantum numbers $(52,53)$ and $(53,54)$ are about 55~MHz and 51~MHz, respectively. Shifting such a frequency is well within the capability of an acousto-optic modulator.

With a large detuning $\Delta_{f}$ at $\lvert f\rangle$, the state $\lvert f\rangle$ in the transition of Eq.~(\ref{1011transition}) can be adiabatically eliminated~\cite{Shi2014}, leading to an effective two-photon Rabi frequency
\begin{eqnarray}
\Omega_{\text{\tiny{CE}}}&=& \frac{\Omega_{r_1f}\Omega_{r_2f}}{2\Delta_{f}}.\label{1011transition2}
\end{eqnarray}
 If a 1013~nm laser of high power as Ref.~\cite{Evered2023} is available, then splitting it into two results in a reduction by a factor of $\sqrt{2}$, i.e., $\Omega_{r_1f}$ and $\Omega_{r_2f}$ would be around $2\pi\times200$~MHz. The derivation from Eq.~(\ref{1011transition}) to Eq.~(\ref{1011transition2}) will lead to an extra detuning $\frac{\Omega_{r_2f}^2-\Omega_{r_1f}^2}{4\Delta_{f}}$, which should be considered in experiments as done in Refs.~\cite{Jiao2024,Li2025}. If the detuning at $\lvert f\rangle$ is ten times over the larger one among $\Omega_{r_1f}$ and $\Omega_{r_2f}$, then $\Omega_{\text{\tiny{CE}}}$ is about $2\pi\times10$~MHz for our theory.

We would have $v\approx3$cm$/$s for a rubidium medium with a motional temperature $10~\mu$K. This means that for cases 6, 8, x, and xii of Table~\ref{table1} if the principal quantum number around $n\sim53$ was used, $\Omega_{\text{\tiny{CE}}}/k_{\text{\tiny{CE}}}v$ would be over 168. Even if one chooses Rydberg states of principal quantum numbers around $100$ with similar laser power, $\Omega_{\text{\tiny{CE}}}$ would be around $2\pi\times1.5~$MHz, leading to $\Omega_{\text{\tiny{CE}}}/k_{\text{\tiny{CE}}}v>25$ at $10~\mu$K, while later on in Sec.~\ref{secVIE2} it will be shown that with $\Omega_{\text{\tiny{CE}}}/k_{\text{\tiny{CE}}}v>5$ the theory works already quite well.

\section{Numerical Study}\label{sec05}
In this section, we use numerical simulation to validate the theory.

\subsection{Method of numerical calculation}
For a Rydberg polariton stored in a cold atomic medium, the drift of different atoms does not interrupt each other as long as they do not collide. We consider the case in the experimental setups of Refs.~\cite{Peyronel2012,Firstenberg2013} where the atomic density of the medium was about $\rho=10^{18}/m^3$, and such a density has been useful in the study of Rydberg polariton~\cite{Fan2023,Fan2023oe2}. For $T=10~\mu$K, the average collision rate of the atom is $4\rho d^2\sqrt{\pi k_BT/m}\approx 0.2$~Hz if we take the diameter of a ground-state rubidium atom to be $d=1$~nm. In some experiments reported in Refs.~\cite{Schmidt-Eberle2020,Padron-Brito2021,Xu2021}, samples with densities $(0.08,~0.4,~0.2)\times10^{18}/m^3$ were used, which will lead to even smaller collision probability. This means that for a timescale up to a millisecond, the collision of two atoms in typical samples can be ignored. The Rydberg excitation is shared by all the atoms in the medium, thus one can ignore the collision between Rydberg atoms for a first approximation. Then, we consider atoms with constant velocity in the atomic medium. Too low density in the atomic medium can result in difficulty in the loading of signal photon, but one can elongate the atom cloud so as to attain a larger optical depth.

Because the drift of different atoms do not interrupt each other, the dephasing of the Rydberg polariton is an average over all the dephasing of the excitations in the atomic medium. So, we can approximate the usual dephasing of the Rydberg polariton by considering the retrieval efficiency
\begin{eqnarray}
  \eta_0 &=&\left| \int \mathscr{G}(v_j) \langle r_1| e^{-ik (z_j+v_jt)}e^{ik z_j} |r_1\rangle dv_j \right|^2 \nonumber\\
  &=& \left| \int \mathscr{G}(v_j) e^{-ik v_jt}  dv_j \right|^2  ,\label{errorTr}
\end{eqnarray}
where $\mathscr{G}(v_j) $ is the velocity distribution of the atom along the quantization axis $\mathbf{z}$
\begin{eqnarray}
  \mathscr{G}(v_j) &=& \sqrt{\frac{m}{2\pi k_BT}}\text{exp}\left(-\frac{mv_j^2}{2k_BT}\right),\label{gaussianDis}
\end{eqnarray}
where $k_B$ is the Boltzmann constant. The integration in Eq.~(\ref{errorTr}) is exactly Eq.~(\ref{eta01}) with $T_2^\ast= 1/(kv)$, which was studied in Refs.~\cite{Wilk2010,Saffman2011} for single atoms. The numerical code essentially has three loops, the first is about the different atoms, where each atom has a certain $z_j$, the second is about the possible $v_j$ for each atom which should be sampled according to the velocity distribution $\mathscr{G}(v_j)$, and the third is about the time evolution governed by a time-dependent Hamiltonian since at different time steps the phase of the Rabi frequencies differ.

In our scheme by the $2\mathbb{N}\pi$ protocol of Sec.~\ref{sec03A}, the $\pi$-wait-$\pi$ protocol in Sec.~\ref{sec03B}, or the wait-$\pi$ protocol in Sec.~\ref{sec03C}, the dephasing can be eliminated if a Rabi frequency that satisfies $|\Omega_{\text{\tiny{CE}}}|\gg k_{\text{\tiny{CE}}}v_j$ is used. However, $\Omega_{\text{\tiny{CE}}}$ is not infinite and in fact can be only on the order of megahertz, and meanwhile $v_j$ spreads to infinity although its mean is zero. So there will still be dephasing. The retrieval efficiency for the $2\mathbb{N}\pi$ or the $\pi$-wait-$\pi$ protocol is then given by
\begin{eqnarray}
  \eta &=&\left|  \int \mathscr{G}(v_j) \langle r_1| e^{-ik (z_j+v_jt)}\mathcal{T}e^{-i \int_0^t \hat{H}(t')dt' }   e^{ik z_j} |r_1\rangle dv_j \right|^2,\nonumber\\\label{errorOur}
\end{eqnarray}
where $\mathcal{T}$ is a time-ordering operator, $\hat{H}$ is the Hamiltonian of the CE fields of Eq.~(\ref{infraH}). However, note that for the $2\mathbb{N}\pi$ protocol of Sec.~\ref{sec03A}, the application of the CE fields should be $2\mathbb{N}\pi/\Omega_{\text{\tiny{CE}}}$ so that the Rydberg state in the Rydberg polariton returns to $|r_1\rangle$ at the end of the storage; similarly, for the $\pi$-wait-$\pi$ protocol in Sec.~\ref{sec03B}, two $\pi$ pulses are used with a wait time in between so that the Rydberg state is restored to $|r_1\rangle$. A derivation for Eq.~(\ref{errorOur}) is given in Appendix~\ref{appendixA}.

For the wait-$\pi$ protocol in Sec.~\ref{sec03C}, the $\pi$ pulse transfers the Rydberg state from $\lvert r_1\rangle$ to $\lvert r_2\rangle$, and the laser configuration during the retrieval stage follows after Fig.~\ref{figure-wait-pi-direction} or Fig.~\ref{figure-wait-pi-direction-2}, which shows that the wavevector of Rydberg polariton during the retrieval stage is given by $k-k_{\text{\tiny{CE}}}$; here, note that $k-k_{\text{\tiny{CE}}}<0$, and the magnitude of the wavevector of the Rydberg polariton is shown in Figs.~\ref{figure-wait-pi-direction} and~\ref{figure-wait-pi-direction-2}. The retrieval efficiency of the signal photon with the wait-$\pi$ protocol is given by
\begin{eqnarray}
  \left|  \int \mathscr{G}(v_j) \langle r_2| e^{-i(k-k_{\text{\tiny{CE}}} ) (z_j+v_jt)}\mathcal{T}e^{-i \int_0^t \hat{H}(t')dt' }   e^{ik z_j} |r_1\rangle dv_j \right|^2.\nonumber\label{errorOur-WaitPi}
\end{eqnarray}

The above analyses have assumed that the atom loss during the storage is negligible, while in practice loss can occur. If the coupling laser for retrieval can't cover the Rydberg atoms that already escape away from the laser spot, there should be a loss of retrieval efficiency. A simple estimate is that the retrieval efficiency due to this will be smaller than $R_{\text{c}}^2/(R_{\text{c}}+\overline{v}t_{\text{s}})^2$, where $\overline{v}$ is the average speed of the atom and $R_{\text{c}}$ is the waist of the coupling laser.

\begin{figure}
\includegraphics[width=3.0in]
{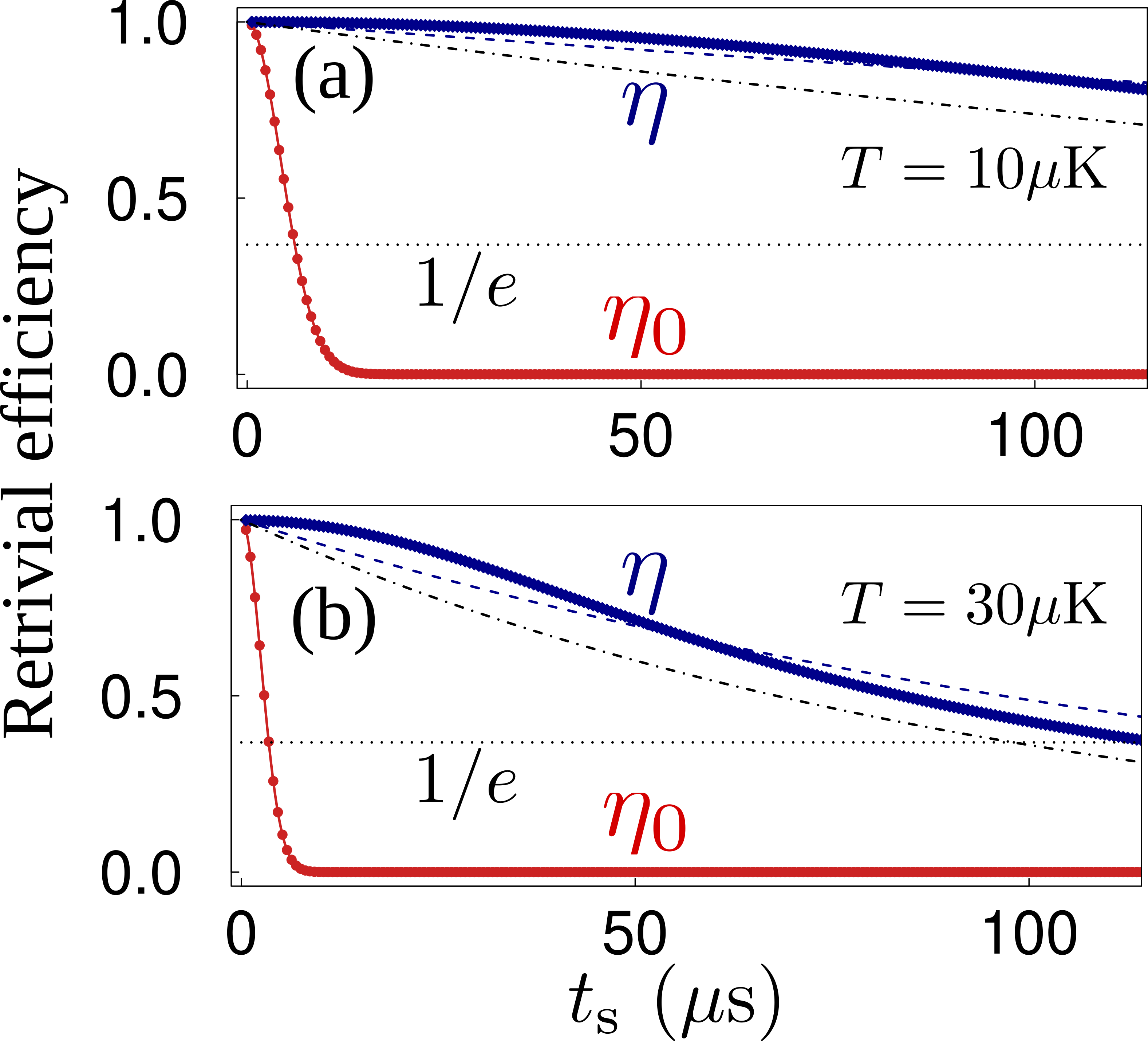}
 \caption{Coherence enhancement of rubidium Rydberg polariton by the $2\mathbb{N}\pi$ protocol of Sec.~\ref{sec03A} with case 6 of Table~\ref{table1} as an example. Shown is the retrieval efficiency of the stored Rydberg polariton with a storage time of $t_{\text{s}}=t_{\text{\tiny{CE}}}k_{\text{\tiny{CE}}}/(2k)$, where $t_{\text{\tiny{CE}}}=2\mathbb{N}\pi/\Omega_{\text{\tiny{CE}}}$, and $\Omega_{\text{\tiny{CE}}}/2\pi=2$~MHz. The thick blue curve shows the CE result evaluated with Eq.~(\ref{errorOur}), while the red dots show results without using the CE protocol, evaluated by Eq.~(\ref{errorTr}). The solid red curve is from Eq.~(\ref{eta01}). We also use the dashed blue curve, $e^{-t/T_2^\ast}$, a fit parameter of $T_2^\ast=100/(vk)$ and $T_2^\ast=40/(vk)$ in (a) and (b), respectively, to fit the CE effect. The atomic temperature is $10$~$\mu$K and $30$~$\mu$K in (a) and (b), respectively. The dotted horizontal line shows $1/e$. Note that the Rydberg-state decay is ignored in Eq.~(\ref{errorOur}), so the actual decoherence, as captured by the dash-dotted curve, can be estimated with an effective coherence time $(vk/100+1/\tau)^{-1}$ in (a) and $(vk/40+1/\tau)^{-1}$ in (b), respectively, where the lifetime of Rydberg state $100S_{1/2}$ state at room temperature is about $\tau=330~\mu$s~\cite{Beterov2009}. \label{figure-Rb-2Npi} }
\end{figure}

\begin{figure}
\includegraphics[width=3in]{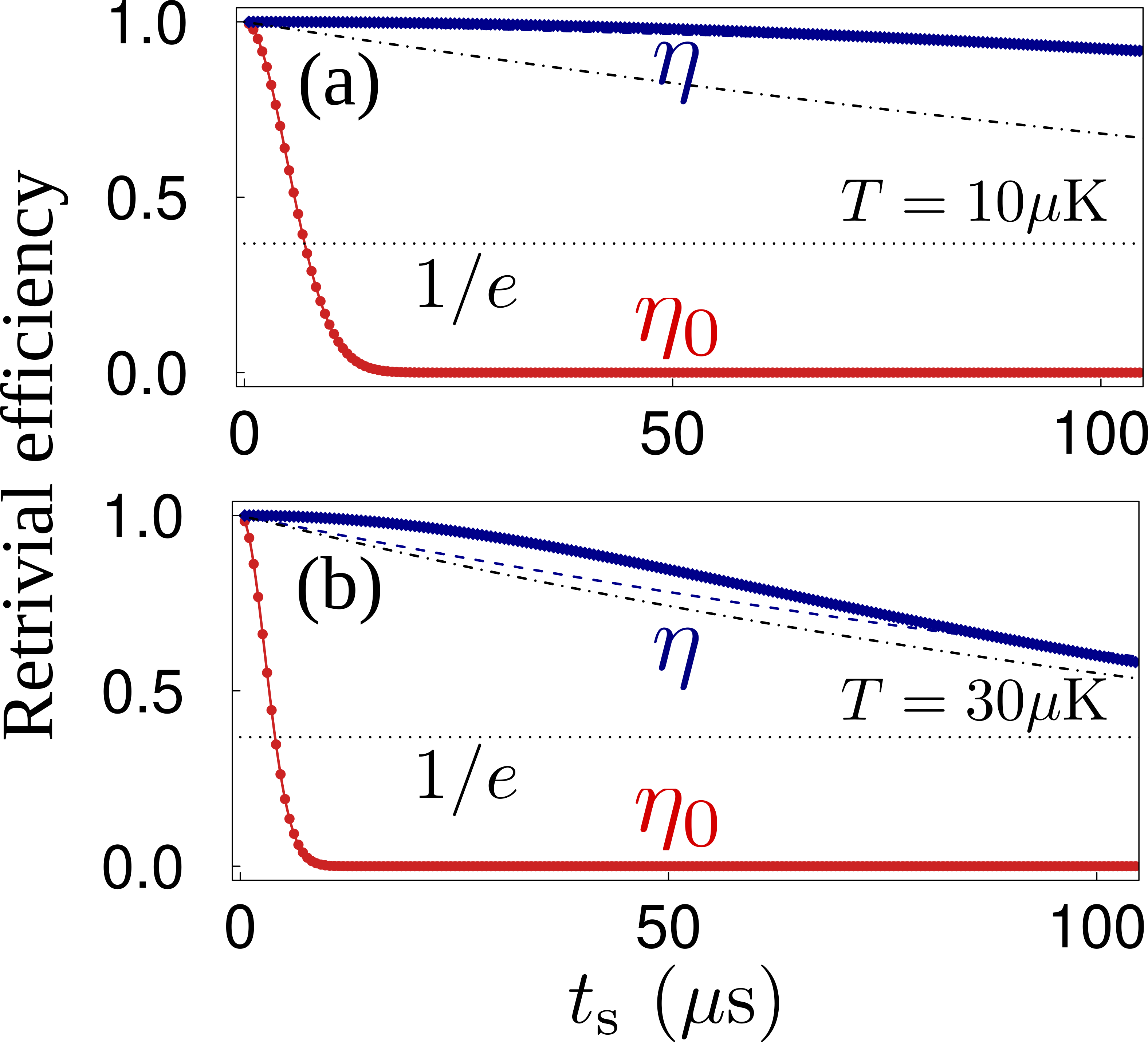}
 \caption{Coherence enhancement of cesium Rydberg polariton by the $2\mathbb{N}\pi$ protocol of Sec.~\ref{sec03A} with case 6 of Table~\ref{table2} as an example. Curves and symbols are defined similarly as in Fig.~\ref{figure-Rb-2Npi} except that cesium-133 is used with the configuration of case 6 of Table~\ref{table2}. The dashed curve is $e^{-t/T_2^\ast}$ with a fit parameter of $T_2^\ast=200/(vk)$ and $T_2^\ast=50/(vk)$ in (a) and (b), respectively; the dashed curve is barely observable for it almost overlaps with the solid blue curve. The dash-dotted curve shows the estimated actual decoherence with a coherence time $(vk/200+1/\tau)^{-1}$ in (a) and $(vk/50+1/\tau)^{-1}$ in (b), respectively.   \label{figure-Cs-2Npi} }
\end{figure}

\subsection{Numerical results of the $2\mathbb{N}\pi$ protocol}
We use numerical method to verify the theory. Below, we show numerical results for the two case 6 in both Tables~\ref{table1} and~\ref{table2}, although we have checked the other cases.

We first look at case 6 of Table~\ref{table1}, where the wavelengths of laser fields for the transitions $|g\rangle\leftrightarrow|e\rangle,~|e\rangle\leftrightarrow|r_1\rangle$ and $|r_{1(2)}\rangle\leftrightarrow|f\rangle$ are $795.0,473.9$, and $1011.4$~nm, respectively, leading to $(k,~k_{\text{\tiny{CE}}})=(5.35,~12.42)\times10^6$m$^{-1}$. According to Sec.~\ref{sec03A}, we must impose
\begin{eqnarray}
\Omega_{\text{\tiny{CE}}} &=& \frac{k_{\text{\tiny{CE}}}}{2k}\frac{2\pi\mathbb{N}}{t_{\text{s}} },
\end{eqnarray}
i.e., there is a correspondence between $\Omega_{\text{\tiny{CE}}}$, $\mathbb{N}$, and $t_{\text{s}}$.

As shown above Eq.~(\ref{condition01}), the suppression of the dephasing depends on the quality of the phase following with the excitation of the CE laser fields. For a storage time $t_{\text{s}}$, the excitation by the CE fields lasts for a duration of $t_{\text{\tiny{CE}}}=2\pi\mathbb{N}/ \Omega_{\text{\tiny{CE}}}$, and the phase change of the Rydberg state $|r_1\rangle$ should be $k_{\text{\tiny{CE}}}v_jt_{\text{\tiny{CE}}}/2$ with an even number of Rabi cycles; but if $\mathbb{N}$ is an odd integer, the phase change of the Rydberg state $|r_1\rangle$ should be $\pi+k_{\text{\tiny{CE}}}v_jt_{\text{\tiny{CE}}}/2$. Because of the above condition between $\Omega_{\text{\tiny{CE}}}$, $\mathbb{N}$, and $t_{\text{s}}$, we have a phase change of $kv_jt_{\text{s}}$ for an even $\mathbb{N}$, and $kv_jt_{\text{s}}+\pi$ for an odd $\mathbb{N}$. We have used the data in case 6 of Table~\ref{table1} and numerically find negligible mismatch between the simulated phase change and $kv_jt_{\text{s}}(+\pi)$ for $t_{\text{s}}$ up to $100~\mu$s when $\Omega_{\text{\tiny{CE}}}/2\pi=2$~MHz and $T=30~\mu$K.

Figure~\ref{figure-Rb-2Npi} shows the suppression of the motional dephasing by the $2\mathbb{N}\pi$ protocol with the parameters from case 6 of Table~\ref{table1}. The atomic temperature is $10$ and $30$~$\mu$K in Fig.~\ref{figure-Rb-2Npi}(a) and Fig.~\ref{figure-Rb-2Npi}(b), respectively. In Fig.~\ref{figure-Rb-2Npi}, the thick blue curve shows the CE effect evaluated with Eq.~(\ref{errorOur}). When using Eq.~(\ref{errorOur}), it seems that the choice of $z_j$ can matter; however, we numerically find that different values of $z_j$ result in almost the same numerical results. Then, we use $z_j=0$ in the integration of Eq.~(\ref{errorOur}). For comparison, the red dots show results without using the CE protocol, evaluated by Eq.~(\ref{errorTr}). The horizontal dotted line shows $1/e$. The solid red curve in Fig.~\ref{figure-Rb-2Npi} shows the dephasing without using our protocol, which is well approximated by $\eta=e^{-(kvt_{\text{s}} )^2}$ in both Figs~\ref{figure-Rb-2Npi}(a) and~\ref{figure-Rb-2Npi}(b).

Importantly, we find that the CE effect of 
our $2\mathbb{N}\pi$ protocol can not be fitted by $e^{-(t/T_2^\ast)^2}$; a better fit is to use $\eta=e^{-t/T_2^\ast}$. The dashed blue curve in Fig.~\ref{figure-Rb-2Npi} shows $e^{-t/T_2^\ast}$, which is a fit with an estimated parameter of $T_2^\ast=100/(vk)$ and $T_2^\ast=40/(vk)$ in (a) and (b), respectively, so as to fit the CE effect. This suggests that the $1/e$ coherence time is enhanced by about 100 and 40 times with $T=10$ and $30~\mu$K, respectively. Of course, for long storage times, the motional decoherence is much less significant compared to the dephasing caused by the Rydberg-state decay. So, we can estimate an effective coherence time $(vk/100+1/\tau)^{-1}$ in (a) and $(vk/40+1/\tau)^{-1}$ in (b), respectively, where $\tau$ denotes the lifetime of Rydberg state $100S_{1/2}$ state at room temperature. According to Ref.~\cite{Beterov2009}, we have $\tau=330~\mu$s.

We then look at case 6 of Table~\ref{table2} with numerical results shown in Fig.~\ref{figure-Cs-2Npi}, where the curves denote quantities similar to those in Fig~\ref{figure-Rb-2Npi}. Due to the larger mass and smaller $k_{\text{\tiny{CE}}}$ for cesium, we expect the CE effect to be better for cesium than for rubidium. Indeed, Fig.~\ref{figure-Cs-2Npi} shows that for the case 6 of Table~\ref{table2} our
$2\mathbb{N}\pi$ protocol can enhance the $1/e$ coherence time from $1/(kv)$ to $200/(kv)$ and $50/(kv)$ when $T=10$ and $30~\mu$K, respectively. Looking at Fig.~\ref{figure-Rb-2Npi} and Fig.~\ref{figure-Cs-2Npi} together, the better enhancement compared to the case of $^{87}$Rb comes from that the mass of $^{133}$Cs is about 1.5 times that of $^{87}$Rb, thus with the same atomic temperature, both $v$ and the velocity spread is smaller, leading to a better fulfillment of the condition $|\Omega_{\text{\tiny{CE}}}|/|k_{\text{\tiny{CE}}}v|\gg1$. Similarly, for long storage times, the Rydberg-state decay will have a major contribution to the dephasing, as captured by the dash-dotted curves in Fig.~\ref{figure-Cs-2Npi}.

\begin{figure}
\includegraphics[width=3.0in]
{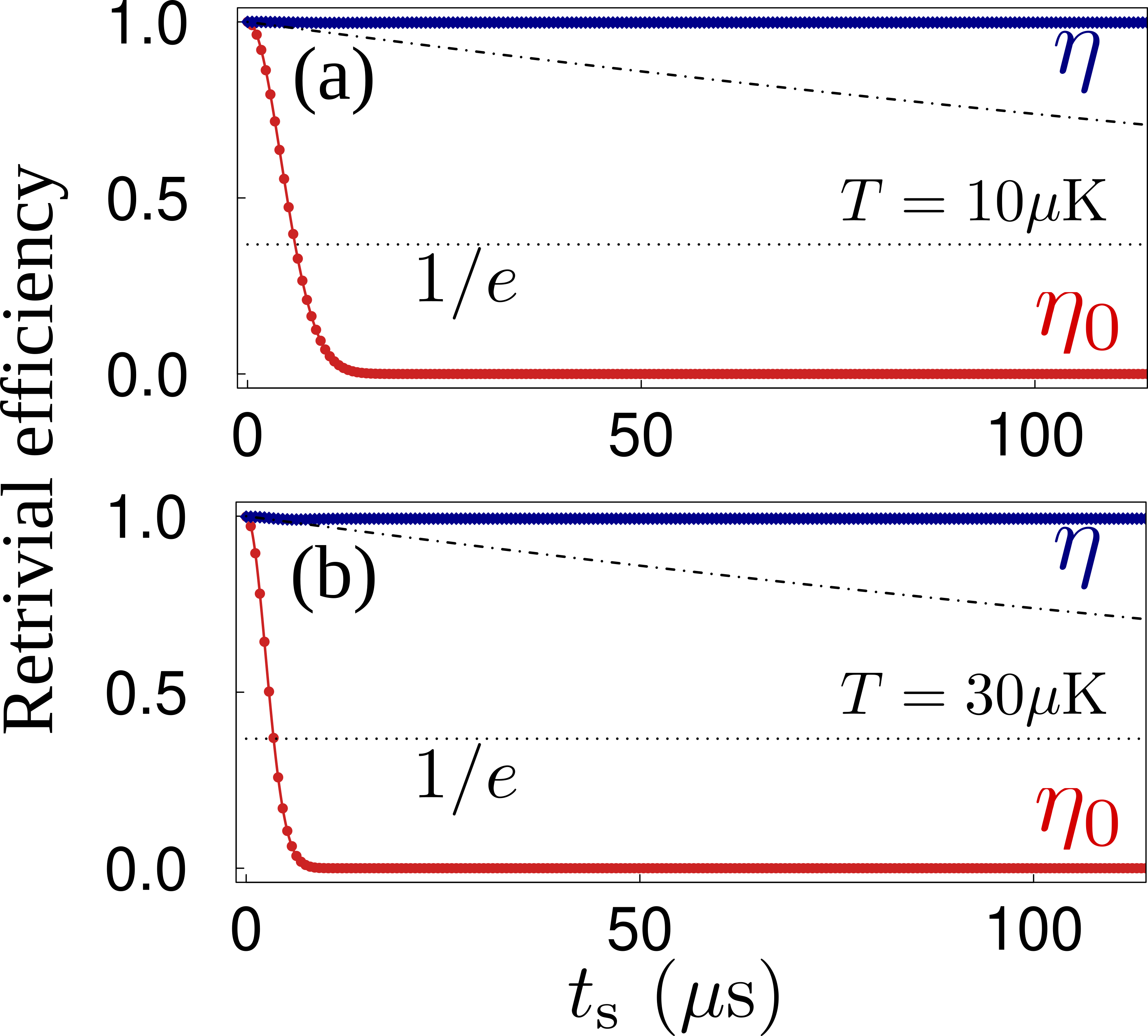}
 \caption{Coherence enhancement of rubidium Rydberg polariton by the $\pi$-wait-$\pi$ protocol of Sec.~\ref{sec03B} with case 6 of Table~\ref{table1} as an example. The curves and points show similar quantities as in Fig.~\ref{figure-Rb-2Npi}. The thick blue curve which denotes retrieval efficiency $\eta$ by our protocol is about 0.998 and 0.993 in (a) and (b), respectively, namely, it doesn't decay when the storage time increases, or, there is no dephasing from the motional effect. Then, the dephasing of Rydberg polariton is from the Rydberg-state decay denoted by the dash-dotted curve, $e^{-t_{\text{s}}/\tau}$, where $\tau$ is the lifetime of the Rydberg state, which is about $330~\mu$s for the $100S_{1/2}$ state at room temperature~\cite{Beterov2009}.   \label{figure-Rb-piwpi} }
\end{figure}

\begin{figure}
\includegraphics[width=3.0in]
{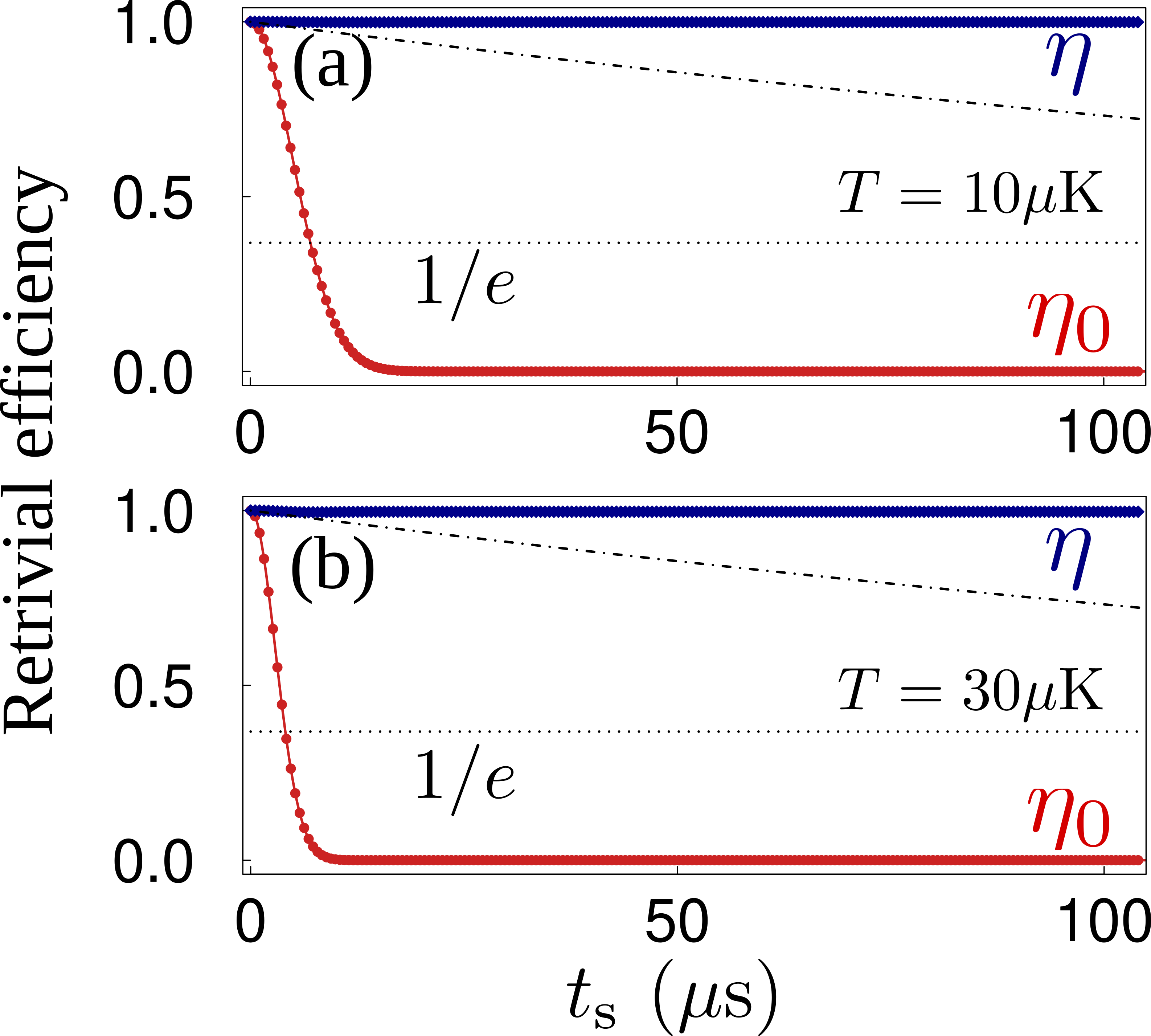}
 \caption{Coherence enhancement of cesium Rydberg polariton by the $\pi$-wait-$\pi$ protocol of Sec.~\ref{sec03B} with case 6 of Table~\ref{table2} as an example. The curves and points show similar quantities as in Fig.~\ref{figure-Rb-2Npi}. The thick blue curve which denotes retrieval efficiency $\eta$ by our protocol is about 0.999 and 0.997 in (a) and (b), respectively, namely, it doesn't decay when the storage time increases. The dash-dotted curve, $e^{-t_{\text{s}}/\tau}$, shows the Rydberg-state decay with $\tau=320~\mu$s for the $100S_{1/2}$ state at room temperature~\cite{Beterov2009}.   \label{figure-Cs-piwpi} }
\end{figure}

\subsection{Numerical results of the $\pi$-wait-$\pi$ protocol}
We use the same parameters as used in Figs.~\ref{figure-Rb-2Npi} and~\ref{figure-Cs-2Npi} to examine the CE effect by the $\pi$-wait-$\pi$ protocol presented in Sec.~\ref{sec03B}, with results shown in Figs.~\ref{figure-Rb-piwpi} and~\ref{figure-Cs-piwpi}, respectively. Importantly, the retrieval efficiency $\eta$ are 0.998 and 0.999 in Figs.~\ref{figure-Rb-piwpi}(a) and~\ref{figure-Cs-piwpi}(a), respectively. In other words, it appears that the $\pi$-wait-$\pi$ protocol can almost completely eliminate the motion-induced dephasing, as shown in Figs.~\ref{figure-Rb-piwpi}(a) and~\ref{figure-Cs-piwpi}(a).
It is  shown that 
there is no decay of the retrieval efficiency if not including the Rydberg-state decay. Similarly for the data with $T=30~\mu$K shown in Figs.~\ref{figure-Rb-piwpi}(b) and~\ref{figure-Cs-piwpi}(b), where the values of $\eta$ with the $\pi$-wait-$\pi$ protocol are equal to $0.993$ and 0.997, respectively. That the value of $\eta$ is not exactly 1 is due to that the condition $\Omega_{\text{\tiny CE}}\gg kv$ can't be perfectly satisfied. Even so, the numerical results here indicate that our CE theory in the form of the $\pi$-wait-$\pi$ protocol can nearly completely suppress the motional dephasing in Rydberg polariton.

The results in Figs.~\ref{figure-Rb-piwpi} and~\ref{figure-Cs-piwpi} indicate that by using our 
$\pi$-wait-$\pi$ protocol, Rydberg polariton will dephase in the way of a single Rydberg atom where the blackbody radiation and spontaneous emission are the cause as indicated by the dash-dotted curves in Figs.~\ref{figure-Rb-piwpi} and~\ref{figure-Cs-piwpi}.

\begin{table}
  \centering
  \begin{tabular}{c|c|cccccc}
    \hline
    \hline   Case &       $\frac{k_{\text{\tiny{CE}}}}{2k}$ &  k$\mu$m    & $\lambda_1/$nm  & $\lambda_2/$nm  & $\lambda_3/$nm  & $(\theta_1,\theta_2) $ & setup \\ \hline
1&	    2.48&  5.35 & 795.0 & 473.9   & 473.9  &	(0,0) & Fig.~\ref{figure-wait-pi-direction}\\
2&	    2.45&  5.35 & 795.0 & 473.9   &479.3   & (0.22,0.13)&	Fig.~\ref{figure-wait-pi-direction} \\
3&	    2.62&  5.06 & 780.2 &  479.3  &473.9   & $\diagdown$ &\\
4&	    2.59&  5.06 & 780.2 & 479.3   & 479.3   & (0,0)&Fig.~\ref{figure-wait-pi-direction}\\  \hline
5&	    1.17& 5.35 & 795.0  & 473.9    &1003.6   & (2.15,0.52)	& Fig.~\ref{figure-wait-pi-direction}\\
6&	    1.16&  5.35 &  795.0  & 473.9   &1011.4   &	(2.17,0.51)	& Fig.~\ref{figure-wait-pi-direction}\\
7&	    1.24&  5.06 & 780.2 &  473.9   &1003.6  & (2.01,0.59)&Fig.~\ref{figure-wait-pi-direction}	\\
8&	    1.23&  5.06 &  780.2 &  473.9  & 1011.4   & (2.03,0.58)&Fig.~\ref{figure-wait-pi-direction}	\\  \hline
9&	    1.53&  8.64 & 421.7   & 1003.6  &473.9  & (0.34,0.91) &	Fig.~\ref{figure-wait-pi-direction-2}	 \\
10&	    1.52&  8.64 &  421.7  &  1003.6 &479.3   & 	(0.35,0.96)  &Fig.~\ref{figure-wait-pi-direction-2}	\\
11&	    1.52&  8.74 & 420.3   & 1011.4  &473.9   &  (0.34,0.93) &Fig.~\ref{figure-wait-pi-direction-2}\\
12&	    1.50&  8.74 &  420.3  & 1011.4   &479.3   &	(0.35,0.98) &Fig.~\ref{figure-wait-pi-direction-2}\\ \hline\hline
  \end{tabular}
  \caption{  \label{table3} Possible level configurations for the wait-$\pi$ protocol with rubidium-87. Here, the cases (1-12) are the same as those in Table~\ref{table1}. However, case 3 here can't fulfill the condition in Eq.~(\ref{waitPiCondition}), so case 3 here can't be used in the wait-$\pi$ protocol if the laser fields and the CE fields propagate along the directions shown in Figs.~\ref{figure-wait-pi-direction}(a) and~\ref{figure-wait-pi-direction}(b). }
  \end{table}

\begin{table}
  \centering
  \begin{tabular}{c|c|cccccc}
    \hline
    \hline   Case &       $\frac{k_{\text{\tiny{CE}}}}{2k}$ &  k$\mu$m    & $\lambda_1/$nm  & $\lambda_2/$nm  & $\lambda_3/$nm  & $(\theta_1,\theta_2) $ & setup \\ \hline
1&	    2.24&  5.69 & 894.6 & 494.4 & 494.4   &	(0,0)&	Fig.~\ref{figure-wait-pi-direction}  \\
2&	    2.17&  5.69 & 894.6 & 494.4 & 508.3   &	(0.36,0.20) &	Fig.~\ref{figure-wait-pi-direction} \\
3&	    2.55&  4.99 & 852.3  & 508.3 & 494.4   & $\diagdown$ &\\
4&	    2.48&  4.99 & 852.3  & 508.3 & 508.3   & (0,0)&	Fig.~\ref{figure-wait-pi-direction}   \\  \hline
5&	    1.07&  5.69 & 894.6 &494.4  & 1037.2   & 	 (2.47,0.35)&	Fig.~\ref{figure-wait-pi-direction} \\
6&	    1.05&  5.69 & 894.6 &494.4  & 1057.1   &  	(2.58,0.30)&	Fig.~\ref{figure-wait-pi-direction}  \\
7&	    1.21&  4.99 & 852.3 & 508.3& 1037.2   & (2.04,0.56) &	Fig.~\ref{figure-wait-pi-direction}  \\
8&	    1.19&  4.99 &  852.3 & 508.3& 1057.1   & (2.09,0.54) &	Fig.~\ref{figure-wait-pi-direction} \\ \hline
9&	    1.67&  7.62 & 459.4 & 1037.2 & 494.4   & (0.29,0.69) &	Fig.~\ref{figure-wait-pi-direction-2} 	 \\
10&	    1.62&  7.62 & 459.4 & 1037.2 & 508.3   & (0.33,0.81)&	Fig.~\ref{figure-wait-pi-direction-2} 	 \\
11&	    1.62& 7.85  & 455.7 & 1057.1& 494.4   &  (0.30,0.74) &	Fig.~\ref{figure-wait-pi-direction-2}  \\
12&	    1.58&  7.85 & 455.7  & 1057.1& 508.3  & (0.33,0.87) &	Fig.~\ref{figure-wait-pi-direction-2}  \\ \hline
13&	    1.01&  12.6 & 389.0 & 1755.1& 494.4   & (0.10,2.70)	&	Fig.~\ref{figure-wait-pi-direction-2}  \\
14&	    1.00& 12.7  & 387.7 & 1781.0& 494.4   & (0.05,2.92) &	Fig.~\ref{figure-wait-pi-direction-2} \\ \hline\hline
  \end{tabular}
  \caption{  \label{table4} Possible level configurations for the wait-$\pi$ protocol with cesium-133. Here, the cases (1-14) are the same as those in Table~\ref{table2}. Case 3 here can't fulfill the condition in Eq.~(\ref{waitPiCondition}), so it can't be used in the wait-$\pi$ protocol if the laser fields and the CE fields propagate along the directions shown in Figs.~\ref{figure-wait-pi-direction}(a) and~\ref{figure-wait-pi-direction}(b).}
  \end{table}


\subsection{Numerical results of the wait-$\pi$ protocol}\label{secvid}
An interesting feature of the wait-$\pi$ protocol is that upon retrieval, the signal photon can come out in a direction that is nearly opposite to that of the loaded signal. Therefore we first explain how this happens.

\subsubsection{Propagation direction of signal photons}\label{sec05E1}
The wait-$\pi$ protocol presented in Sec.~\ref{sec03C} differs from the previous two protocols in that during the retrieval, the coupling laser and the signal photon no longer propagate along the respective directions as those during the loading stage. In order to send the coupling laser to the atoms that carry the Rydberg polariton, it is better to let the coupling laser field propagate along $\pm\mathbf{z}$ because usually the atomic cloud has a longer longitudinal axis compared to its radial axis, and, more importantly, the signal photon is stored in an elongated space along the propagation direction of the coupling laser fields during the loading. Moreover, sometimes the gas cell in the vacuum chamber is set in a configuration so that laser fields are easier to travel along or near a certain direction~(which is $\pm\mathbf{z}$ for our case). So, if the signal field also travels near $\pm\mathbf{z}$, it would be easier to collect the retrieved signal photon during the retrieval stage. In Tables~\ref{table3} and~\ref{table4}, one can see that there are two cases with $\theta_1 =\theta_2=0$, so it is possible to send the coupling laser field opposite to that during the loading while the retrieved signal will come out in the direction as in the loading, where the angles $\theta_1$ and $\theta_2$ are illustrated in Fig.~\ref{figure-wait-pi-direction}.

Another interesting case is that the retrieved signal can come out in a direction nearly opposite to that of the signal photon loaded in the gas. There are two general conditions for this. First, with $\lambda_1>\lambda_2$, as shown in Fig.~\ref{figure-wait-pi-direction}, if there is any case with $(\theta_1 , \theta_2)\approx(\pi, 0)$, then the propagation directions of the coupling laser and the signal field during the retrieval will be nearly opposite to those during the loading stage, where $\theta_1$ and $\theta_2$ are defined in Fig.~\ref{figure-wait-pi-direction}. As shown in Tables~\ref{table3} and~\ref{table4}, the cases nearest to $(\theta_1 , \theta_2)\approx(\pi, 0)$ are case 6 for rubidium and case 6 for cesium, in Tables~\ref{table3} and~\ref{table4}, respectively. Interestingly, cases 5-8 in Table~\ref{table3} and Table~\ref{table4} all have $\theta_1$ near to $\pi$, which means that for these cases the signal photon will come out in a direction nearly opposite to that of the signal photon during the loading stage.

Besides the above cases that the signal photon can come out in a direction nearly opposite to that of the loaded signal, Fig.~\ref{figure-wait-pi-direction-2} shows that when $\lambda_1<\lambda_2$, cases with conditions $(\theta_1 , \theta_2)\approx(0, \pi)$ also means the propagation directions of the coupling laser and the signal field during the retrieval will be nearly opposite to those during the loading stage, where $\theta_1$ and $\theta_2$ are defined in Fig.~\ref{figure-wait-pi-direction-2}. In this case, there is no appropriate atomic configurations with rubidium as shown in Table~\ref{table3}. But for cesium, there are two cases, cases 13 and 14. Of course, cases 9-12 in Table~\ref{table3} and cases 9-14 in Table~\ref{table4} all have a $\theta_1$ near to 0, which means that for these cases the signal photon will propagate in a direction nearly opposite to that of the signal photon during the loading stage. This indicates that our theory can enable quantum routers on the single-photon level.

We note that in Ref.~\cite{Murray2017}, a theoretical scheme was proposed to realize a Rydberg-mediated conversion between distinct types of dark-state polaritons with different propagation characteristics, resulting in signal photons going in and coming out in opposite directions. Compared to Ref.~\cite{Murray2017}, here the signal photon comes out in an opposite direction not due to Rydberg blockade, but because of the change of the direction of the wavevector of Rydberg polariton.

\subsubsection{Retrieval efficiency }
The retrieval efficiency with the wait-$\pi$ protocol is very similar to that with the $\pi$-wait-$\pi$ protocol because only one $\pi$ pulse is used with the Rydberg atoms, resulting in little population loss due to the failure of the condition $\Omega_{\text{\tiny CE}}\gg k_{\text{\tiny CE}}v$. Because of this, we don't show the simulation data by figures for the results will be very similar to those shown in Figs.~\ref{figure-Rb-piwpi} and~\ref{figure-Cs-piwpi}.

We have used numerical simulation to test typical cases presented in Tables~\ref{table3} and~\ref{table4}. However, as discussed in Sec.~\ref{sec05E1}, it would be easier to send to the atomic medium with a control laser field along the direction that is parallel or anti-parallel to the control laser used during the loading stage. Of course, as long as the angle $\theta_2$ in Fig.~\ref{figure-wait-pi-direction} or the angle $\theta_1$ in Fig.~\ref{figure-wait-pi-direction-2} are near 0 or $\pi$, it is not challenging to retrieve the signal photon with a high efficiency.

With the above consideration, we consider the easiest cases, i.e., cases 1 or 4 in Tables~\ref{table3} and~\ref{table4} for in these cases the control laser will be sent anti-parallel to the control laser used during the loading stage. More specifically, we choose case 4 in Table~\ref{table3} for rubidium, and found that the retrieval efficiency without considering Rydberg-state decay is 0.986~(0.957) with atomic temperature $10~(30)~\mu$K. This means that for long-time storage, the retrieval efficiency of the signal photon would be appropriately given by $e^{-t_{\text{s}}/\tau}$ with $\tau$ the radiative lifetime of Rydberg states, as shown by the dash-dotted curve in Fig.~\ref{figure-Rb-piwpi}. Here, the retrieval efficiency without the Rydberg-state decay is slightly smaller than that in Fig.~\ref{figure-Rb-piwpi}. This is because the wavevector $k_{\text{\tiny CE}}$ here is more than two times larger than the value of $k_{\text{\tiny CE}}$ in Fig.~\ref{figure-Rb-piwpi}. As a result, the condition $\Omega_{\text{\tiny CE}}\gg k_{\text{\tiny CE}}v$ is not well satisfied compared to the case of Fig.~\ref{figure-Rb-piwpi}. For cesium with case 4 of Table~\ref{table4},
our numerical simulation shows that the retrieval efficiency without considering Rydberg-state decay is 0.992~(0.975) with atomic temperature $10~(30)~\mu$K. Here the efficiency is higher than the case 4 of Table~\ref{table3}, which is mainly due to that the mass of cesium is larger than that of rubidium, so that the velocity has a narrower distribution, resulting in a better fulfillment of the condition $\Omega_{\text{\tiny CE}}\gg k_{\text{\tiny CE}}v$.

\subsection{Influence by finite $\Omega_{\text{\tiny{CE}}}/k_{\text{\tiny{CE}}}v$ }
Since the protocols rely on $\Omega_{\text{\tiny{CE}}}/k_{\text{\tiny{CE}}}v$, it is useful to quantitatively analyze how the retrieval efficiency will be affected by the ratio $\Omega_{\text{\tiny{CE}}}/k_{\text{\tiny{CE}}}v$. Because the velocities of the atoms in the media spread through a large interval, a rigorous analysis should consider numerical integration of the retrieval efficiency over the distribution of the atomic velocity.

\subsubsection{Population loss with finite $\Omega_{\text{\tiny{CE}}}$ }
The condition $|\Omega_{\text{\tiny{CE}}}|\gg k_{\text{\tiny{CE}}}v$ is required for our theory. This is because as analytically proven in Appendix A of Ref.~\cite{Shi2020}, only if $|\Omega_{\text{\tiny{CE}}}|\gg k_{\text{\tiny{CE}}}v$ can the phase accumulation in a $\pi$ pulse be given by the ones used in the derivation of the three protocols. Further, Fig.~3 of Ref.~\cite{Shi2020} showed that the population loss can be severe if $|\Omega_{\text{\tiny{CE}}}|\gg k_{\text{\tiny{CE}}}v$ is not satisfied, meaning that without having this condition in our protocols there can be significant population loss, i.e., the photon retrieval can have a low efficiency.
We can first consider case 6 of Table~\ref{table2} and a Rydberg atom in the state $|r_1\rangle$ at $t=0$. A $\pi$ pulse with the Hamiltonian in Eq.~(\ref{infraH}) will take the atom to the Rydberg state $|\psi(t)\rangle=e^{i\varphi_1}|r_2\rangle$ at $t=t_{\pi}\equiv \pi/\Omega_{\text{\tiny{CE}}}$. It is easy to show that $\varphi_1=kz_0-\pi/2$ if $v_j=0$, where $z_0$ is the initial coordinate of the atom. For a nonzero $v_j$ bounded by $|v_j|\ll\Omega_{\text{\tiny{CE}}}/k_{\text{\tiny{CE}}}$, Appendix A of Ref.~\cite{Shi2020} showed that $\varphi_1$ is given by
\begin{eqnarray}
\varphi_1 &=&kz_0+k_{\text{\tiny{CE}}}(z_0+v_jt_{\pi}/2)-\pi/2. \label{varphi01}
\end{eqnarray}
The value of $k_{\text{\tiny{CE}}}$ is $4\pi/1057.1$nm$^{-1}\approx1.2\times10^7$m$^{-1}$ for case 6 of Table~\ref{table2}, which means that the condition $|\Omega_{\text{\tiny{CE}}}|\gg k_{\text{\tiny{CE}}}v_j$ requires $\Omega_{\text{\tiny{CE}}}/2\pi\gg0.3$~MHz if $v_j=\sqrt{k_BT/m}$ and $T=10~\mu$K. By numerical simulation, we find that when $\Omega_{\text{\tiny{CE}}}$ is $ 2\pi\times0.5$~MHz the population loss to the $|r_2\rangle$ is large for $v_j$ over $1$cm/s. Fortunately, we also find that with $\Omega_{\text{\tiny{CE}}}= 2\pi\times2$~MHz, the population loss is marginal. However, because the speed of the atom in the atomic medium is centered around zero, a value of $\Omega_{\text{\tiny{CE}}}$ lower than $2\pi\times2$~MHz can still yield the CE effect quite well which we have tested numerically.

The above picture focuses on an individual atom. We then consider a delocalized Rydberg excitation in the Rydberg polariton,
\begin{eqnarray}
|s_1\rangle&=& \frac{1}{\sqrt{N}}\sum_{j=1}^{N}e^{ikz_j} |gg\cdots r_1^{(j)}ggg\rangle,\nonumber
\end{eqnarray}
where $(j)$ denotes the location of the $j$-th atom. It can be easily shown that the above state is excited to
\begin{eqnarray}
|s_2\rangle&=& \frac{-i}{\sqrt{N}}\sum_{j=1}^{N}e^{i(k-k_{\text{\tiny{CE}}})z_j-ik_{\text{\tiny{CE}}}v_jt_{\pi}/2} |gg\cdots r_2^{(j)}ggg\rangle,\nonumber
\end{eqnarray}
by a Hamiltonian similar to the one in Eq.~(\ref{infraH}). So, the picture for one atom is still valid for the collective Rydberg excitation.

\begin{figure}
\includegraphics[width=3.0in]
{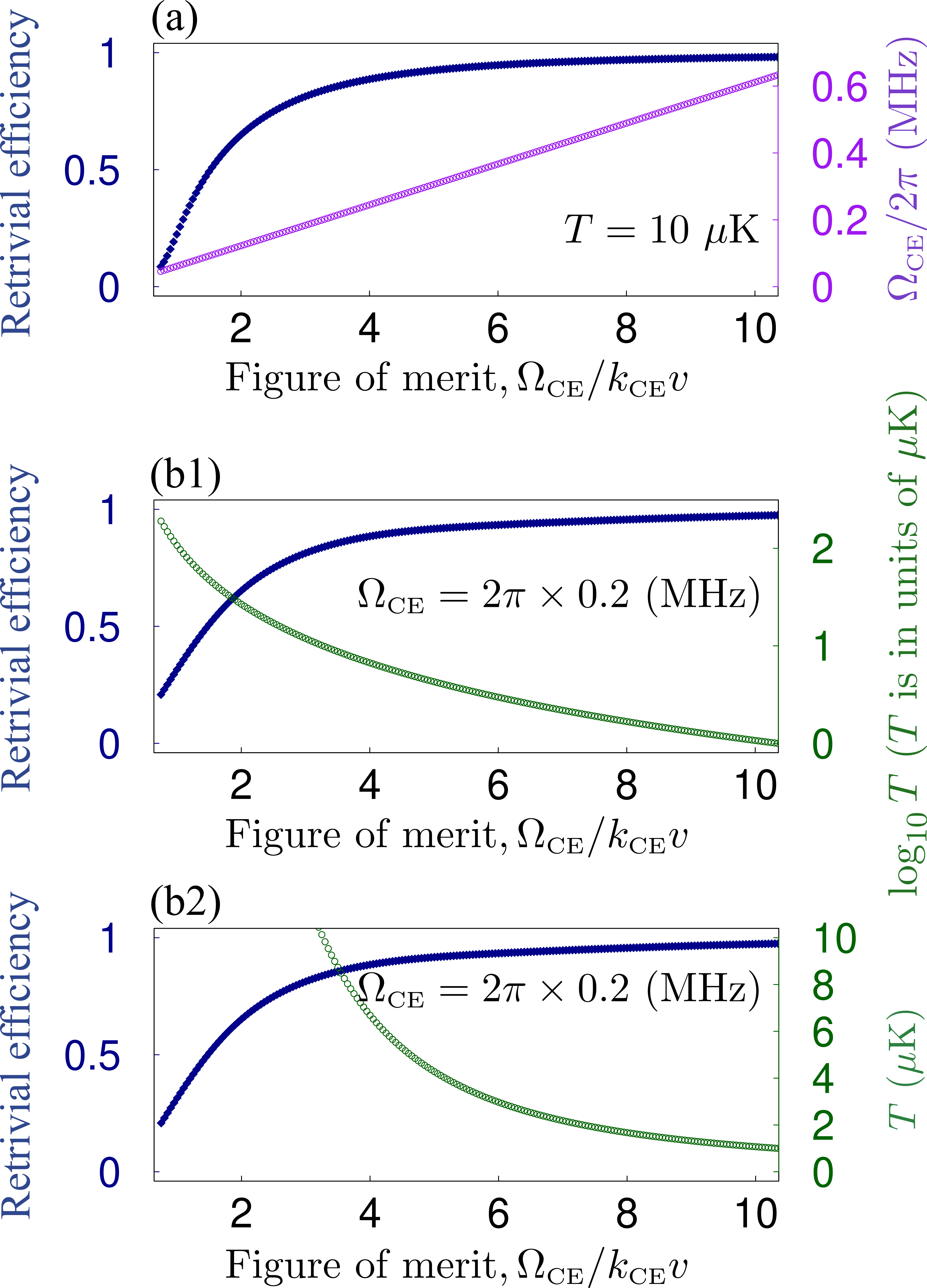}
 \caption{The value of $\eta$ in the retrieval efficiency $\eta e^{-t_{\text{s}}/\tau}$ as a function of $\Omega_{\text{\tiny{CE}}}/k_{\text{\tiny{CE}}}v$ with case 6 of Table~\ref{table1} as an example, with $v$ fixed in (a), i.e., the atomic temperature is fixed, and with $\Omega_{\text{\tiny{CE}}}$ fixed in (b1,b2), respectively. The green symbols in (b1) and (b2) show the motional temperature of the atoms in log$_{10}$ scale and in normal scale, respectively. Here, the simulation is via the $\pi$-wait-$\pi$ protocol with $t_{\text{s}}=30~\mu$s, and the final photonic readout efficiency is equal to the $\eta$ indicated by the blue diamond symbols multiplied by $e^{-t_{\text{s}}/\tau}\approx91\%$ when using Rydberg states of principal quantum numbers around $100$. At $\Omega_{\text{\tiny{CE}}}/k_{\text{\tiny{CE}}}v=(2,4,6,8,10)$, the values of $\eta$ in (a) and (b1) are $(0.65,0.89,0.95,0.97,0.98)$ and $(0.65,0.88,0.93,0.96,0.97)$, respectively. The atomic temperature shown in (b2) is in normal scale, which indicates that if one uses a small Rabi frequency $\Omega_{\text{\tiny{CE}}}=2\pi\times0.2$~MHz, cooling atoms to $1~\mu$K is necessary for achieving $\eta=98\%$ for a storage time $30~\mu$s.   \label{figure-scalingRb} }
\end{figure}

\subsubsection{Retrieval efficiency as a function of $\Omega_{\text{\tiny{CE}}}/k_{\text{\tiny{CE}}}v$ }\label{secVIE2}
$\Omega_{\text{\tiny{CE}}}$ can be tuned by adjusting the power of the external lasers, $v$ is determined by the motional temperature $T$ of the atom once the atomic species is chosen, and $k_{\text{\tiny{CE}}}$ is intrinsic, not adjustable once the level configuration is chosen. As a result, to investigate the retrieval efficiency as a function of $\Omega_{\text{\tiny{CE}}}/k_{\text{\tiny{CE}}}v$, one can either vary $\Omega_{\text{\tiny{CE}}}$ or change $T$. By using the same method as used above with $\eta$ defined in a way of Eqs.~(\ref{errorOur}) and~(\ref{errorOur-WaitPi}), we have numerically studied $\eta$ in the retrieval efficiency $\eta e^{-t_{\text{s}}/\tau}$ with the parameters in, e.g., case 6 of Table~\ref{table1}. Remarkably, the retrieval efficiency shown in Fig.~\ref{figure-scalingRb} indicates that for $\Omega_{\text{\tiny{CE}}}/k_{\text{\tiny{CE}}}v$ over 2, the values of $\eta$ with different $\Omega_{\text{\tiny{CE}}}$ in Fig.~\ref{figure-scalingRb}(a) are quite near to those in Fig.~\ref{figure-scalingRb}(b1,b2) when varying $v$~(or, equivalently, $T$). The range of $\Omega_{\text{\tiny{CE}}}$ in Fig.~\ref{figure-scalingRb}(a) is $2\pi\times(0.046,0.67)$~MHz. To have $\eta>90\%$, we need $\Omega_{\text{\tiny{CE}}}/k_{\text{\tiny{CE}}}v>4.3$, or $\Omega_{\text{\tiny{CE}}}>2\pi\times0.26$~MHz in Fig.~\ref{figure-scalingRb}(a). These show that without using large $\Omega_{\text{\tiny{CE}}}$ the theory is well behaved.

The results in Fig.~\ref{figure-scalingRb} show that moderate laser powers and not too cold atom medium are fine with the theory. For example, Refs.~\cite{Jiao2024,Li2025} tested case 4 of Table~\ref{table2} with cesium media of $T\approx40~\mu$K, where $\Omega_{\text{\tiny{CE}}}/2\pi$ was about $1~(0.9)$~MHz in Refs.~\cite{Jiao2024,Li2025}, leading to $\Omega_{\text{\tiny{CE}}}/k_{\text{\tiny{CE}}}v\approx5.1~(4.6)$. According to Fig.~\ref{figure-scalingRb}, the value of $\eta$ for $\Omega_{\text{\tiny{CE}}}/k_{\text{\tiny{CE}}}v\sim5$ can yield $\eta>90\%$. This was why the results in Refs.~\cite{Jiao2024,Li2025} showed that the theory worked well. Microscopically, this is because as proven in Ref.~\cite{Shi2020} and shown in Fig.~3 therein, the deviation of $\Omega_{\text{\tiny{CE}}}/k_{\text{\tiny{CE}}}v$ from $\infty$ is mainly a population loss, which is quite marginal for $\Omega_{\text{\tiny{CE}}}/2\pi\geq1$~MHz even with the atomic speed $v_j$ up to $20$~cm$/$s. But with $T\sim40~\mu$K as in Refs.~\cite{Jiao2024,Li2025}, the probability for an atom to have $v_j>$$20$~cm$/$s is less than $10^{-5}$.

\section{Application in quantum information processing}\label{sec06}
A direct application of the CE theory is in quantum information processing with single photons. As an example, we show a simple but new protocol to realize a CZ quantum gate between single photons via the $\pi$-wait-$\pi$ protocol. Below, we present the gate protocol, an example for the level configuration, and an analysis for the achievable gate fidelity which can be over 0.9 with attainable experimental conditions.

\subsection{A photon quantum gate}\label{sec06A}
Consider two atomic media with their longitudinal axes parallel to each other and their centers separated by a distance $d$ which is large compared to the longitudinal length $L$ of each atom medium. Later on, we show that high-fidelity gate is achievable with $L\ll d$.

We further consider a photonic qubit defined by the circular polarizations of a single photon, and label the right and left-hand polarized qubit by $\lvert \circlearrowright\rangle$ and $\lvert \circlearrowleft\rangle$, respectively. By using EIT, the two different photonic modes are mapped to two Rydberg polaritons of different electron angular momenta $\lvert n S_{1/2}, m_J =m_{\text{\tiny r}}\rangle$ and $\lvert n S_{1/2}, m_J =m_{\text{\tiny l}}\rangle$,
\begin{eqnarray}
\lvert \circlearrowright\rangle &\rightarrow& \lvert e_{\text{\tiny r}}\rangle \rightarrow\lvert n S_{1/2},m_{\text{\tiny r}}\rangle,\nonumber\\
\lvert \circlearrowleft\rangle &\rightarrow& \lvert e_{\text{\tiny l}}\rangle\rightarrow \lvert nS_{1/2},m_{\text{\tiny l}}\rangle.
\end{eqnarray}
In order to have negligible Rydberg interaction between the two Rydberg polariton in the two atomic media, $n$ can be of a moderate value, like $60$, so that for $L$ around $20~\mu$m, the interaction between the Rydberg polaritons during loading is tiny, and the dissipative process of the Rydberg dark-state polariton~\cite{Gorshkov2011,Gorshkov2013} can be safely ignored.

\begin{figure*}
\includegraphics[width=6.0in]
{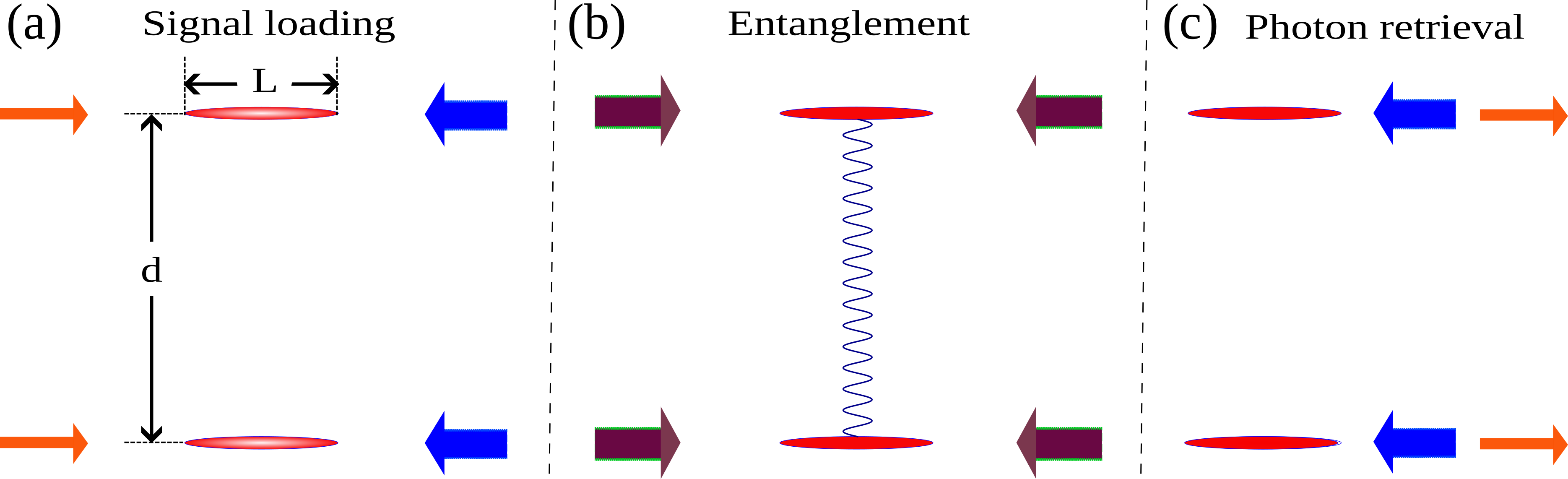}
 \caption{Sequence to induce interaction-induced dynamical phase between two signal photons. With two ultracold atomic media separated by $d$, each with a longitudinal extension $L$, two signal photons are loaded into the two separate media. For $d$ of several microns and $n$ around 60, the interaction between an arbitrary atom in one atomic and another atom in the other atomic medium will be weak enough compared to the Rabi frequency of the coupling laser field or the EIT linewidth if using EIT for loading. Better gate fidelity is conditioned on $L/d$ being much smaller than 1 so that the interaction between one Rydberg atom in one atomic medium and another Rydberg atom in the other atomic medium is nearly uniform.    \label{figure-photon-gate} }
\end{figure*}

During the storage, a $\pi$ pulse is applied to each type of Rydberg polariton, so that
\begin{eqnarray}
\lvert n S_{1/2},m_{\text{\tiny r}}\rangle &\rightarrow& \lvert n_{\text{\tiny r}} S_{1/2},m_{\text{\tiny r}}\rangle ,\nonumber\\\lvert n
  S_{1/2},m_{\text{\tiny l}}\rangle&\rightarrow&\lvert n_{\text{\tiny l}}
  S_{1/2},m_{\text{\tiny l}}\rangle,\label{photongate01}
\end{eqnarray}
where $n_{\text{\tiny r}}$ and $n_{\text{\tiny l}}$ are two different principal quantum numbers which satisfy the condition
\begin{eqnarray}
V(n_{\text{\tiny r}}n_{\text{\tiny r}})\gg V(n_{\text{\tiny l}}n_{\text{\tiny l}}),V(n_{\text{\tiny l}}n_{\text{\tiny r}}),V(n_{\text{\tiny r}}n_{\text{\tiny l}}), \label{largeV}
\end{eqnarray}
where $V(xy)$ denotes the Rydberg interaction between two atoms in the state $\lvert xS_{1/2}\rangle\otimes \lvert yS_{1/2}\rangle$. Moreover, in order to efficiently realize Eq.~(\ref{photongate01}), we impose the condition
\begin{eqnarray}
\Omega_{\text{\tiny CE}}\gg V(n_{\text{\tiny r}}n_{\text{\tiny r}}).
\end{eqnarray}
Then, we wait for a time specified by Eq.~(\ref{protocol2}), and, simultaneously, 
$L$ and $n_{\text{\tiny r}}$ shall be chosen so that
\begin{eqnarray}
  V(n_{\text{\tiny r}}n_{\text{\tiny r}})t_{\text{\tiny wait}} =(2\mathbb{N}+1)\pi, \label{wait01}
\end{eqnarray}
where $\mathbb{N}$ is an integer. In principle, we need only to have $\mathbb{N}=0$ so that the shortest storage time is used so as to suppress dephasing. However, we find that in order to have nearly uniform phase accumulation in the single-photon Rydberg polariton, it is useful to have a short atomic cloud whose length $L$ should be much smaller than the inter-cloud distance $d$, illustrated in Fig.~\ref{figure-photon-gate}. With a small $L$, $d$ can also be small, so that the gate duration can be small due to the strong Rydberg interaction with small $d$. However, during the experimental implementation, timing is essential to have the desired phase, and to avoid large timing error with a small gate duration, one can choose $\mathbb{N}\geq1$ if a laboratory doesn't have the capability to rapidly switch laser field.

After the retrieval of the photonic modes, the phase of the Rydberg polariton maps to the photon. By this, we have realized the following quantum gate,
\begin{eqnarray}
\lvert \circlearrowright\circlearrowright\rangle &\rightarrow&- \lvert \circlearrowright\circlearrowright\rangle  ,\nonumber\\
\lvert \circlearrowright\circlearrowleft\rangle &\rightarrow& \lvert \circlearrowright\circlearrowleft\rangle,\nonumber\\
\lvert \circlearrowleft\circlearrowright\rangle &\rightarrow& \lvert \circlearrowleft\circlearrowright\rangle,\nonumber\\
\lvert \circlearrowleft \circlearrowleft\rangle &\rightarrow& \lvert \circlearrowleft \circlearrowleft\rangle. \label{IdealGate}
\end{eqnarray}
where the gate error will be mainly from the Rydberg-state decay.

\subsection{Residual interactions}\label{sec06B}
The gate map in Eq.~(\ref{IdealGate}) hinges on Eq.~(\ref{largeV}). In practice, it is impossible to have the three interactions on the right side of Eq.~(\ref{largeV}) to be zero, so that there will be undesired phase accumulations in the three lower input states in Eq.~(\ref{IdealGate}). In other words, we shall instead consider an adjusted wait time so that the gate map becomes,
\begin{eqnarray}
\lvert \circlearrowright\circlearrowright\rangle &\rightarrow&e^{i\varphi_{1}}\lvert \circlearrowright\circlearrowright\rangle  ,\nonumber\\
\lvert \circlearrowright\circlearrowleft\rangle &\rightarrow& e^{i\varphi_{2}}\lvert \circlearrowright\circlearrowleft\rangle,\nonumber\\
\lvert \circlearrowleft\circlearrowright\rangle &\rightarrow& e^{i\varphi_{2}}\lvert \circlearrowleft\circlearrowright\rangle,\nonumber\\
\lvert \circlearrowleft \circlearrowleft\rangle &\rightarrow& e^{i\varphi_{3}}\lvert \circlearrowleft \circlearrowleft\rangle,\label{photonGate}
\end{eqnarray}
where~\cite{PhysRevLett.85.2208}
\begin{eqnarray}
\varphi_1:\varphi_2:\varphi_3 = V(n_{\text{\tiny r}}n_{\text{\tiny r}}):V(n_{\text{\tiny l}}n_{\text{\tiny r}}):V(n_{\text{\tiny l}}n_{\text{\tiny l}}).\label{phi123-v}
\end{eqnarray}
By using single-qubit gates,
\begin{eqnarray}
 \lvert \circlearrowright\rangle &\longmapsto&e^{i\varphi_{3}/2-i\varphi_{2}}\lvert \circlearrowright\rangle  ,\nonumber\\
 \lvert \circlearrowleft\rangle &\longmapsto&e^{-i\varphi_{3}/2}\lvert \circlearrowleft\rangle  ,
\end{eqnarray}
the gate map will become diag$\{e^{\varphi_1-2\varphi_2+\varphi_3},1,1,1\}$. So, Eq.~(\ref{wait01}) shall be updated to
\begin{eqnarray}
 [V(n_{\text{\tiny r}}n_{\text{\tiny r}})-2V(n_{\text{\tiny l}}n_{\text{\tiny r}})+ V(n_{\text{\tiny l}}n_{\text{\tiny l}}) ]t_{\text{\tiny wait}} =(2\mathbb{N}+1)\pi. \label{wait02}
\end{eqnarray}
This means that in principle by correctly updating Eq.~(\ref{wait01}) to Eq.~(\ref{wait02}), the gate can attain a high fidelity. Short gate durations come with smaller Rydberg-state decay, so the highest fidelity for any set of atomic states is achieved with $\mathbb{N}=0$, which we assume later on.

During the idle time specified by Eq.~(\ref{prot2gap}), there will be a global phase accumulation $-V(nn)t_{\text{\tiny gap}} $. However, this phase appears for all the four independent input states, therefore it is trivial.

\subsection{An example for the gate}\label{sec06C}
The gate protocol can be realized with any choice in Table~\ref{table1} and Table~\ref{table2}. In a practical implementation, however, we would like the total storage time of Rydberg polariton to be as short as possible so as to avoid the residual decoherence of Rydberg polariton due to Rydberg-state decay. Therefore, it is better to choose a case in Table~\ref{table1} and Table~\ref{table2} with a small enough $\frac{k_{\text{\tiny{CE}}}}{2k}$ so that the wait time in Eq.~(\ref{protocol2}) is near to the total storage time $t_{\text{\tiny s}}$. Then, the idle time in Eq.~(\ref{prot2gap}) is minimal so as to avoid extra Rydberg-state decay.

Armed with the above understanding, we analyze a photon gate with the level configuration of case (vii) in Table~\ref{table2} and with parameters $(n,n_{\text{\tiny l}}, n_{\text{\tiny r}}) = (60, 63, 100)$. By using second-order perturbation~\cite{Walker2008, Shi2014,Shi2021qst} and quantum defects from Refs.~\cite{Lorenz1984,PhysRevA.35.4650}, we find that
\begin{eqnarray}
&&[V(n_{\text{\tiny r}}n_{\text{\tiny r}}),  V(n_{\text{\tiny l}}n_{\text{\tiny l}}),V(n_{\text{\tiny l}}n_{\text{\tiny r}}), V(nn) ]/2\pi \nonumber\\&&\approx [47000,190, -24 ,110]\frac{\mu\text{m}^6\text{GHz}}{d^6}, \label{RydbergV}
\end{eqnarray}
where $V(n_{\text{\tiny r}}n_{\text{\tiny l}})$ is equal to $V(n_{\text{\tiny l}}n_{\text{\tiny r}})$, and the calculation is performed by assuming that the two Rydberg atoms are separated along a direction perpendicular to the quantization axis. Moreover, Eq.~(\ref{RydbergV}) means that when a conditional $\pi$ phase appears for the input state
$\lvert \circlearrowright\circlearrowright\rangle$, the residual phases for the input states $\lvert \circlearrowright\circlearrowleft\rangle, \circlearrowleft\circlearrowright\rangle,$ and $
\lvert \circlearrowleft \circlearrowleft\rangle$, being proportional to $V(n_{\text{\tiny r}}n_{\text{\tiny r}}), V(n_{\text{\tiny l}}n_{\text{\tiny r}})$ and $ V(n_{\text{\tiny l}}n_{\text{\tiny l}})$, are tiny, namely,
\begin{eqnarray}
\varphi_1:\varphi_2:\varphi_3 \approx 1: -5\times10^{-4}: 4\times10^{-3},
\end{eqnarray}
which means that the wait times in Eq.~(\ref{wait01}) and~(\ref{wait02}) differ by about $0.5\%$.

\subsection{Accuracy of the gate}\label{sec06D}
The infidelity of the photonic entanglement comes from three sources. First, the state of Rydberg polariton can't perfectly accumulate the dynamical phase as predicted in Eq.~(\ref{photonGate}) because Rydberg polaritons are stored in elongated atomic media as shown in Fig.~\ref{figure-photon-gate}. Second, when the polariton is retrieved, the fact that different atoms in each of the atomic media can accumulate different dynamical phases can lead to imperfect conversion from Rydberg polariton to photon, resulting in an extra gate error. These two sources of gate infidelity come from the same issue, i.e., different atom pairs in the two Rydberg polaritons hosted in the two atomic media accumulate different interaction-induced phases, and both of them will affect the final gate fidelity: the first directly hampers the fidelity, the second influences the transfer efficiency from Rydberg polariton to photon. Finally, the Rydberg-state decay also causes error. We will sequentially study them below.

\begin{figure}
\includegraphics[width=3.5in]
{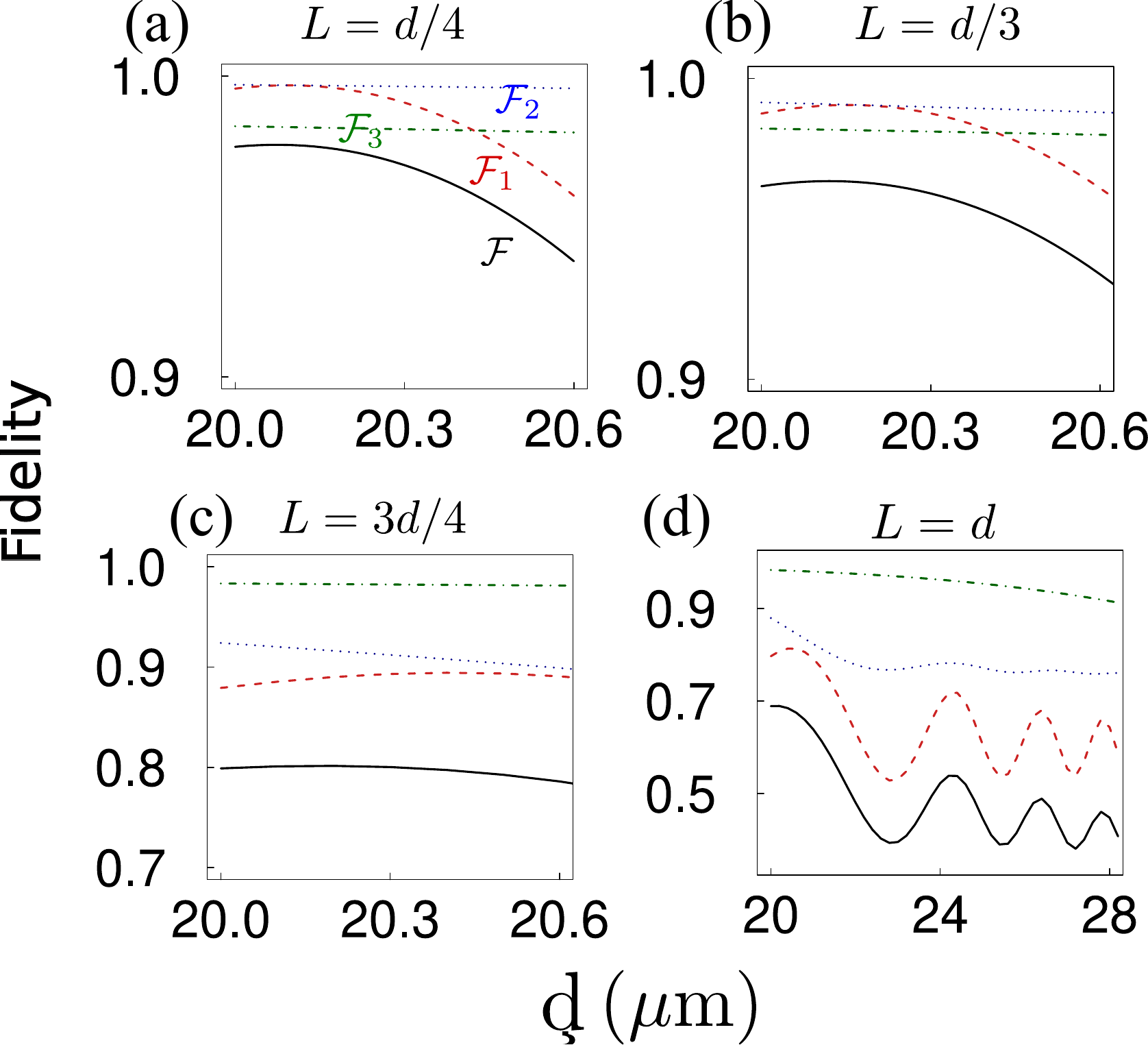}
 \caption{Fidelity of the photonic quantum gate with $d=20~\mu$m as a function of $\c{d}$~[see text around Eq.~(\ref{Lwait})] with $L/d$ equal to $1/4, 1/3, 3/4$, and 1 in (a,b,c) and (d), respectively. The gate-map fidelity $\mathcal{F}_1$, the read-out fidelity $\mathcal{F}_2$, and the Rydberg-state-decay limited fidelity $\mathcal{F}_3$ are shown by the dashed, dotted, and dash-dotted curves, respectively. The total fidelity $\mathcal{F}$ is shown by the solid curve, whose largest value occurs at $\c{d}= 20.08,~20.12,~20.2$, and $20.2~\mu$m in (a,~b,~c) and (d), respectively. The Rydberg-state decay is a major error source, where we have chosen $(n,n_{\text{\tiny l}}, n_{\text{\tiny r}}) = (60, 63, 100)$ for each case here. \label{figure-fidelity} }
\end{figure}

\subsubsection{Gate-map fidelity with Rydberg polariton }\label{sec06C1}
We consider the geometry in Fig.~\ref{figure-photon-gate} to evaluate the direct fidelity, i.e., the fidelity of the gate on the Rydberg polariton. If we absorb the phase of the phase coherent state of Rydberg polariton into the definition of Rydberg state, then the wavefunction associated with Eq.~(\ref{State-S1}) can be rewritten as
\begin{eqnarray}
\lvert s\rangle&=& \frac{1}{\sqrt{N}} \sum_{j=1}^{N} |gg\cdots r_1^{(j)}\cdots ggg\rangle,\label{Rydberg polariton-state-1}
\end{eqnarray}
based on which we consider two-photon polariton state associated with different photonic modes as from Eq.~(\ref{photonGate}),
\begin{eqnarray}
\lvert ss(\circlearrowright\circlearrowright)\rangle ,~
\lvert  ss(\circlearrowright\circlearrowleft)\rangle ,~
\lvert  ss(\circlearrowleft\circlearrowright)\rangle ,~
\lvert  ss(\circlearrowleft \circlearrowleft)\rangle.
\end{eqnarray}
These states, as in Eq.~(\ref{photonGate}), shall transform according to
\begin{eqnarray}
\lvert ss(\circlearrowright\circlearrowright)\rangle &\rightarrow&e^{i\varphi_{1}}\lvert  ss(\circlearrowright\circlearrowright)\rangle  ,\nonumber\\
\lvert  ss(\circlearrowright\circlearrowleft)\rangle &\rightarrow& e^{i\varphi_{2}}\lvert ss(\circlearrowright\circlearrowleft)\rangle,\nonumber\\
\lvert  ss(\circlearrowleft\circlearrowright)\rangle &\rightarrow& e^{i\varphi_{2}}\lvert ss(\circlearrowleft\circlearrowright)\rangle,\nonumber\\
\lvert  ss(\circlearrowleft \circlearrowleft)\rangle &\rightarrow& e^{i\varphi_{3}}\lvert ss(\circlearrowleft \circlearrowleft)\rangle,\label{photonGate2}
\end{eqnarray}
which can be labelled $\hat{U}$. The state transform in Eq.~(\ref{photonGate2}) can be exact only when $d\gg L$. We consider the choice of storage times discussed in Sec.~\ref{sec06B} by assuming that the distance between any two atom pairs in the two-Rydberg polariton state is $d$, then this means that for the input state $\lvert ss(\circlearrowright\circlearrowright)\rangle$ expanded as
\begin{eqnarray}
 \frac{1}{N} \sum_{j=1}^{N} \sum_{k=1}^{N}|gg\cdots r_1^{(j)}\cdots ggg\rangle\otimes|gg\cdots r_1^{(k)}\cdots ggg\rangle,\label{Rydberg polariton-state-2}
\end{eqnarray}
where we have not explicitly shown the angular-momentum dependence in the Rydberg state, we have the state transform,
\begin{eqnarray}
&&\lvert ss(\circlearrowright\circlearrowright)\rangle\rightarrow\lvert ss(\circlearrowright\circlearrowright)\rangle'\nonumber\\
&&=\frac{1}{N}\sum_{j=1}^{N} \sum_{k=1}^{N}e^{i\varphi_1^{(jk)}}|gg\cdots r_1^{(j)}\cdots ggg\rangle\otimes|gg\cdots r_1^{(k)}\cdots ggg\rangle\nonumber\\
&&=\frac{e^{i\varphi_1}}{N}\sum_{j=1}^{N} \sum_{k=1}^{N}e^{i\varphi_1^{(jk)}-i\varphi_1}|gg\cdots r_1^{(j)}\cdots ggg\rangle\nonumber\\&&~~~~\otimes|gg\cdots r_1^{(k)}\cdots ggg\rangle
\label{Rydberg polariton-state-3}
\end{eqnarray}
where
\begin{eqnarray}
\varphi_1^{(jk)} = \frac{d^6}{[d^2+(z_j-z_k)^2]^3}\varphi_1,
\end{eqnarray}
where $z_{j(k)}$ is the z-component of the coordinate of the $j$th~($k$th) atom in the first~(second) atomic medium, where $z_{j(k)}\in[0,L]$. The above condition means that among the $N^2$ atom pairs, only $N$ pairs with $z_j=z_k$ will induce the correct phase $\varphi_1$, while all the other phases will be smaller. This can cause gate infidelity. Therefore we suppose that there can be an optimal storage time that is a little longer than the one determined by Eq.~(\ref{wait02}), in other words, we would consider the correct phase $\varphi_1$ is accumulated by atom pairs separated by a certain $\c{d}$ which is larger than $d$, so that the above equation becomes,
\begin{eqnarray}
\varphi_1^{(jk)} = \frac{\c{d}^6}{[d^2+(z_j-z_k)^2]^3}\varphi_1,\label{Lwait}
\end{eqnarray}
where the value of $\c{d}$ can be determined numerically as shown later, with a corresponding wait time given by $t_{\text{\tiny wait}} = \varphi_1/V(n_{\text{\tiny r}}n_{\text{\tiny r}})$, and the storage time given via Eq.~(\ref{protocol2}). With such a choice, we can have the other three state transformations for the input states $\lvert ss(\circlearrowright\circlearrowleft)\rangle ,~
\lvert  ss(\circlearrowleft\circlearrowright)\rangle ,~
\lvert  ss(\circlearrowleft \circlearrowleft)\rangle$, each like that in Eq.~(\ref{Rydberg polariton-state-3}). These state transform can be labelled as a gate map $\hat{\mathcal{U}}$. As a consequence, the direct fidelity is given by~\cite{Pedersen2007}
\begin{equation}
\mathcal{F}_1 = \left[|\text{Tr}(\hat{U}^\dag \hat{\mathcal{U}})|^2 + \text{Tr}(\hat{U}^\dag\hat{\mathcal{U}}\hat{\mathcal{U}}^\dag\hat{U} )\right]/20.\label{fidelity-definition}
\end{equation}
To evaluate Eq.~(\ref{fidelity-definition}), we shall consider the inner product
\begin{eqnarray}
f_1
&=&\frac{1}{N}\langle ss(\circlearrowright\circlearrowright)\rvert \sum_{j=1}^{N} \sum_{k=1}^{N}e^{i\varphi_1^{(jk)}-i\varphi_1}|gg\cdots r_1^{(j)}\cdots ggg\rangle\nonumber\\&&~~~~\otimes|gg\cdots r_1^{(k)}\cdots ggg\rangle\nonumber\\
&=& \frac{1}{N^2}\sum_{j=1}^{N} \sum_{k=1}^{N}e^{i\varphi_1^{(jk)}-i\varphi_1}\nonumber\\
&=& \frac{1}{N^2}\sum_{j=1}^{N} \sum_{k=1}^{N}\exp\left[ i\varphi_1  \left(\frac{\c{d}^6}{[d^2+(z_j-z_k)^2]^3}-1 \right)\right],\nonumber
\end{eqnarray}
where $\varphi_1 $ is determined by Eqs.~(\ref{phi123-v}) and~(\ref{wait02}). With a calculation similar to the above one, we can have $f_2$ and $f_3$ for the input states $\lvert  ss(\circlearrowright\circlearrowleft)\rangle$ and $
\lvert  ss(\circlearrowleft \circlearrowleft)\rangle$. Equation~(\ref{fidelity-definition}) shows that the fidelity is
\begin{eqnarray}
 \mathcal{F}_1 = (\left|f_1+2f_2+f_3\right|^2 + |f_1|^2+2|f_2|^2+|f_3|^2)/20.\label{F1}
\end{eqnarray}
We can therefore use the $C_6$ coefficients in Eq.~(\ref{RydbergV}) to search the optimal $\c{d}$.

\subsubsection{Fidelity of read-out }\label{sec06C2}
The gate operation in the Rydberg polariton does not guarantee that the state can be perfectly transferred to the photonic state. Therefore we should include another error: the failure of the state transfer from Rydberg polariton to photon. If the output state of Rydberg polariton should be $\lvert ss(\circlearrowright\circlearrowright)\rangle$, then the retrieval efficiency is given by $|\langle ss(\circlearrowright\circlearrowright)\rvert  ss(\circlearrowright\circlearrowright)\rangle'|^2=|f_1|^2$. Similarly, the retrieval efficiency would be $|f_2|^2,|f_2|^2,$ or $|f_3|^2$ if the input state is $\lvert  ss(\circlearrowright\circlearrowleft)\rangle ,~
\lvert  ss(\circlearrowleft\circlearrowright)\rangle$, or $\lvert  ss(\circlearrowleft \circlearrowleft)\rangle$, respectively. On average, the read-out fidelity is captured by
\begin{eqnarray}
\mathcal{F}_2=  ( |f_1|^2+2|f_2|^2+|f_3|^2)/4.\label{F2}
\end{eqnarray}

\subsubsection{Rydberg-state decay}
The Rydberg polariton will experience Rydberg-state decay during the storage time. For the input state $\lvert \circlearrowright\circlearrowright\rangle$, the two Rydberg polaritons will experience Rydberg-state decay with a probability exp$[-2\frac{t_{\text{wait}}}{\tau(n_{\text{r}})}-2 \frac{t_{\text{s}}-t_{\text{wait}}}{\tau(n)}]$, where $\tau(n),\tau(n_{\text{\tiny l}})$, and $\tau(n_{\text{\tiny r}})$ denote the radiative lifetimes of Rydberg states of principal quantum numbers $n,n_{\text{\tiny l}}$, and $n_{\text{\tiny r}}$, respectively. Because $n$ and $n_{\text{\tiny l}}$ are near to each other, we use $\tau(n)$ for $\tau(n_{\text{\tiny l}})$ in the numerical study. For the input states $\lvert \circlearrowright\circlearrowleft\rangle$ and $
\lvert \circlearrowleft \circlearrowleft\rangle$, the Rydberg-state decay will have probabilities exp$[-t_{\text{wait}}\left(\frac{1 }{\tau(n_{\text{r}})} + \frac{1 }{\tau(n_{\text{l}})} \right)-2 \frac{t_{\text{s}}-t_{\text{wait}}}{\tau(n)}]$ and exp$[-2\frac{t_{\text{wait}}}{\tau(n_{\text{l}})}-2 \frac{t_{\text{s}}-t_{\text{wait}}}{\tau(n)}]$, respectively. The case for $\lvert \circlearrowleft\circlearrowright\rangle$ is identical to that of $\lvert \circlearrowright\circlearrowleft\rangle$. On average, the Rydberg-state decay will induce an error characterized by the fidelity
\begin{eqnarray}
\mathcal{F}_3 &=&  \frac{1}{4}\left[ e^{ -2\frac{t_{\text{wait}}}{\tau(n_{\text{r}})}-2 \frac{t_{\text{s}}-t_{\text{wait}}}{\tau(n)}}+ e^{ -2\frac{t_{\text{wait}}}{\tau(n_{\text{l}})}-2 \frac{t_{\text{s}}-t_{\text{wait}}}{\tau(n)}}\right] \nonumber\\
&&+\frac{1}{2} e^{ -t_{\text{wait}}\left(\frac{1 }{\tau(n_{\text{r}})} + \frac{1 }{\tau(n_{\text{l}})} \right)-2 \frac{t_{\text{s}}-t_{\text{wait}}}{\tau(n)} } .  \label{F3}
\end{eqnarray}
Summarizing the three sources of error, the final gate fidelity is given by \begin{eqnarray}
\mathcal{F} &=&  \mathcal{F}_1 \mathcal{F}_2 \mathcal{F}_3.
\end{eqnarray}
According to the above condition, one can see that there is no obvious clue about which $\c{d}$ can give the largest fidelity. Therefore we resort to a numerical search for the optimal value of $\c{d}$, each with a specific storage time. For different $d$, the value of Rydberg interactions would be different, so the time required to have the desired phases in different cases would be different. This means that $\mathcal{F}_j$ would be different for different $d$ for all $j=1,2$ and $3$.

To verify the above physical picture, we consider $\Omega/2\pi=2$~MHz, $(\tau(n_{\text{r}}, \tau(n))=(320,95)~\mu$s~\cite{Beterov2009} at room temperature when $(n_{\text{r}},n)=(100,60)$ and various values of $L$ when $d=20~\mu$m. From Fig.~\ref{figure-fidelity}(d) one can find that for different choices of $\c{d}$, $\mathcal{F}_1$ can oscillate due to the interfering nature of different phases in Eq.~(\ref{F1}) which originates from Eq.~(\ref{Rydberg polariton-state-3}). As a result, Fig.~\ref{figure-fidelity}(d) shows that the total fidelity also oscillates. However, for small $L$, the oscillation is less obvious, and the fidelity can be quite large. The largest fidelity for $L/d=1/4, 1/3, 3/4$, and 1 are $0.98, 0.97, 0.80$, and 0.69, respectively.

Though the data in Fig.~\ref{figure-fidelity} indicate that high-fidelity photon gates can be realized with our protocol, very short $L$ may lead to small optical depth. But it is relatively easy to experimentally demonstrate a moderately short $L$, e.g., $L/d=1/3$ as in Fig.~\ref{figure-fidelity}(b), where the optimal total fidelity $\mathcal{F}\approx0.97$ occurs. Thus, it appears that high-fidelity gate is possible with our protocols.

\section{Discussion}\label{sec07}
Among the three protocols proposed in this work, the wait-$\pi$ protocol
is of special interest since it not only can allow the suppression of the motional dephasing, but also can result in opposite read-out of the signal which potentially can enable single-photon router. However, it requires two sets of coupling laser, one for the loading of the signal, one for the retrieval. Therefore, the wait-$\pi$ protocol seems more challenging to be realized in experiments~\cite{Li2025}.

When we compare the $2\mathbb{N}\pi$ protocol and the $\pi$-wait-$\pi$ protocol by looking at Figs.~\ref{figure-Rb-2Npi} and~\ref{figure-Cs-2Npi}, and Figs.~\ref{figure-Rb-piwpi} and~\ref{figure-Cs-piwpi}, it seems that the $\pi$-wait-$\pi$ protocol looks superior. In theory, it is so. However, the switch on and off of laser fields can cause extra noise to enter into the atomic system as analyzed and discussed for experiments on Rydberg-mediated quantum logic gates~\cite{Maller2015}. This effect is not significant in experiments on Rydberg polariton for the accuracy has not yet risen to such a level. But for Rydberg-mediated quantum logic gates where the fidelity has risen to levels over 0.995~\cite{Evered2023}, such effect is significant, which is why for high-fidelity realizations of Rydberg quantum gates usually one pulse of laser excitation was used~\cite{Evered2023,Scholl2023,Ma2023}. Therefore, it is unclear in real experiments whether the $\pi$-wait-$\pi$ protocol or the $2\mathbb{N}\pi$ protocol can bring us a better CE effect.

Besides Rydberg-mediated quantum optics, our theory brings fresh insights into the study of suppressing motional dephasing in atomic, molecular, and optical systems. To understand this, it is useful to briefly compare the theory in this work and previous works on similar topics.

In Ref.~\cite{Rui2015}, suppression of motional dephasing in ground-state polaritons in two ground states $\lvert g\rangle$ and $\lvert s\rangle$ was experimentally demonstrated, where the authors proposed an interesting way to simultaneously realize the transitions $\lvert g\rangle\rightarrow\lvert s\rangle$ and $\lvert g\rangle\leftarrow\lvert s\rangle$ by one Raman laser so as to generate a correct phase change to the polaritons. If we tried to use the method of Ref.~\cite{Rui2015} for Rydberg polaritons, then we should image that the following state
\begin{eqnarray}
\frac{1}{\sqrt{N}} \sum_{j=1}^{N}e^{ik z_j } |r_1r_1\cdots g^{(j)}\cdots r_1r_1r_1 \rangle,\label{comparison}
\end{eqnarray}
must be generated from, e.g, Eq.~(\ref{protocol3-3}). This requires exciting nearly all atoms in the atomic gas from ground to Rydberg states which, unfortunately, may be challenging in the presence of Rydberg blockade. Even if it can be done, severe many-body-induced dephasing~\cite{Bariani2012pra} will occur due to the almost universal two-atom Rydberg interaction between nearly any atom pair in the state of Eq.~(\ref{comparison}). Even so, the method verified in Ref.~\cite{Rui2015} can be very useful for stabilizing polaritons in ground states which is important for the study of quantum optics.

In Ref.~\cite{Firstenberg2021}, motional dephasing in Rydberg polaritons of wave length $\Lambda$ defined in a ground state $\lvert g\rangle$ and a low-lying state $\lvert s\rangle$ in an atomic vapor was experimentally studied by dressing $\lvert s\rangle$ with a low-lying Rydberg state $\lvert R\rangle$ with a dressing wavevector $k_{\text{d}}$, where the suppression occurs when $k_{\text{d}}\Lambda<0$ and $|k_{\text{d}}/\Lambda|\gg1$, i.e., compared to the Doppler shift in $\lvert g\rangle\leftrightarrow\lvert s\rangle$, the Doppler shift in $\lvert s\rangle\leftrightarrow\lvert R\rangle$ must be both opposite and much much larger; this was why in Ref.~\cite{Firstenberg2021} a pair of states $\lvert g\rangle$ and $\lvert s\rangle$ were chosen so that a small $\Lambda|\sim 0.04~\mu m^{-1}$ is with the Rydberg polariton. Unfortunately, for Rydberg polaritons where $|\Lambda|$ is usually over $4\mu m^{-1}$ as in Tables~\ref{table1} and~\ref{table2}, it looks challenging to get the method of Ref.~\cite{Firstenberg2021} to work. However, the work in Ref.~\cite{Firstenberg2021} is interesting~\cite{li2024}.

In Ref.~\cite{kurzyna2024}, an ac Stark lattice modulation was used to suppress the dephasing of Rydberg polariton. The mechanism in Ref.~\cite{kurzyna2024} is motivated by that if a sub-microsecond linear phase modulation were available, then the position-dependent phase term $e^{ik z_j }$ in Eq.~(\ref{protocol3-3}) would be removed. However, when understood as removable by an ac Stark shift upon the application of an external laser field, $k z_j$ would be extremely large since the wavevector of a Rydberg polariton is usually over $4\mu m^{-1}$ as in Tables~\ref{table1} and~\ref{table2}. Therefore Ref.~\cite{kurzyna2024} proposed off-resonant ac-Stark lattice modulation by using a sinusoidal phase modulation, where the second-order term in the modulated Rydberg Rydberg polariton can be canceled, leading to an upper bound of the retrieval efficiency below 0.06. It would be interesting to explore the novel method of Ref.~\cite{kurzyna2024} to see whether the dephasing of the full Rydberg polariton can be canceled.

\section{Conclusions}\label{sec08}
We have presented three protocols to almost completely suppress the motional decoherence of the Rydberg polariton. All the three protocols depend on appropriately choosing a configuration of the loading of the signal photon, and that the wavevector, $k$, of the Rydberg polariton is smaller than the wavevector, $k_{\text{\tiny{CE}}}$, of a two-photon Rabi frequency for the Raman transition between $|r_1\rangle$ and $|r_{2}\rangle$.  
$|r_1\rangle$ is the Rydberg state in the initially loaded Rydberg polariton, while $|r_2\rangle$ is a nearby Rydberg state. Our protocol
requires large enough Rabi frequency $\Omega_{\text{\tiny CE}}$ for the Raman transition. But for both cesium and rubidium, we find that the motional dephasing is nearly completely suppressed with a moderate, experimentally realizable $\Omega_{\text{\tiny CE}}/2\pi=2$~MHz in a cold gas of temperature around $10~\mu$K.

The essence of the protocols is that by using two counter-propagating laser fields to transfer the state between $|r_1\rangle$ and $|r_{2}\rangle$ via an appropriate largely-detuned low-lying state, the phase change of the Rydberg polariton can approach the correct value, and simultaneously the Rydberg blockade persists between the Rydberg polariton and another Rydberg atom or Rydberg polariton. An adiabatic argument, supported by numerical calculations, suggests that the proposed scheme can enhance the motional coherence time by orders of magnitude, effectively leaving the Rydberg-state decay as the only fundamental source of dephasing of the Rydberg polariton. Therefore, our theory does not bring in any extra decoherence effect to the system, but only induces a nearly correct phase to the Rydberg polariton.

Because many applications with Rydberg polariton are sensitive to its dephasing, a long coherence time can greatly improve their performance. Our theory brings hope to conquer the issue of fast dephasing of Rydberg polariton, and sheds light on the study of single-photon quantum nonlinear optics with Rydberg atoms.

\section*{ACKNOWLEDGMENTS}
X.F.S. and Y.L. are grateful to Li You and T. A. B. Kennedy for fruitful discussions, and are supported by the National Natural Science Foundation of China under Grant No. 12074300 and the Innovation Program for Quantum Science and Technology under Grant No. 2021ZD0302100. Y.J. and J.Z. are supported by
the National Natural Science Foundation of China under Grant Nos. 62175136, 12241408, 12120101004, and U2341211, and the
Fundamental Research Program of Shanxi Province under Grant No. 202303021224007.

\appendix

\section{Definition of operators in Eq.~(\ref{Heisenberg01})}\label{appendixA0}
In this Appendix we show the detailed definition of the operators in Eq.~(\ref{Heisenberg01}) following Ref.~\cite{Gorshkov2007}. In a quasi one dimensional atomic medium along $\mathbf{z}$ of length $L$, we can divide the atomic medium into thin slices along $\mathbf{z}$, and the number of atoms in each slice is $N_z$. The operators in Eq.~(\ref{Heisenberg01}) are given by
\begin{eqnarray}
 \hat{p}(z,t) &=& \frac{N}{N_z} \sum_{j=1}^{N_z} (\lvert e\rangle\langle g\rvert)_j e^{-i\omega_{eg}(t-z_j/c)},\nonumber\\
 \hat{s}(z,t) &=& \frac{N}{N_z} \sum_{j=1}^{N_z} (\lvert r_1\rangle\langle g\rvert)_j e^{-i\omega_{r_1g}(t-z_j/c)},\nonumber\\
 \hat{\mathcal{E}}(z,t) &=& \sqrt\frac{L}{2\pi c} e^{-i\omega_{eg}(t-z_j/c)}\int d\omega \hat{a}_\omega(t)   e^{i\omega z_j/c},\nonumber\\
\end{eqnarray}
where $N$ is the total number of atoms in the atomic medium, and $N_z\gg1$ though the slice is thin so that the operators are continuous and slowly varying, $\omega_{eg}$~($\omega_{r_1g}$)
are the energy difference between $\lvert e (r_1)\rangle$ and $\lvert g\rangle$ divided by $\hbar$. The Langevin noise operators $\hat{F}$ characterizes the vacuum and thermal field which couple to the atomic transitions. Because the noise is from vacuum, we have $\langle\hat{F(z,t)}\rangle=0$ and $\langle\hat{F}(z,t)\hat{F}^\dag(z',t')\rangle=L\delta(z-z')\delta(t-t')$. In principle there is also a Langevin noise in the third equation of Eq.~(\ref{Heisenberg01}), but we have ignored it since the coupling of vacuum and thermal noise to the Rydberg state is much weaker compared to the coupling to the low-lying state $\lvert e\rangle$.

\section{Retrieval efficiency of Rydberg polariton }\label{appendixA}
Here, we derive Eq.~(\ref{errorOur}) by using the method in Ref.~\cite{Jenkins2012}. For a spatially uniform sample with $N$ atoms and when the beam waist is much larger than the transverse size of the sample, the creation operator of the prepared Rydberg polariton is
\begin{eqnarray}
\hat{S}^\dag &=& \frac{1}{\sqrt{N}}\sum_{j=1}^{N}e^{ikz_j} (|r_1\rangle\langle g|)_j,
\end{eqnarray}
so that the state of the Rydberg polariton is $\hat{S}^\dag |g,g,\cdots,g\rangle$ after the loading, where $|g,g,\cdots,g\rangle\equiv|g\rangle_1\otimes|g\rangle_2\otimes\cdots \otimes|g\rangle_N$. Here, we did not show the coordinates of atoms in the wavefunction since the phase matching concerns the internal state of the atoms. Because of the random motion of the atoms in the sample, the phase-coherent Rydberg polariton upon retrieval is given by
\begin{eqnarray}
\hat{\mathscr{S}}^\dag(t) &=&\frac{1}{\sqrt{N}} \sum_{j=1}^{N}e^{ik(z_j+v_jt)} (|r_1\rangle\langle g|)_j,
\end{eqnarray}
where $v_j$ is the speed of the atom along the propagation direction of the fields used in the preparation of the Rydberg polariton. According to Eq.~(10) of Ref.~\cite{Jenkins2012}, the retrieval efficiency, if not using our theory by the CE laser fields, is given by
\begin{eqnarray}
  \eta_0(t) &=&\left|\langle[ \hat{\mathscr{S}}(t),~\hat{S}^\dag  ]\rangle \right|^2\nonumber\\
  &=&\frac{1}{N}\sum_{j=1}^N \left|\langle e^{-ikv_jt}\left( |g\rangle\langle g|-|r_1\rangle\langle r_1|\right)_j\rangle \right|^2,\label{eqA3}
\end{eqnarray}
where $\langle{\cdots}\rangle$ denotes ensemble average. Because there is only one excitation in the system, the expectation value of $|g\rangle\langle g|-|r_1\rangle\langle r_1|$ for any $j$ is $1-1/N$. For $N\gg1$, we have
\begin{eqnarray}
  \eta_0(t) &\approx&\frac{1}{N}\sum_{j=1}^N \left|\langle e^{-ikv_jt}\rangle \right|^2.
\end{eqnarray}
For all atoms in the system, the temperature is the same, thus the above equation simplifies to 
\begin{eqnarray}
  \eta_0(t) &=& \left|\langle e^{-ikv_jt}\rangle \right|^2\nonumber\\
  &=& \left| \int \mathscr{G}(v) e^{-ik vt}  dv \right|^2,\label{eqAppendxieta0}
\end{eqnarray}
 where the second line is derived in that the statistical average over the $N$ atoms in the atomic medium at the same time is equal to averaging over a single atom by the velocity distribution. This works because the thermal motion of the atoms in the atomic medium is statistically ergodic. In numerical simulation, we have tested that sampling the atom locations $z_j$ and the velocity distributions yield the same result from sampling only the velocity distribution.

In our method, the excitation by the CE laser fields effectively changes the creation operator of the prepared Rydberg polariton to
\begin{eqnarray}
\hat{\mathcal{S}}^\dag &=& \frac{1}{\sqrt{N}}\sum_{j=1}^{N}\mathcal{T}e^{-i \int_0^t \hat{H}(t')dt' }e^{ikz_j} (|r_1\rangle\langle g|)_j,\label{eqA6}
\end{eqnarray}
where we should bare in mind that the Hamiltonian $\hat{H}$ for different atoms are different because of the different spatial locations; but the dynamics in response to the CE laser fields is similar. As in the above derivation, the average over the different speeds of the different atoms in the gas can be replaced by the average via the distribution of speed of one atom; insertion of Eq.~(\ref{eqA6}) into Eq.~(\ref{eqA3}) leads to Eq.~(\ref{errorOur}):
\begin{eqnarray}
  \eta(t) &=&\left|\langle [ \hat{\mathscr{S}}(t),~\hat{\mathcal{S}}^\dag  ]\rangle \right|^2\nonumber\\
  &\approx& \left|  \int \mathscr{G}(v) \langle r_1| e^{-ik vt}\mathcal{T}e^{-i \int_0^t \hat{H}(t')dt' }   |r_1\rangle dv \right|^2,\label{eqA7}
\end{eqnarray}
where we have assumed that for different $z_j$, the average over the Maxwell distribution of $v_j$ is the same, which was tested numerically.

%

\end{document}